\documentclass[iop]{emulateapj}
\usepackage[usenames,dvipsnames]{color}

\slugcomment{Accepted to the Astrophysical Journal}

\begin{document}

\shorttitle{Atmospheric circulation of WASP-43b}
\shortauthors{Kataria et al.}

\title{The atmospheric circulation of the hot Jupiter WASP-43b: Comparing three-dimensional models to spectrophotometric data }

\author{Tiffany Kataria}
\affil{\it{Department of Planetary Sciences and Lunar and Planetary Laboratory, The University of Arizona,
Tucson, AZ 85721}}
\affil{\it{Astrophysics Group, School of Physics, University of Exeter, Stocker Road, Exeter EX4 4QL, UK}}
\email{\bf{tkataria@astro.ex.ac.uk}}

\author{Adam P. Showman}
\affil{\it{Department of Planetary Sciences and Lunar and Planetary Laboratory, The University of Arizona,
Tucson, AZ 85721}}

\author{Jonathan J. Fortney}
\affil{\it{Department of Astronomy \& Astrophysics, University of California, Santa Cruz, CA 95064}}

\author{Kevin B. Stevenson}
\affil{\it{Department of Astronomy and Astrophysics, University of Chicago, Chicago, IL 60637}}
\affil{\it NASA Sagan Fellow}

\author{Michael R. Line}
\affil{\it{Department of Astronomy \& Astrophysics, University of California, Santa Cruz, CA 95064}}

\author{Laura Kreidberg}
\affil{\it{Department of Astronomy and Astrophysics, University of Chicago, Chicago, IL 60637}}

\author{Jacob L. Bean}
\affil{\it{Department of Astronomy and Astrophysics, University of Chicago, Chicago, IL 60637}}
\affil{\it Alfred P. Sloan Research Fellow}
\and

\author{Jean-Michel D\'esert}
\affil{\it{CASA, Department of Astrophysical and Planetary Sciences, University of Colorado, Boulder, CO 80309}}

\begin{abstract}

The hot Jupiter WASP-43b (2 $\mathrm{M_{J}}$, 1 $\mathrm{R_J}$, $\mathrm{T_{orb}}=$ 19.5 hr) has now joined the ranks of transiting hot Jupiters HD 189733b and HD 209458b as an exoplanet with a large array of observational constraints. Because WASP-43b receives a similar stellar flux as HD 209458b but has a rotation rate four times faster and a higher gravity, studying WASP-43b probes the effect of rotation rate and gravity on the circulation when stellar irradiation is held approximately constant. Here we present three-dimensional (3D) atmospheric circulation models of WASP-43b, exploring the effects of composition, metallicity, and frictional drag. We find that the circulation regime of WASP-43b is not unlike other hot Jupiters, with equatorial superrotation that yields an eastward-shifted hotspot and large day-night temperature variations ($\sim$600 K at photospheric pressures). We then compare our model results to HST/WFC3 spectrophotometric phase curve measurements of WASP-43b from 1.12-1.65 microns. Our results show the 5$\times$ solar model lightcurve provides a good match to the data, with a peak flux phase offset and planet/star flux ratio that is similar to observations; however, the model nightside appears to be brighter. Nevertheless, our 5$\times$ solar model provides an excellent match to the WFC3 dayside emission spectrum. This is a major success, as the result is a natural outcome of the 3D dynamics with no model tuning. These results demonstrate that 3D circulation models can help interpret exoplanet atmospheric observations, even at high resolution, and highlight the potential for future observations with HST, JWST and other next-generation telescopes.  

\end{abstract}

\keywords{atmospheric effects, methods: numerical, planets and satellites: atmospheres, planets and satellites: composition, planets and satellites: individual (WASP-43b)} 

\section{Introduction}
As the number of discovered extrasolar planets swells to approximately 1800, exoplanet science has begun to move from a phase of exoplanet detection to exoplanet characterization.  From the direct measurement of the rotation rate of $\mathrm{\beta}$ Pictoris b \citep{snellen+2014}, to the calculation of the carbon-to-oxygen ratio of HR 8799c \citep{konopacky+2013},  scientists have begun to explore the quantitative properties of a range of exotic planets.  While observations of directly-imaged planets provide new avenues to characterize exoplanet atmospheres, spectroscopy and phase curve observations of transiting exoplanets continue to yield impactful results.  These observations allow us to better unravel the dynamics of planets with permanent daysides and nightsides---an interesting dynamical regime that direct imaging cannot probe.  

Most follow-up observations of transiting planets to date have been for hot Jupiters HD 209458b \citep[e.g.,][]{charbonneau+2002,desert+2008,knutson+2008,sing+2008,crossfield+2012,deming+2013} and HD 189733b \citep[e.g.,][]{knutson+2007,charbonneau+2008,pont+2008,majeau+2012,barstow+2014} because of their large planet-star radius ratios, bright host stars enabling better signal-to-noise, and short orbital periods.  WASP-43, another hot Jupiter system, has recently emerged as another system that satisfies such criteria.   Discovered by the Wide Angle Search of Planets (WASP) team in 2011, WASP-43b is a 2.0 $\mathrm{M_J}$ mass planet in a 19.5-hour orbit around a K7 star with a mass of 0.72~$\mathrm{M_{\sun}}$, one of the lowest-mass stars to host a hot Jupiter \citep{hellier+2011}.    Like HD 209458b, WASP-43b has an equilibrium temperature of approximately 1400 K.  Therefore, studying the atmosphere of WASP-43b can provide insights into the influence of orbital distance, rotation rate and gravity on the circulation, under conditions of nearly identical stellar flux.  

Many recent hot Jupiter observations have focused on better characterizing the WASP-43 system.  As part of the WASP team, \cite{gillon+2012} refined the WASP-43 system parameters with 27 transit light curves and seven secondary eclipse light curves using the TRAnsiting Planets and PlanetesImals Small Telescope (TRAPPIST), the Very Large Telescope High Acuity Wide field K-band Imager (VLT HAWK-I), and EulerCAM on the {\it Euler} Swiss telescope, as well as eight radial velocity measurements using the {\it Euler} CORALIE spectrograph.  The observations favor a planet with low planetary heat redistribution from dayside to nightside and no dayside temperature inversion.  These results were later corroborated with ground-based secondary eclipse measurements in the $H$ and $K_s$ bands using the Canada-France-Hawaii Telescope (CFHT) Wide-Field Infrared Camera (WIRCam) \citep{wang+2013} and space-based eclipse measurements at 3.6 and 4.5 $\mathrm{\mu m}$ using the Infrared Array Camera (IRAC) on the {\it Spitzer Space Telescope} \citep{blecic+2014}.  

Recently, \cite{stevenson+2014} and \cite{kreidberg+2014b} observed WASP-43b with unprecedented detail using the Hubble Space Telescope WFC3 G141 grism, which spans 1.12 -- 1.65 microns.  WFC3 has proven to be a workhorse for exoplanet transit spectroscopy, providing robust datasets for a variety of transiting exoplanets \citep[e.g.,][]{deming+2013,mandell+2013,line+2013,wakeford+2013,ehrenreich+2014,knutson+2014a,knutson+2014b,kreidberg+2014a,ranjan+2014,wilkins+2014}.  We observed three phase curves of WASP-43b from 4-7 November 2013, and three transit and two eclipse observations between 9 November and 15 December.  In total, we observed six transits and five eclipses of WASP-43b, providing high precision transmission, dayside emission, and phase-resolved emission measurements.  These observations were decomposed into 15 spectrophotometric channels, which probe distinct pressure levels (approximately 10 mbar to 1 bar, see Stevenson et al. 2014).  Therefore, like previous spectrophotometric observations of variable brown dwarfs \citep[e.g.,][]{buenzli+2012,apai+2013}, the dataset provides us with a heretofore-unseen look at the vertical thermal structure of a hot Jupiter, which allows us to place constraints on the wind and temperature structure as a function of height.  The observations also place strict constraints on the water abundance \citep{kreidberg+2014b}, which allows us to better determine the metallicity and recirculation efficiency of the planet.  Overall, such a robust dataset provides unprecedented constraints for circulation models, quantifying the extent to which other dynamical mechanisms (e.g. magnetic effects, chemical mixing) play a role.  Comparison of such high-quality data with detailed three-dimensional (3D) circulation models that include radiative transfer holds the promise of allowing a significant new understanding of the atmospheric structure and circulation behavior of a hot Jupiter.

In this paper, we aim to further characterize the atmosphere of WASP-43b using a circulation model coupled to a non-grey radiative transfer code, dubbed the Substellar Planetary Atmospheric Radiation and Circulation Model (the SPARC/MITgcm, see below).  We will compare our simulation results, including synthetic light curves and spectra, to the observations of \cite{stevenson+2014} and \cite{kreidberg+2014b}.  In doing so, we can place better constraints on the planet's atmospheric composition, horizontal and vertical temperature structure, and overall dynamical regime.  In Section 2, we describe our model setup and simulations.  In Section 3, we describe the dynamical regime of WASP-43b, exploring the effects of frictional drag and metallicity.  In Section 4 we present synthetic light curves and spectra to compare to our WFC3 observations, noting major similarities and differences.  We also use our model results to comment on future prospects with the James Webb Space Telescope (JWST) and other next-generation telescopes.  In Section 5 we discuss our results, and in Section 6 we present our conclusions.

\section{Model}

\subsection{The SPARC/MITgcm}
To model the atmospheric circulation of WASP-43b, we utilize the SPARC/MITgcm \citep{showman+2009}, a model that couples the MITgcm, a general circulation model (GCM) maintained at the Massachusetts Institute of Technology \citep{adcroft+2004}, with a two-stream implementation of the multi-stream, non-grey radiation code of \cite{marley+mckay_1999}.  Our group has utilized the SPARC/MITgcm to model the atmospheric circulation of a variety of extrasolar planets, including hot Jupiters \citep{showman+2009, showman+polvani_2011, kataria+2013, parmentier+2013, showman+2013}, hot Neptunes \citep{lewis+2010}, and super Earths \citep{kataria+2014}.  The MITgcm solves the primitive equations, valid for stably stratified atmospheres where the horizontal length scales greatly exceed vertical length scales, on a cubed-sphere grid.  The radiative transfer code solves for the upward and downward fluxes through each vertical atmospheric column in the GCM, whose values are used to calculate the heating rate and update the temperature and wind fields in the dynamical core.  We divide the opacities for each atmospheric composition into 11 frequency bins using the correlated-$k$ method \citep{goody+1989,kataria+2013}.  Each bin contains $10^5$ to $10^6$ frequency points.  The SPARC model remains one of the most sophisticated three-dimensional coupled circulation/radiation models used to investigate hot Jupiter circulation and their observational implications; other groups use only dual-band radiative transfer \citep[e.g.,][]{heng+2011,rauscher+menou_2012} or average opacities within bins using only a Planck mean rather than full correlated-$k$ \citep[e.g.,][]{dobbsdixon+agol_2013}.

\begin{deluxetable}{lc}
%\tabletypesize{\scriptsize}
\tablecaption{Properties of WASP-43A/b system, taken from \cite{gillon+2012}.}
\tablewidth{0pt}
\tablehead{
\colhead{Parameter} & \colhead{Value}
}
\startdata
$\mathrm{M_{\star}~(M_{\odot})}$ & 0.717 \\
$\mathrm{R_{\star}~(R_{\odot})}$  & 0.667 \\
$\mathrm{M_P~(M_J)}$ & 2.034 \\
$\mathrm{R_{P}~(R_J)}$  & 1.036 \\
$a$~(AU)  & 0.01526 \\
$\mathrm{T_{orb}=T_{rot}~(days)}$  & 0.81 \\
$g~\mathrm{(m~s^{-2})}$ & 47.0
\enddata
\label{params_table}
\end{deluxetable}

\subsection{WASP-43b model setup}
\begin{figure}
\begin{centering}
\epsscale{.80}
\includegraphics[trim = 0.0in 0.0in 0.0in 0.0in, clip, width=0.45\textwidth]{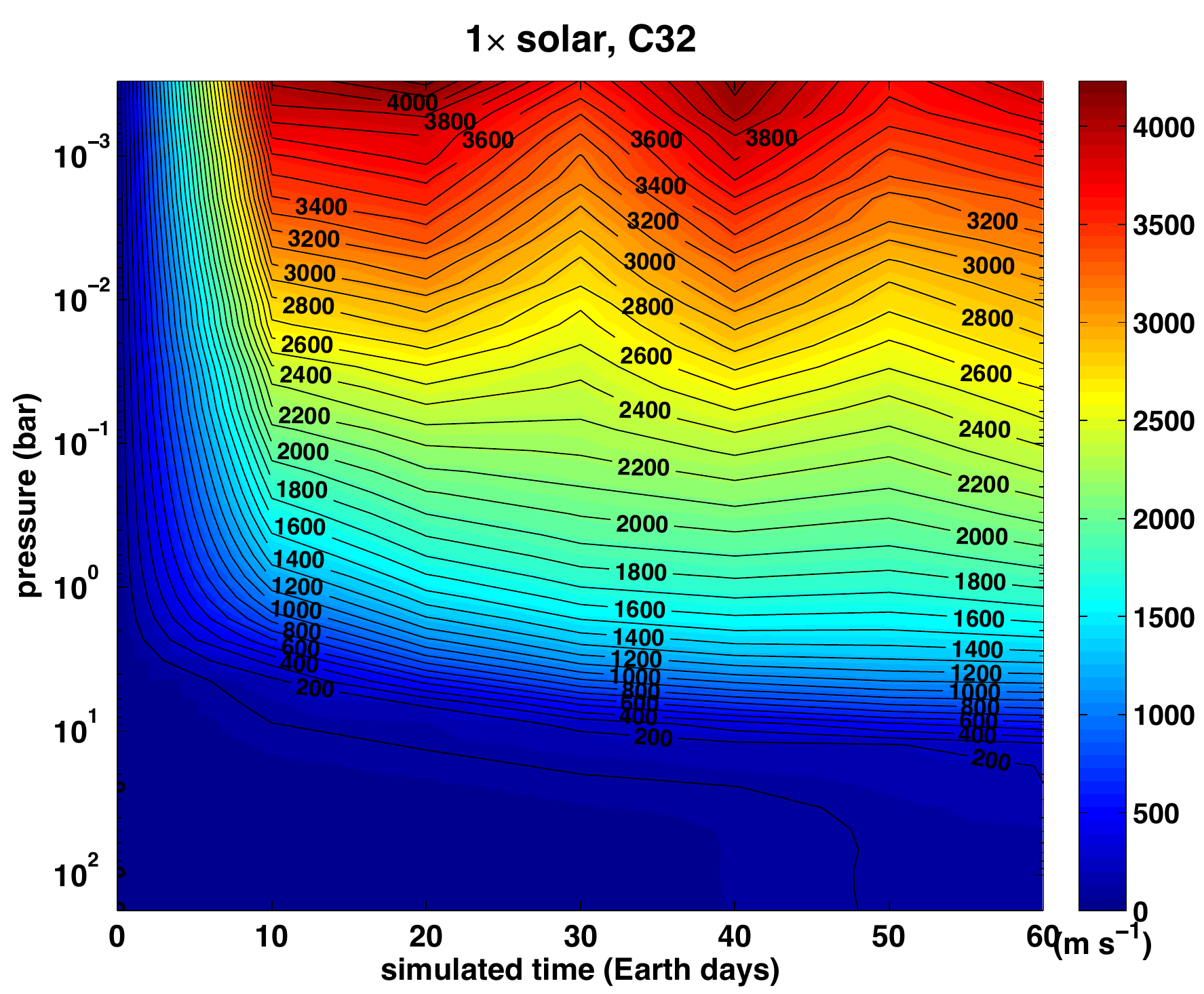}
\caption{Root-mean-squared velocity versus integrated time for the 1$\times$ solar model of WASP-43b at a resolution of C32.}
\label{vrms_plots}
\end{centering}
\end{figure}

For each model integration, we adopt the planetary and stellar parameters of WASP-43b from \cite{gillon+2012}, reproduced in Table~\ref{params_table}.  We assume the specific heat at constant pressure, $c_p$, and $\kappa$, the ratio of the specific gas constant, $R$, to $c_p$, to be values appropriate for a $\mathrm{H_2}$-dominated atmosphere ($1.3 \times 10^4~ \mathrm{J~kg^{-1}~K^{-1}}$ and 2/7, respectively).  We vary the atmospheric metallicity from 1$\times$ to 5$\times$ solar.  To calculate the opacities we assume local chemical equilibrium (accounting for rainout of condensates) at any local pressure and temperature using the elemental abundances in \cite{lodders2003}, and we ignore clouds and hazes.  Because the higher resolution models have proven difficult to run for long integration times, we compare both high- and low-resolution models of WASP-43b.  In the cubed-sphere setup these high and low resolutions are referred to as C32 and C16, corresponding to cell faces of 32 $\times$ 32 and 16 $\times$ 16, respectively, and resolutions of 128 $\times$ 64 and 64 $\times$ 32 in longitude and latitude, respectively.  A plot of the root-mean squared velocity versus model integration time for the C32 model shows that the winds are fairly steady at and above 1 bar (Figure \ref{vrms_plots}).  Each model integration has 40 pressure levels, evenly spaced in log pressure, that extend from a mean pressure of 200 bars at the bottom to 0.2 mbar at the top.  We also use a fourth-order Shapiro filter to maintain numerical stability; this smooths the grid-scale variations while minimally affecting the flow at large scales.

To explore the extent to which Lorentz forces or other magnetic effects might influence the dynamics of WASP-43b, we additionally run three C32, solar metallicity models that include a linear, frictional drag term in the horizontal momentum equation.  Namely, we include the term $-u/\tau_{\rm drag}$ in the zonal momentum equation, and $-v/\tau_{\rm drag}$ in the meridional momentum equation, where $u$ is zonal (east-west) wind and $v$ is meridional (north-south) wind.  Within each simulation, we assume the drag time is constant throughout the domain.  We run models with three different frictional drag time constants: $1\times10^6$ s, $3\times10^5$ s, and $1\times10^5$ s.  While this is a crude treatment of magnetic drag, these models can nevertheless shed light on the influence of drag on the dynamics (see further discussion in Section \ref{drag_section}).  Lastly, we compare our nominal 1$\times$ and 5$\times$ solar models to those with thermal inversions, including TiO and VO in chemical equilibrium.  
 
Each simulation is initialized with zero winds, noting that this choice of initial conditions is not expected to affect the resultant flow field \citep{liu+showman_2013}. We initialize each vertical column of atmosphere using the global-mean radiative-equilibrium temperature-pressure profile, which is calculated using the one-dimensional radiative transfer code of \cite{fortney+2005,fortney+2006,fortney+2008} adapted from \cite{marley+mckay_1999}.  

\begin{figure}
\begin{centering}
\epsscale{.80}
\includegraphics[trim = 0.0in 0.0in 0.0in 0.0in, clip, width=0.45\textwidth]{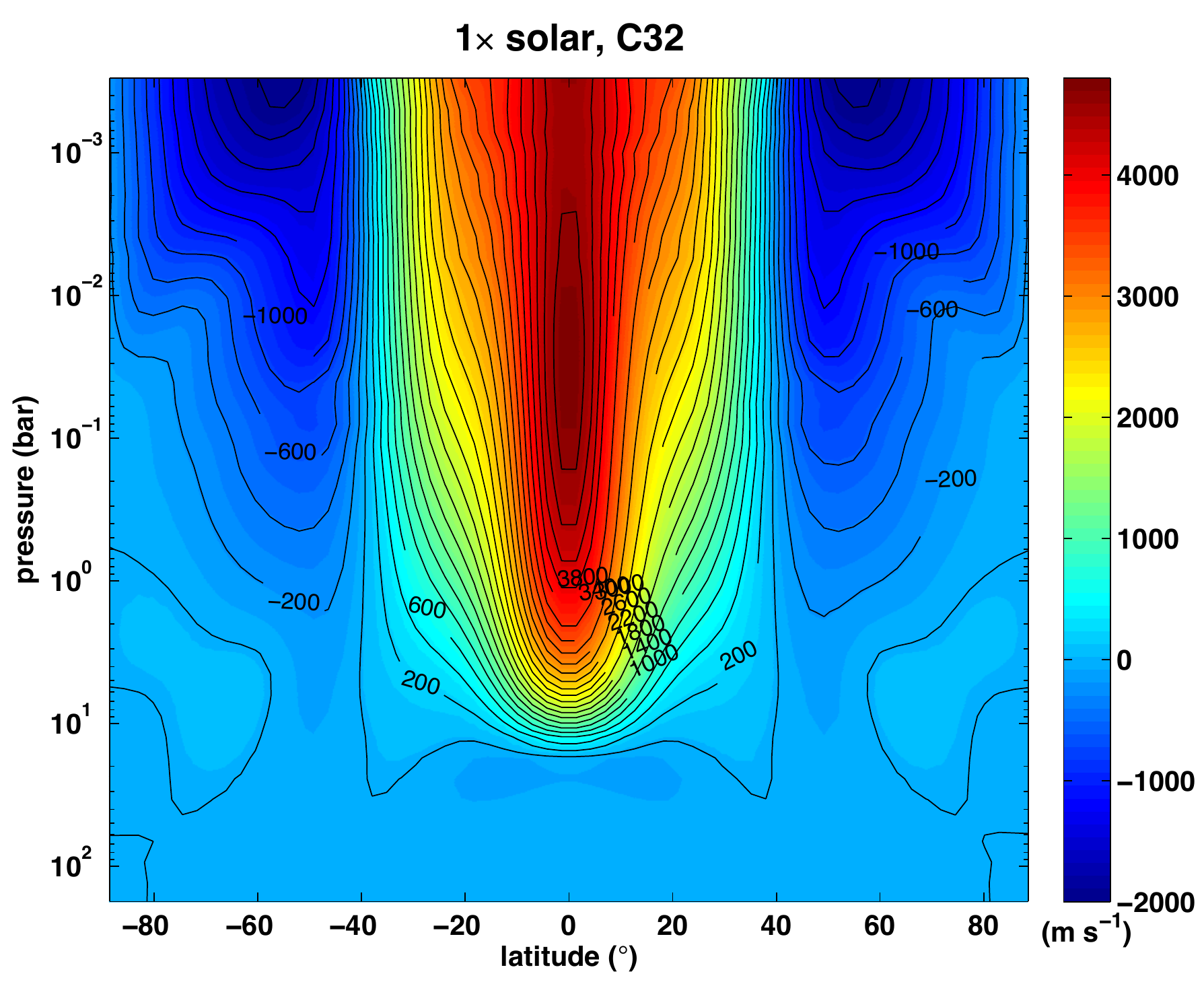}\\
\includegraphics[trim = 0.0in 0.0in 0.0in 0.0in, clip, width=0.45\textwidth]{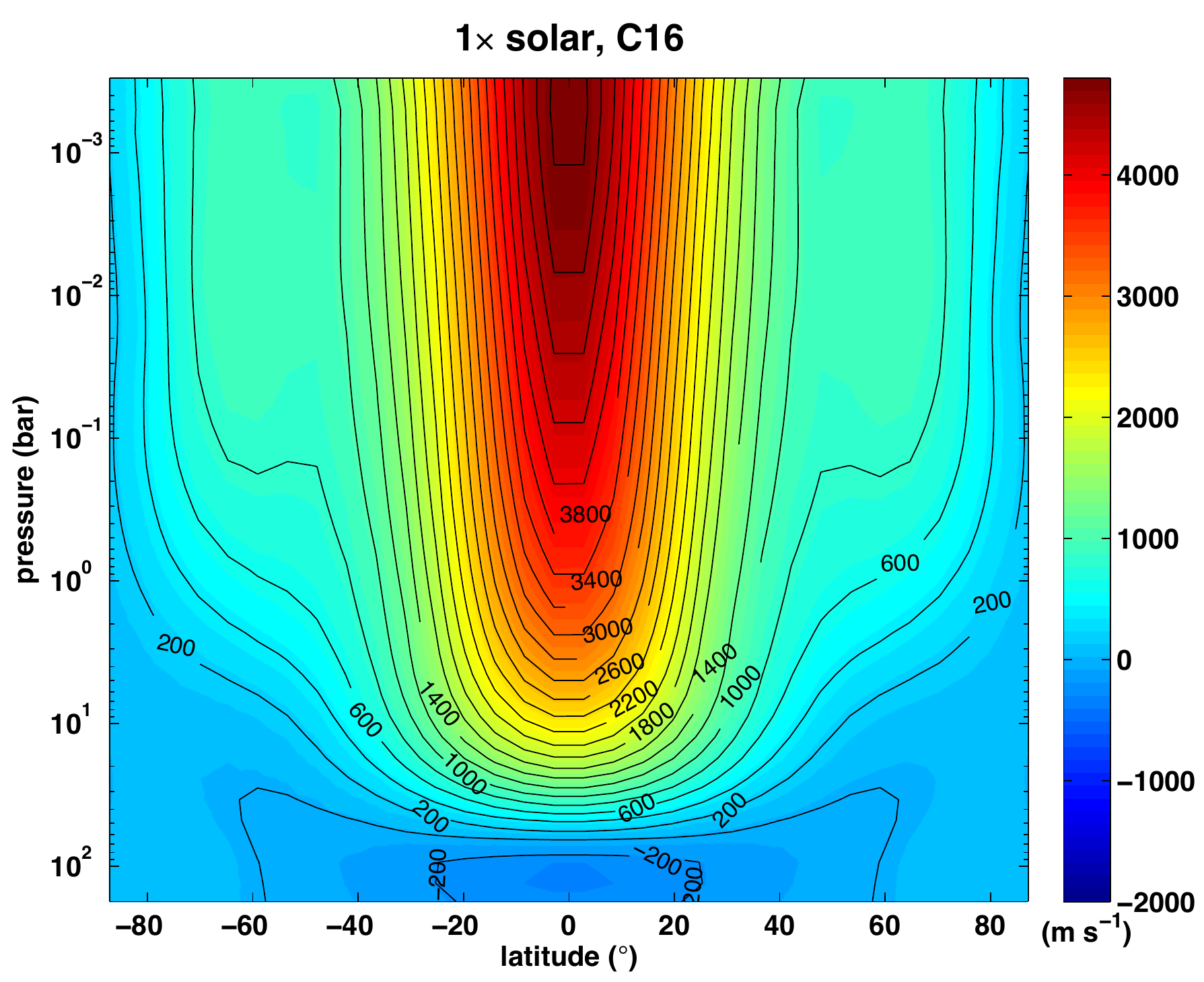}
\caption{Zonal-mean zonal wind profiles of WASP-43b, averaged over an orbit, at two model resolutions: C32 (top) and C16 (bottom).  Both profiles have the same color scale.}
\label{zonalwind_onesolar}
\end{centering}
\end{figure}

\section{Results}
\subsection{1$\times$ solar atmospheric composition}
\subsubsection{Dynamical regime}

Like models of HD 189733b and HD 209458b \citep[e.g.,][]{showman+2009}, eastward equatorial superrotation dominates the circulation of the 1$\times$ solar model of WASP-43b at both high and low resolutions (Figure \ref{zonalwind_onesolar}).  However, the equatorial jet on WASP-43b is much narrower in latitudinal width (only $\sim$40 degrees, extending from +20 to -20 degrees); this is a result of the planet's faster rotation rate, which yields a smaller Rossby deformation radius than the other two planets.  At both high and low resolution, the equatorial jets on WASP-43b extend to a pressure of approximately 10 bars with speeds exceeding 4 $\mathrm{km~s^{-1}}$.  Such fast winds result from the planet's small orbital distance, which yields strong day-night forcing \citep{showman+polvani_2011}.  Additionally, the atmosphere exhibits westward flow near the poles in the C32 models, which is not captured by the C16 models due to the lower resolution (Figure \ref{zonalwind_onesolar}, top panel).  Because the sound speed for WASP-43b is $\sim$2-3 $\mathrm{km~s^{-1}}$, the equatorial jet is therefore supersonic.  While we do not treat the formation of shocks directly in our models, we note that more constraints are needed on this important circulation problem \citep[see discussion in][]{kataria+2013}.

\begin{figure*}
\begin{centering}
\epsscale{.80}
\includegraphics[trim = 0.0in 0.0in 0.0in 0.0in, clip, width=0.4\textwidth]{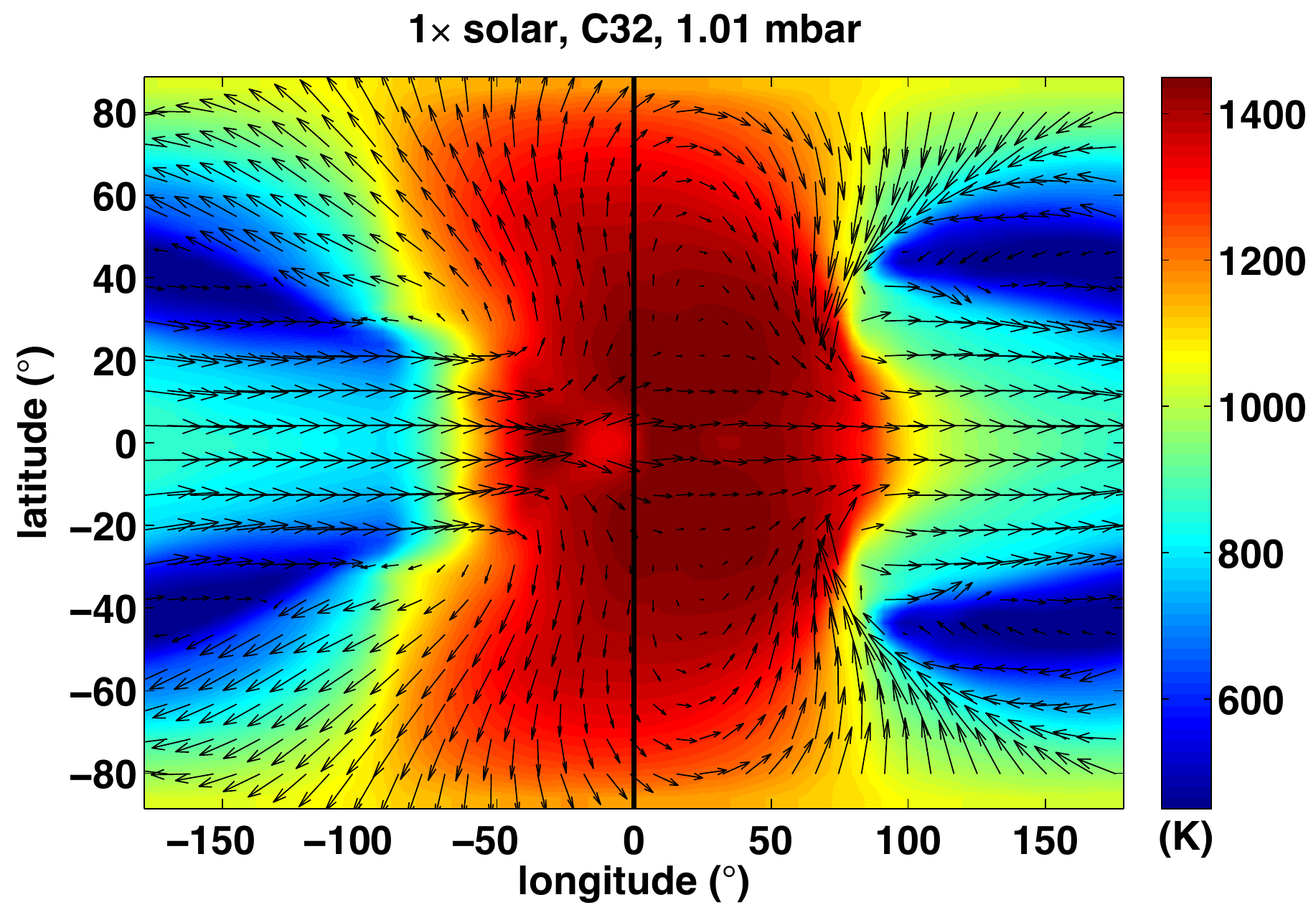}
\includegraphics[trim = 0.0in 0.0in 0.0in 0.0in, clip, width=0.4\textwidth]{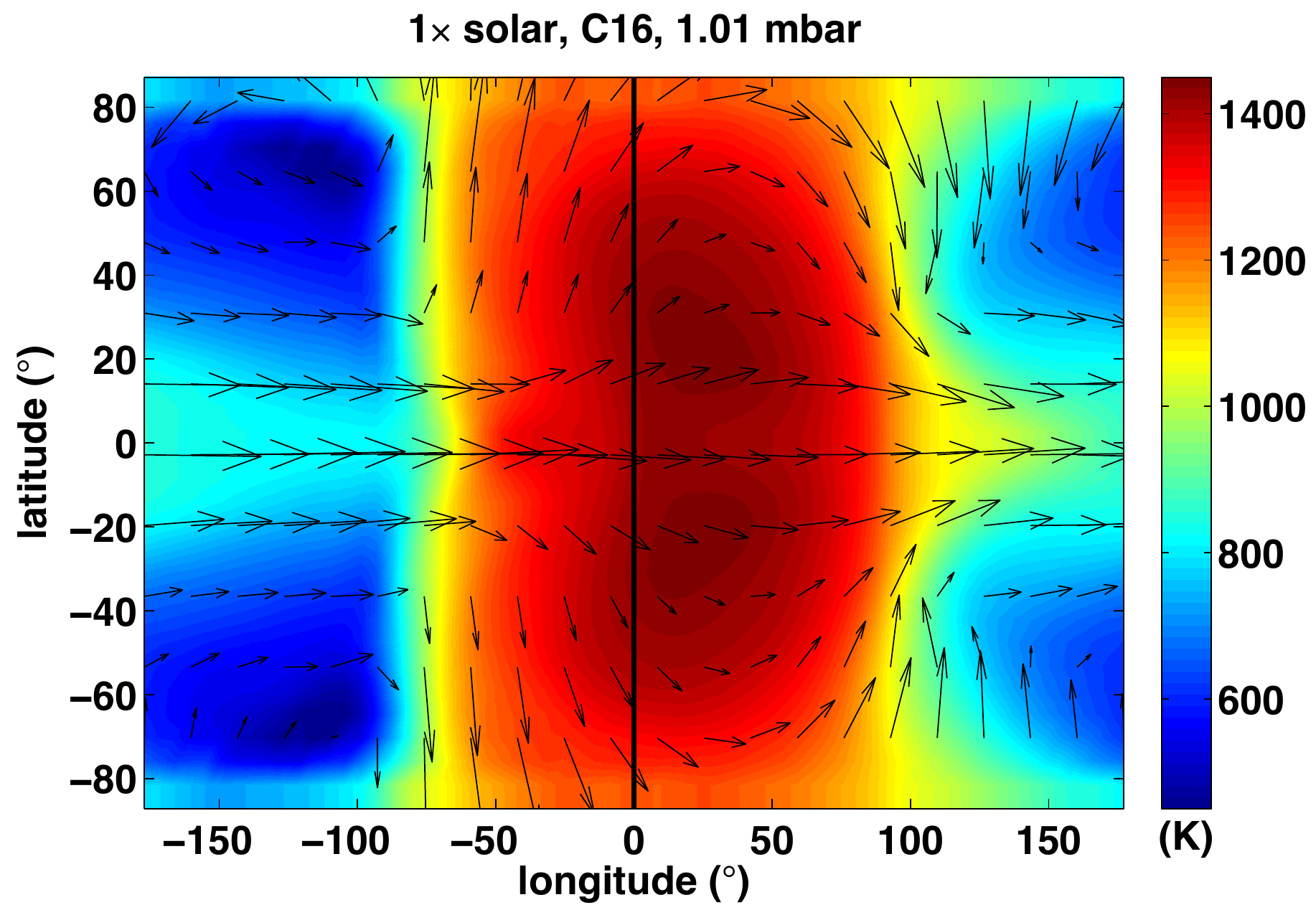}\\
\includegraphics[trim = 0.0in 0.0in 0.0in 0.0in, clip, width=0.4\textwidth]{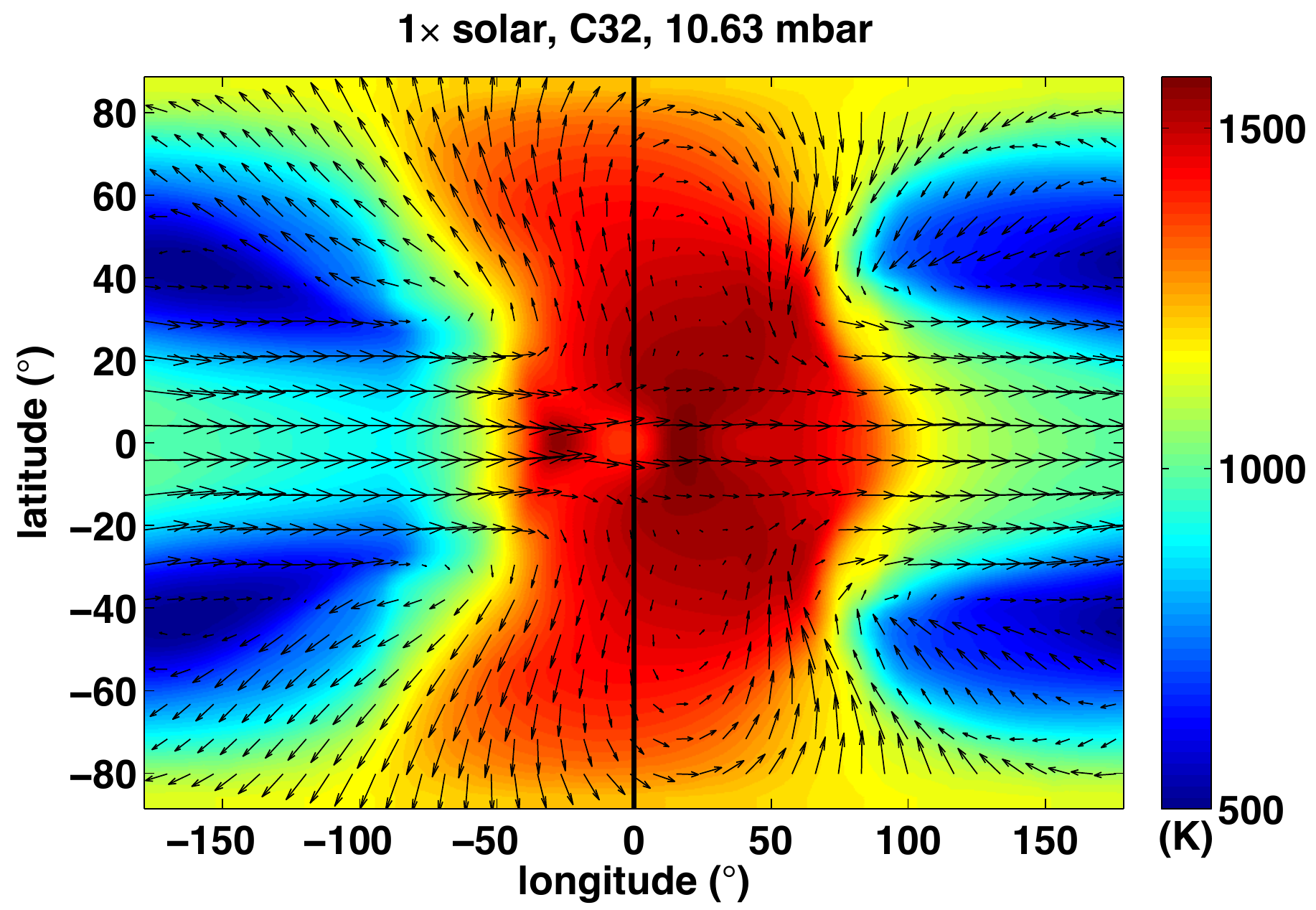}
\includegraphics[trim = 0.0in 0.0in 0.0in 0.0in, clip, width=0.4\textwidth]{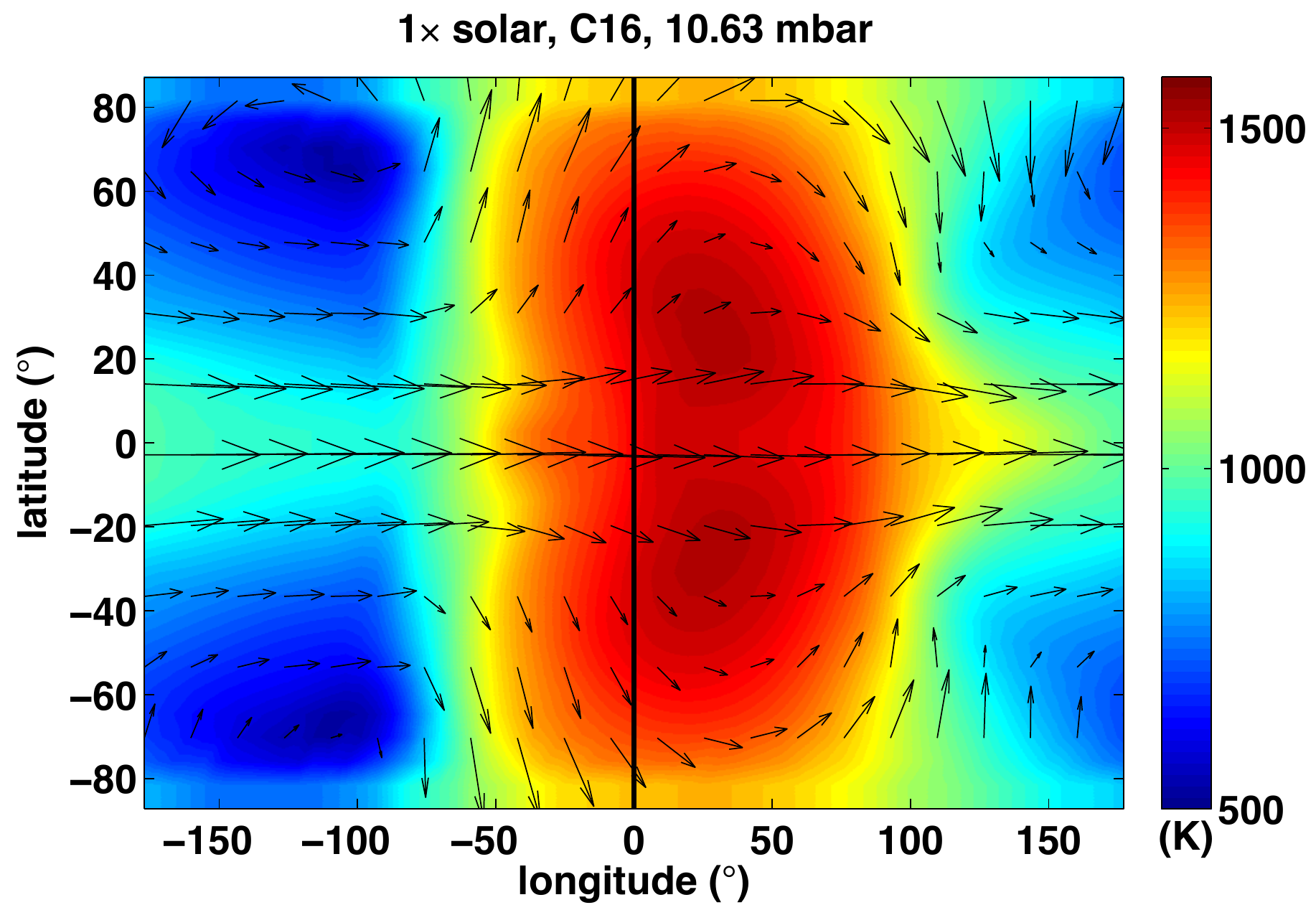}\\
\includegraphics[trim = 0.0in 0.0in 0.0in 0.0in, clip, width=0.4\textwidth]{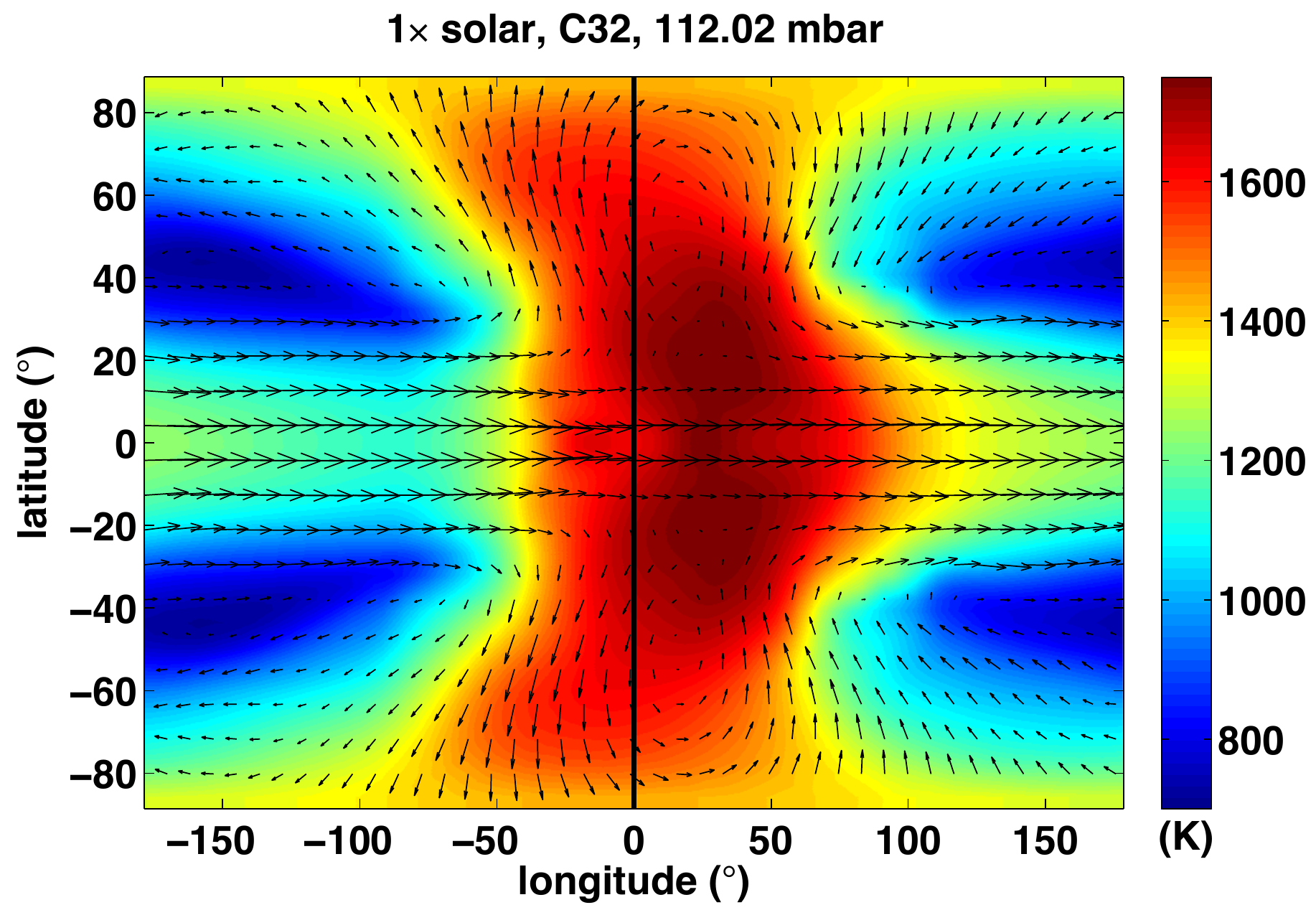}
\includegraphics[trim = 0.0in 0.0in 0.0in 0.0in, clip, width=0.4\textwidth]{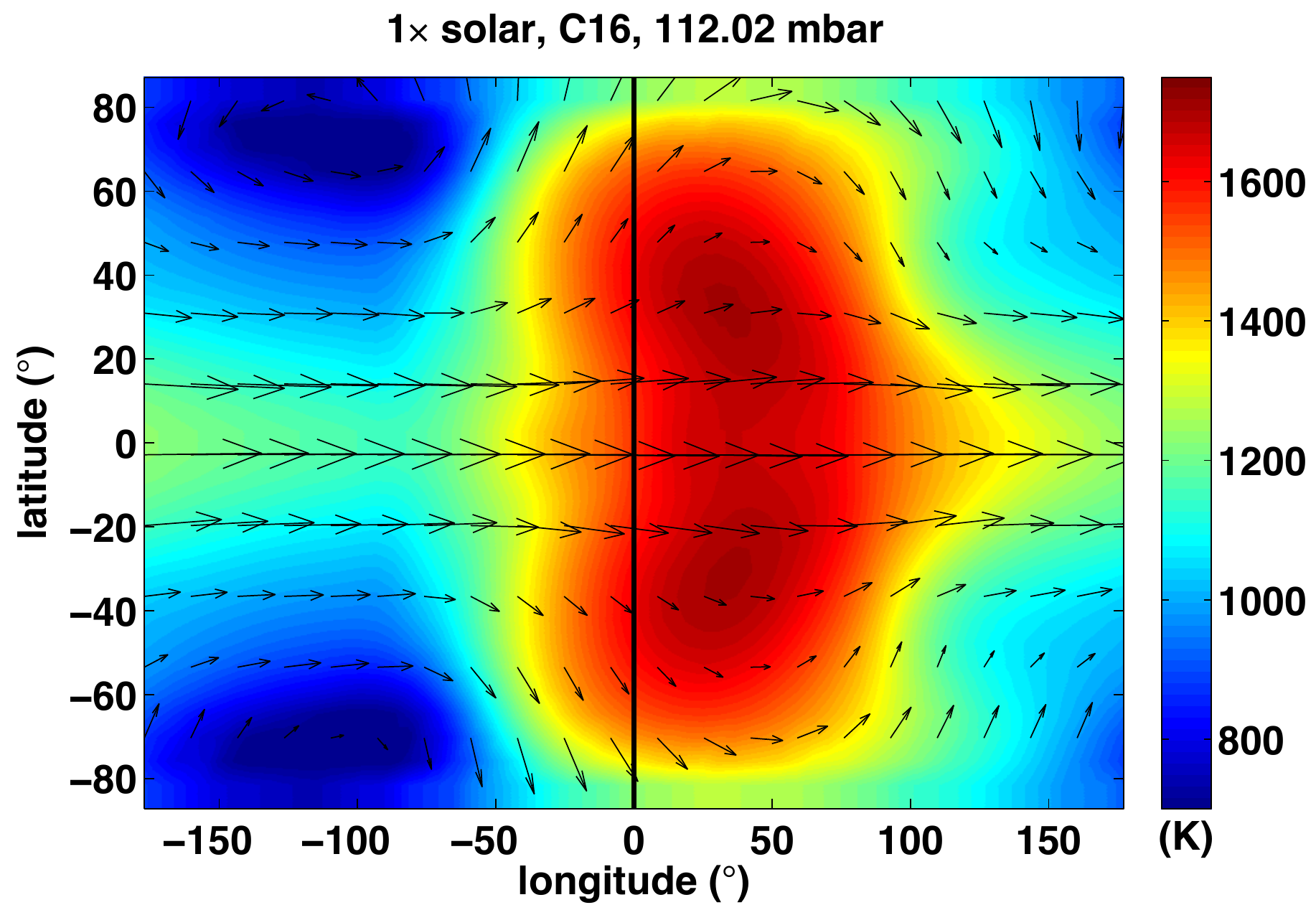}\\
\includegraphics[trim = 0.0in 0.0in 0.0in 0.0in, clip, width=0.4\textwidth]{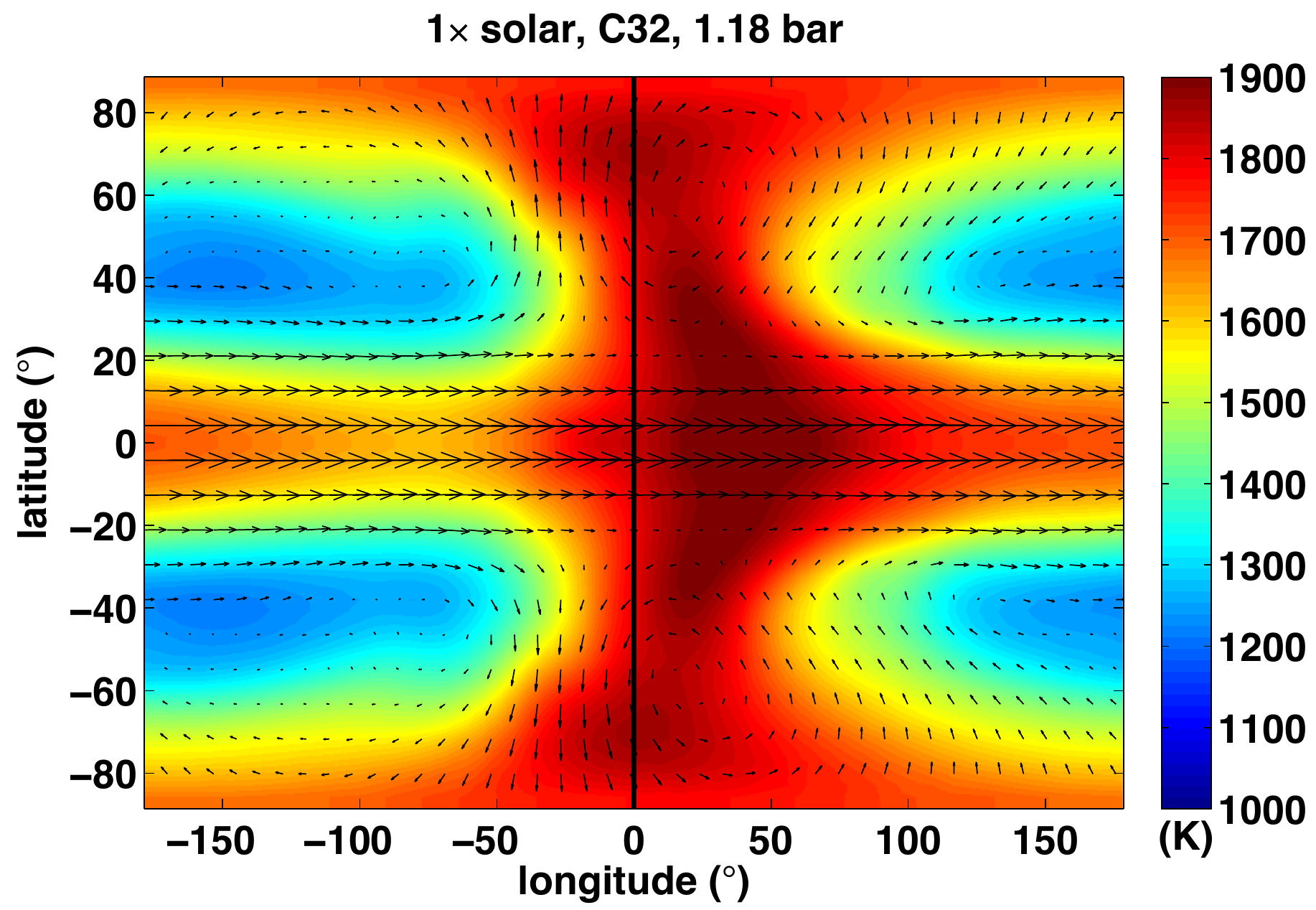}
\includegraphics[trim = 0.0in 0.0in 0.0in 0.0in, clip, width=0.4\textwidth]{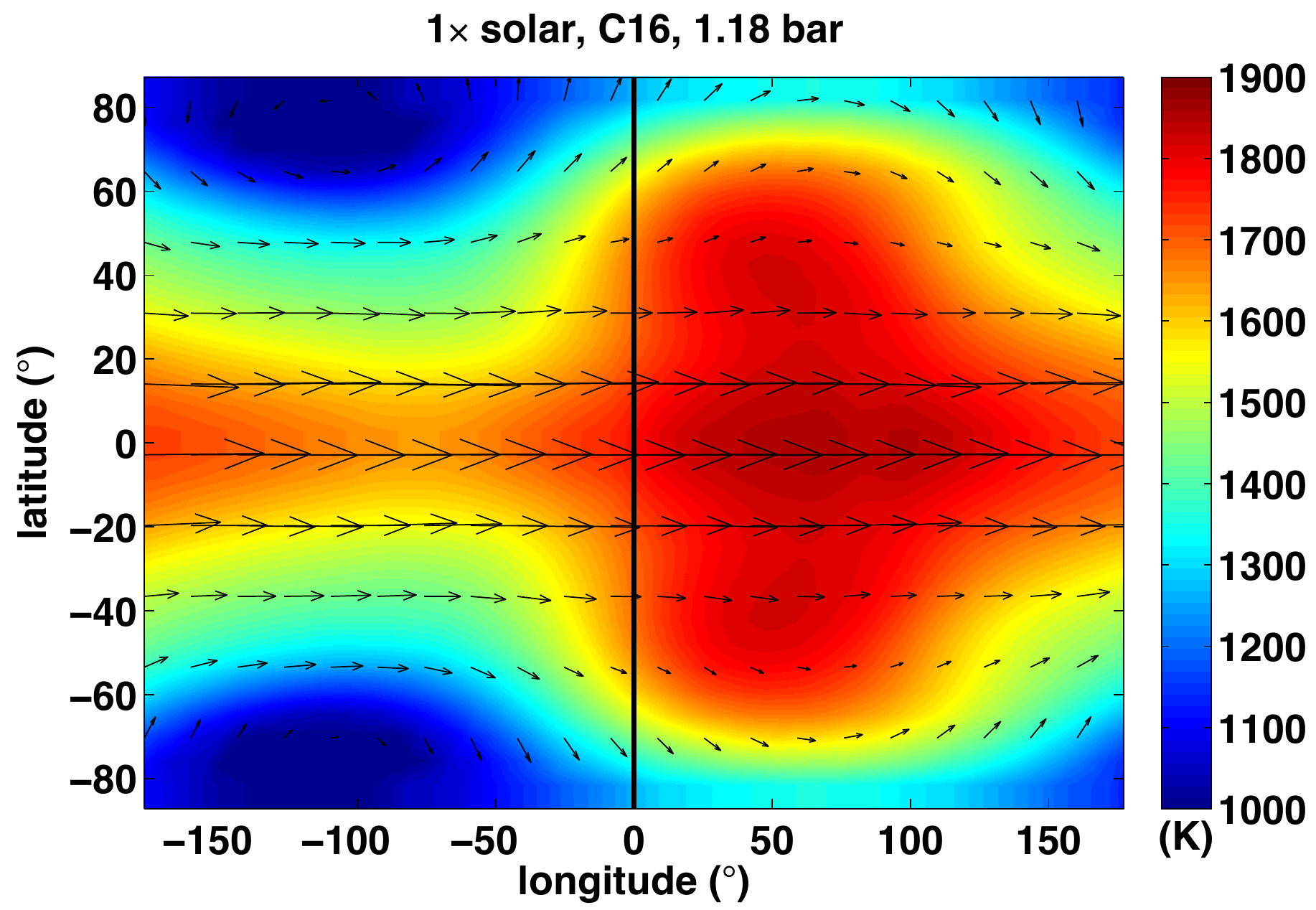}
\caption{Wind and temperature profiles of WASP-43b with an atmospheric composition of 1$\times$ solar, at two model resolutions: C32 (left column) and C16 (right column). The profiles are compared at four different pressure levels, from top to bottom: 1 mbar, 10 mbar, 100 mbar and 1 bar.  The black line in each profile denotes the longitude of the substellar point.  Each row of profiles has the same color scale. }
\label{windtemp_onesolar}
\end{centering}
\end{figure*}

\subsubsection{Vertical and horizontal temperature structure}
The atmospheric equatorial superrotation of WASP-43b causes the hottest regions to be shifted eastward of the substellar point over a wide range of pressures, from 0.2 mbar to $\sim$1-3 bars.  This is seen at both resolutions in wind and temperature profiles at pressures of 1 mbar, 10 mbar, 100 mbar and 1 bar (Figure \ref{windtemp_onesolar}).  At the pressures probed by observations, the temperature contrast from dayside to nightside is $\sim$600-800 K.  This is slightly larger than day-night contrasts at photospheric pressures of HD 209458b.  

When comparing the high and low resolution models, it is clear that the high resolution models capture more small-scale structure, particularly in longitude.  At higher resolution, the chevron shape and eddy velocity tilts of the hot spot are well-defined at all pressure levels at both the leading and trailing edges, while at low resolution the hot spot is much broader and retains less of a chevron shape.  On the nightside, the temperature and wind structure differs most in latitude; cold regions in C32 are narrower and limited to lower latitudes, while the cold regions at C16 are wider and extend to the pole. 

Regardless of the differences between the high and low resolution simulations, both models show that the global circulation regime and temperature structure of WASP-43b is consistent with that of other tidally-locked planets, including hot Jupiters HD 189733b and HD 209458b. Therefore, we would expect that phase curve observations of WASP-43b should resemble those of other planets, with a peak infrared (IR) flux that occurs before secondary eclipse when the dayside rotates into view \citep{showman+guillot_2002}.  We will explore the similarities and differences between the models and our observations in Section 4.

\subsubsection{Sensitivity to drag}
\label{drag_section}

Because WASP-43b orbits so closely to its star, it is possible that magnetic effects may play a role in its atmospheric circulation.  Modeling these effects self-consistently and rigorously is challenging \citep{rogers+showman_2014} and to date has only been attempted in models with other simplifications, in particular the use of Newtonian cooling rather than radiative transfer.  Nevertheless, it has been suggested that the Lorentz force may qualitatively act to brake the winds, thereby limiting wind speeds \citep{perna+2010}.  

\begin{figure}
\begin{centering}
\epsscale{.80}
\includegraphics[trim = 0.0in 0.0in 0.0in 0.0in, clip, width=0.45\textwidth]{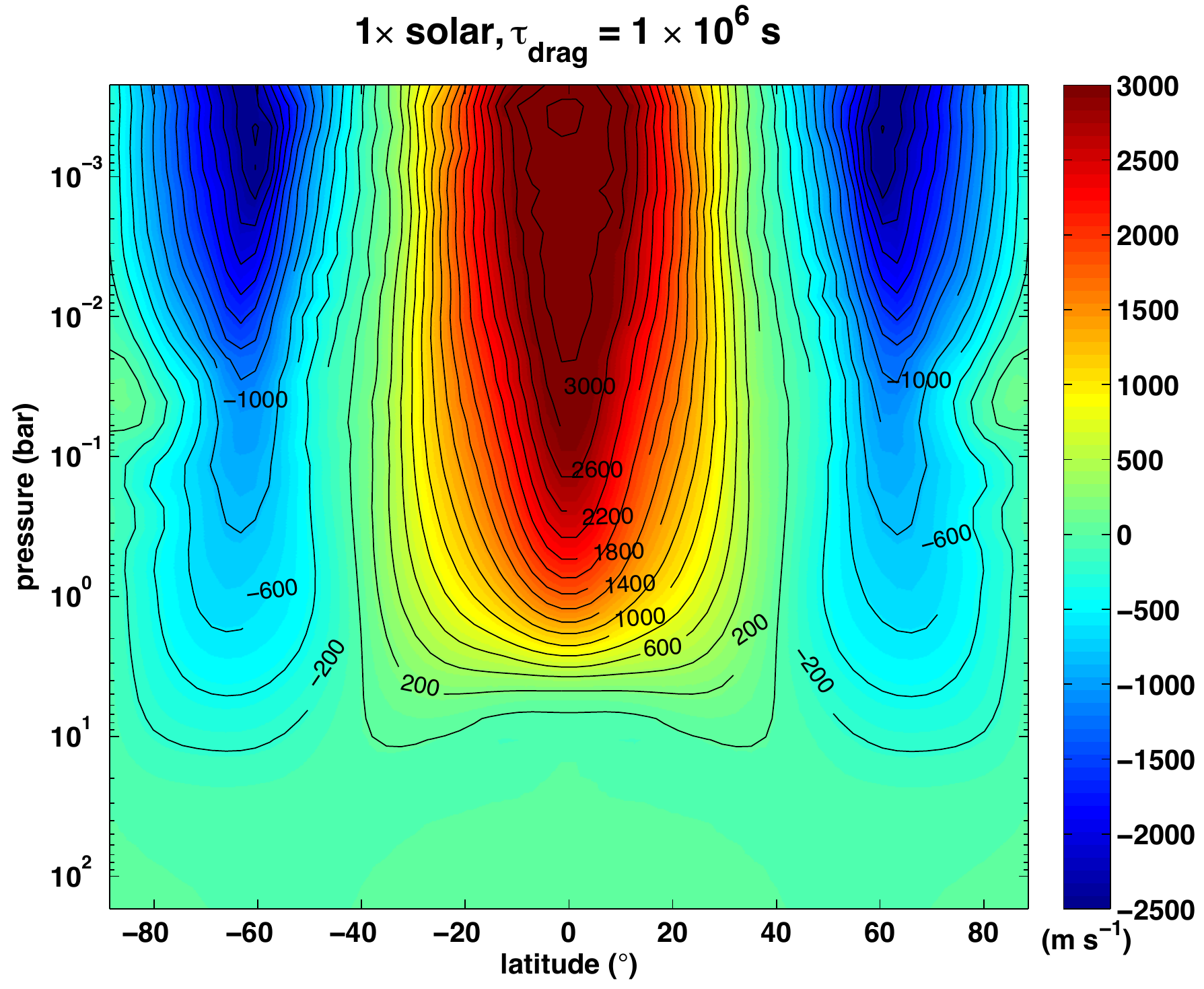}\\
\includegraphics[trim = 0.0in 0.0in 0.0in 0.0in, clip, width=0.45\textwidth]{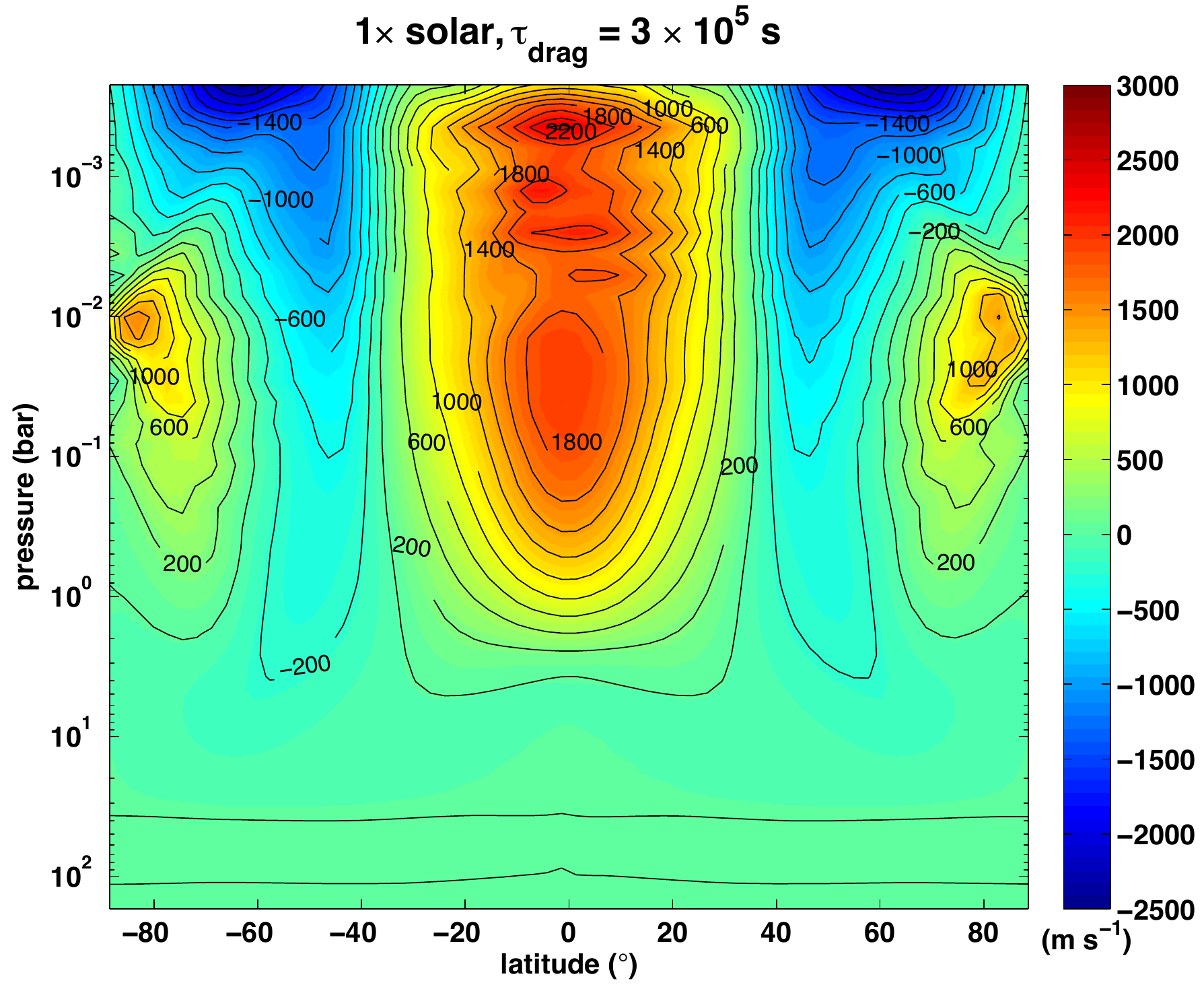}\\
\includegraphics[trim = 0.0in 0.0in 0.0in 0.0in, clip, width=0.45\textwidth]{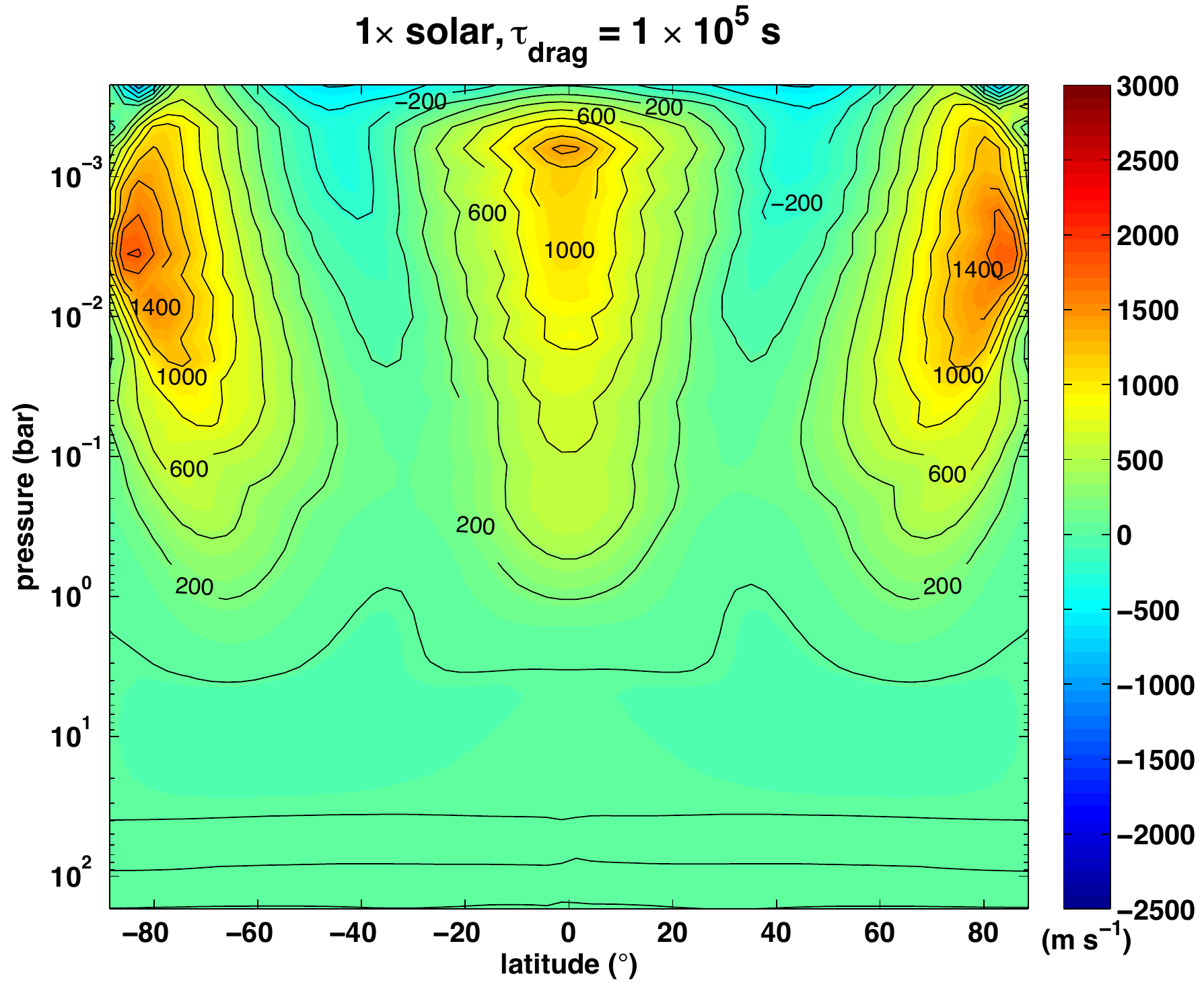}
\caption{Zonal-mean zonal wind profiles of WASP-43b with an atmospheric composition of 1$\times$ solar and a resolution of C32 at varying degrees of frictional drag, increasing from top to bottom: $\tau_{drag}=1\times10^6$ s, top; $\tau_{drag}=3\times10^5$ s, middle; $\tau_{drag}=1\times10^5$ s, bottom.  Each profile has the same color scale. }
\label{zonalmean_drag}
\end{centering}
\end{figure}

Here, we test the sensitivity of this effect by adding frictional drag everywhere throughout the domain, with a specified value of the frictional drag time constant, $\mathrm{\tau_{drag}}$, in each model.  We explore frictional drag time constants ranging from $1\times10^6$ s (weak drag), to $3\times10^5$ s (intermediate drag), to $1\times10^5$ s (strong drag).  This treatment, of course, is not a replacement for performing fully coupled magnetohydrodynamic calculations in models with realistic radiative transfer (a goal for the future), but nevertheless allows us to understand in an idealized context the dynamical effect that frictional drag exerts on the circulation.  In this set of numerical experiments, we keep $\tau_{drag}$ spatially constant throughout each model, which, although not expected to be realistic, enables a more straightforward interpretation of how drag alters the dynamics.  Figure \ref{zonalmean_drag} plots the zonal-mean zonal wind profiles for each model integration.  As the frictional drag is increased (as $\mathrm{\tau_{drag}}$ becomes shorter), the fastest zonal winds shift poleward---they occur near the equator in the weak-drag models but occur at high latitudes in the strong-drag models.  At $\mathrm{\tau_{drag}=1\times10^6}$ s, the flow looks similar to the drag-free model, with an equatorial jet exceeding 3 $\mathrm{km~s^{-1}}$ and westward flow near the poles.  With intermediate drag ($\mathrm{\tau_{drag}=3\times10^5}$ s), the equatorial winds are still stronger than winds elsewhere on the planet ($\mathrm{ > 2~km~s^{-1}}$), but there is also {\it eastward} flow at the poles, with westward flow aloft at those same latitudes.  With the strongest drag applied, there is only eastward flow at the poles, and the equatorial jet is only $\mathrm{\sim}$1 $\mathrm{km~s^{-1}}$.  

\begin{figure*}
\begin{centering}
\epsscale{.80}
\includegraphics[trim = 0.0in 0.0in 0.0in 0.0in, clip, width=0.25\textwidth]{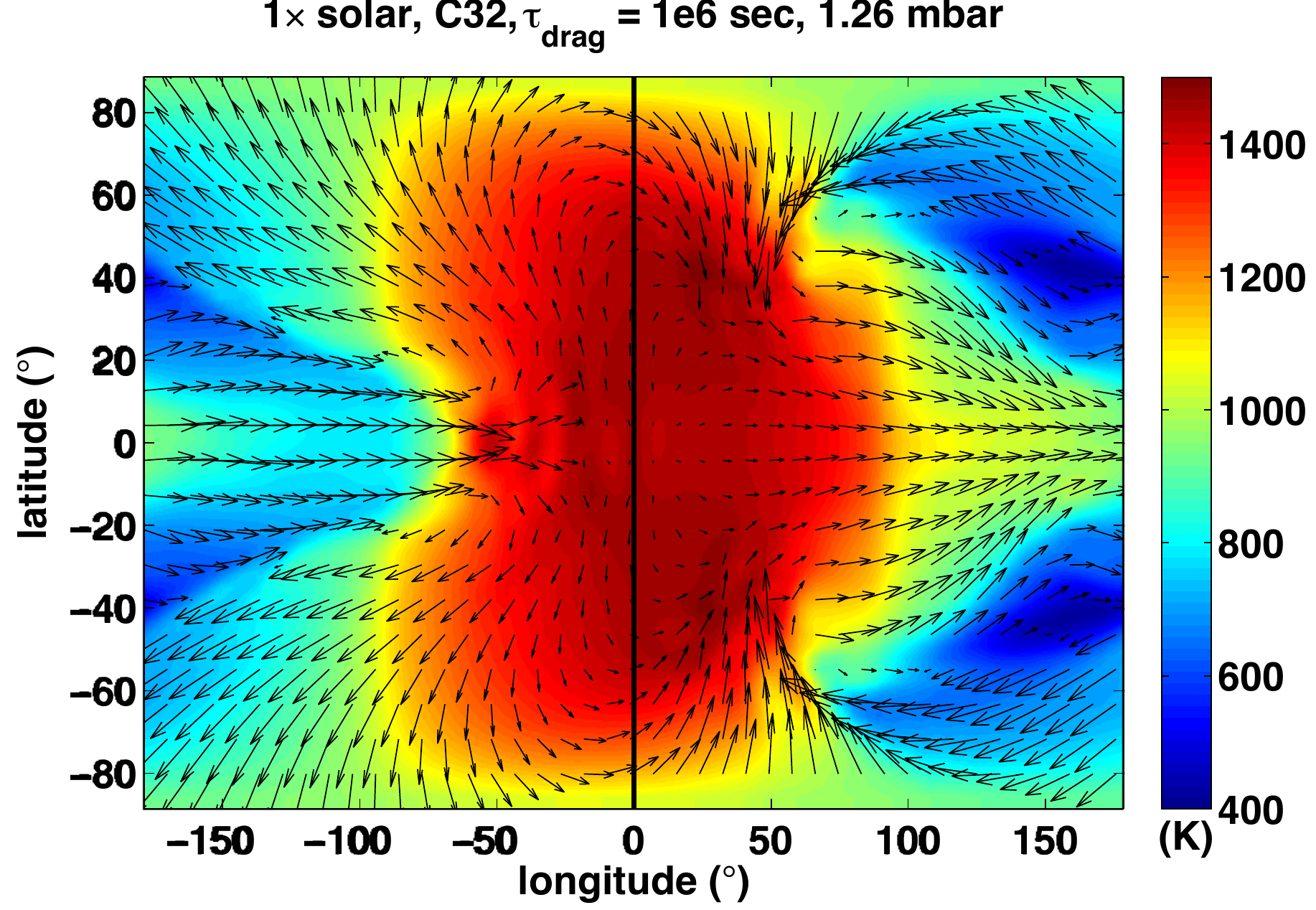}
\includegraphics[trim = 0.0in 0.0in 0.0in 0.0in, clip, width=0.25\textwidth]{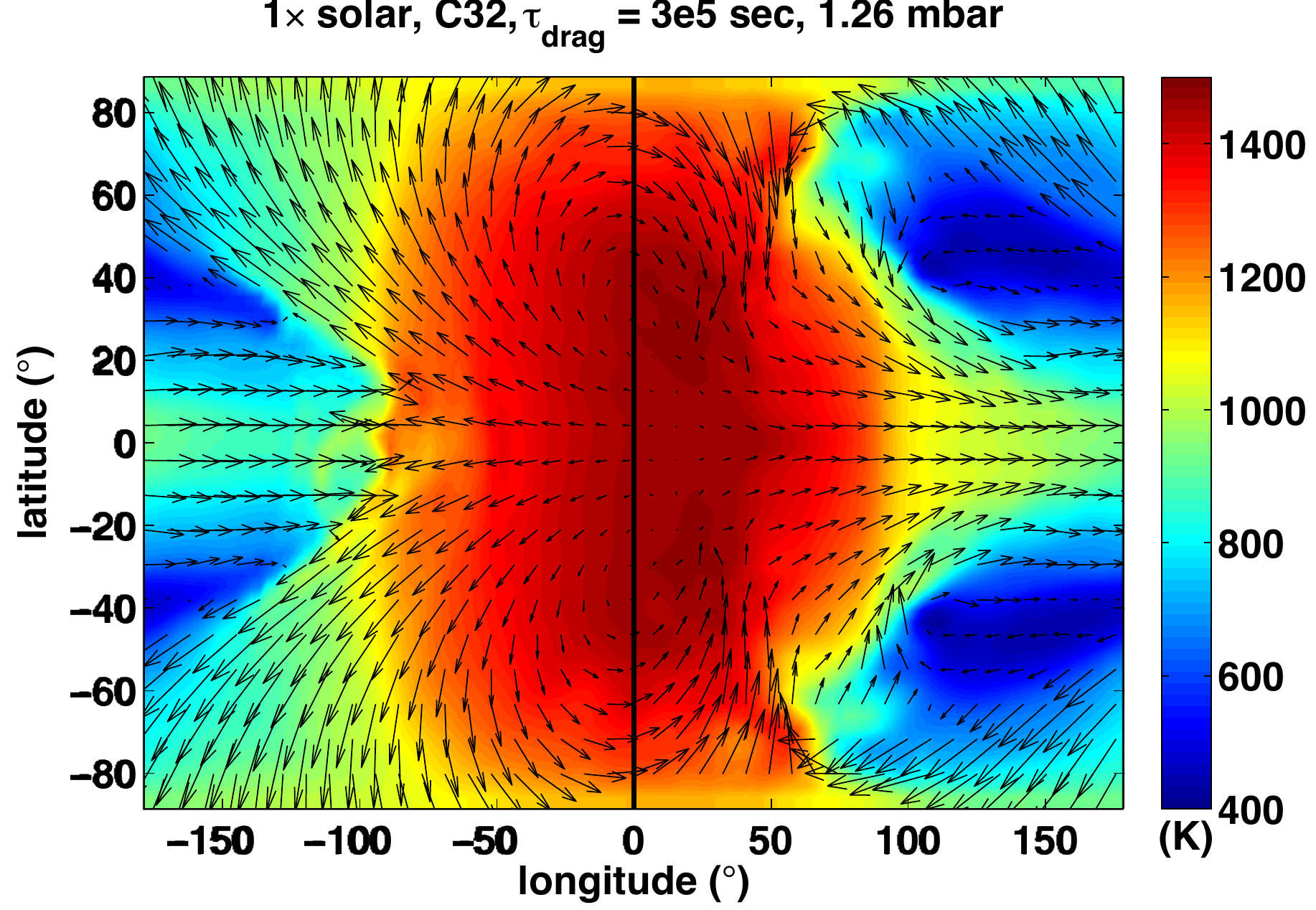}
\includegraphics[trim = 0.0in 0.0in 0.0in 0.0in, clip, width=0.25\textwidth]{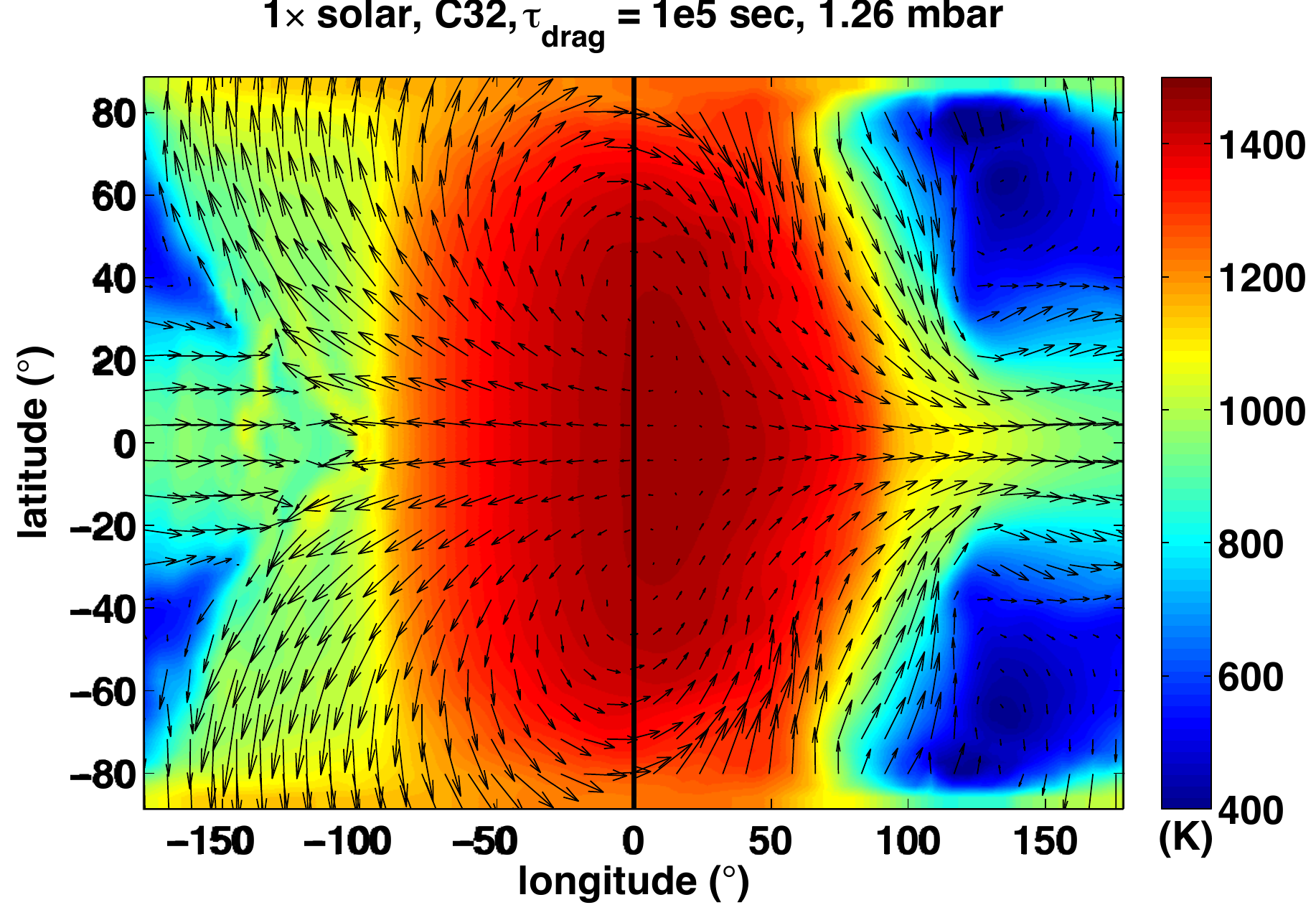}\\
\includegraphics[trim = 0.0in 0.0in 0.0in 0.0in, clip, width=0.25\textwidth]{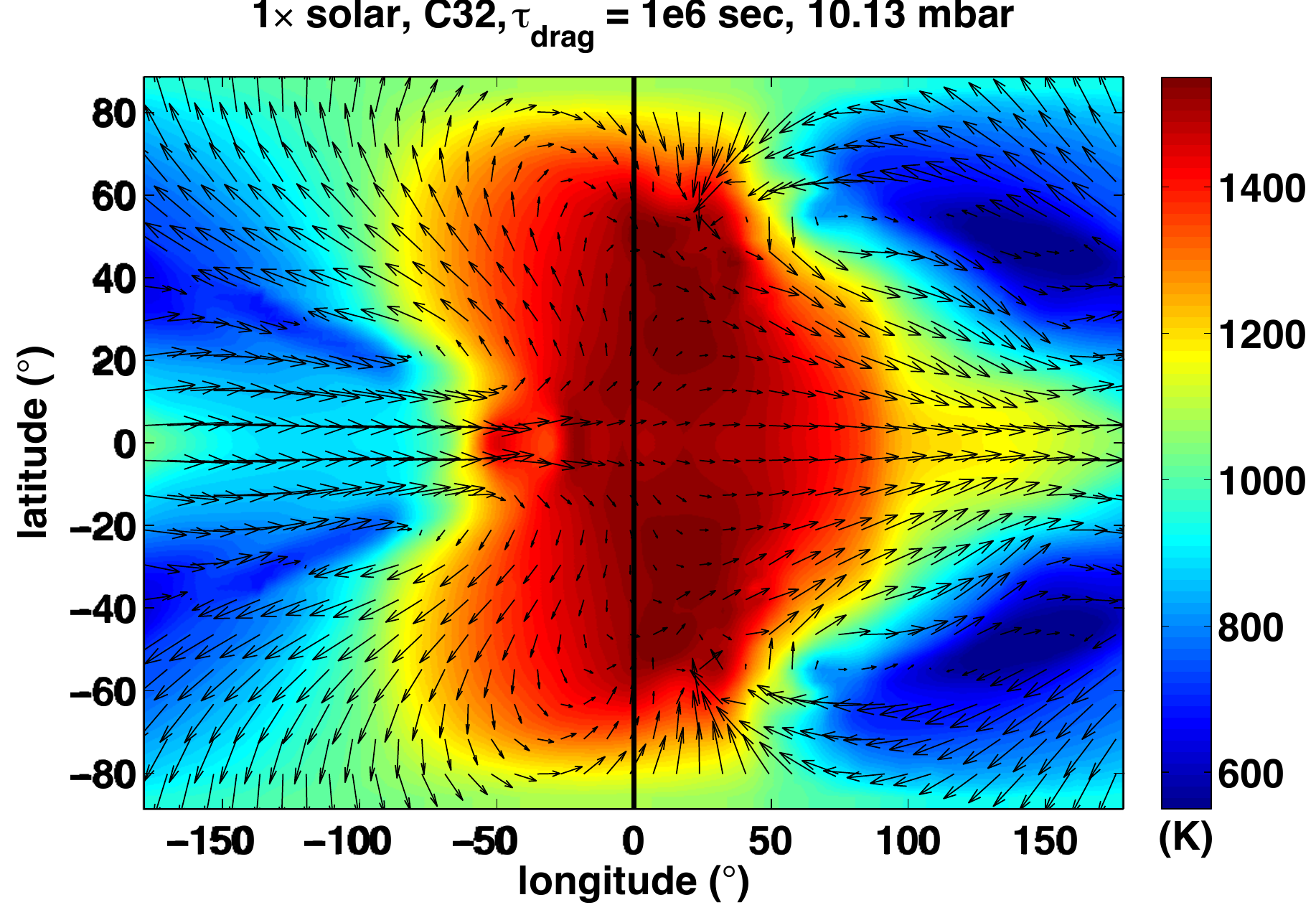}
\includegraphics[trim = 0.0in 0.0in 0.0in 0.0in, clip, width=0.25\textwidth]{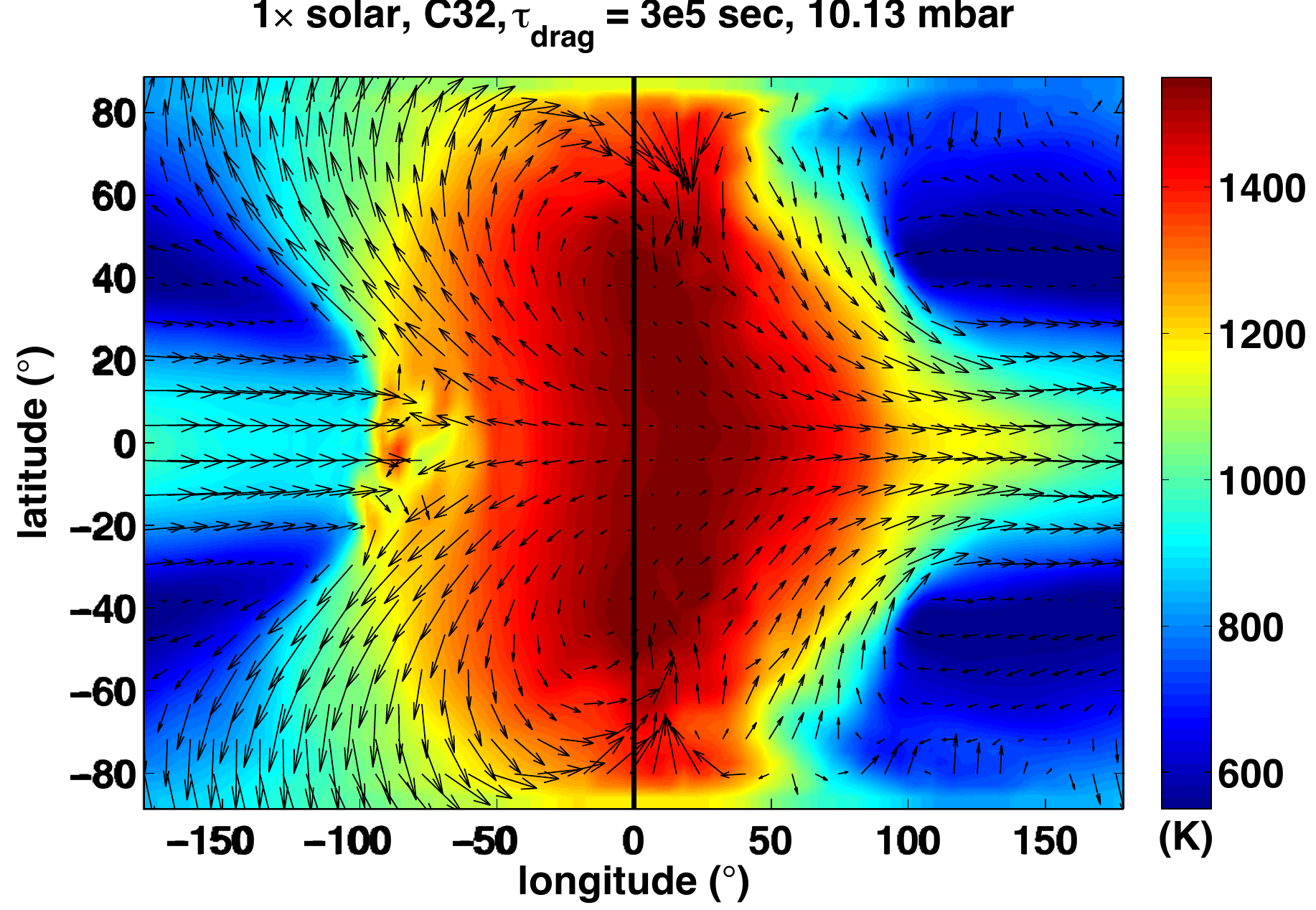}
\includegraphics[trim = 0.0in 0.0in 0.0in 0.0in, clip, width=0.25\textwidth]{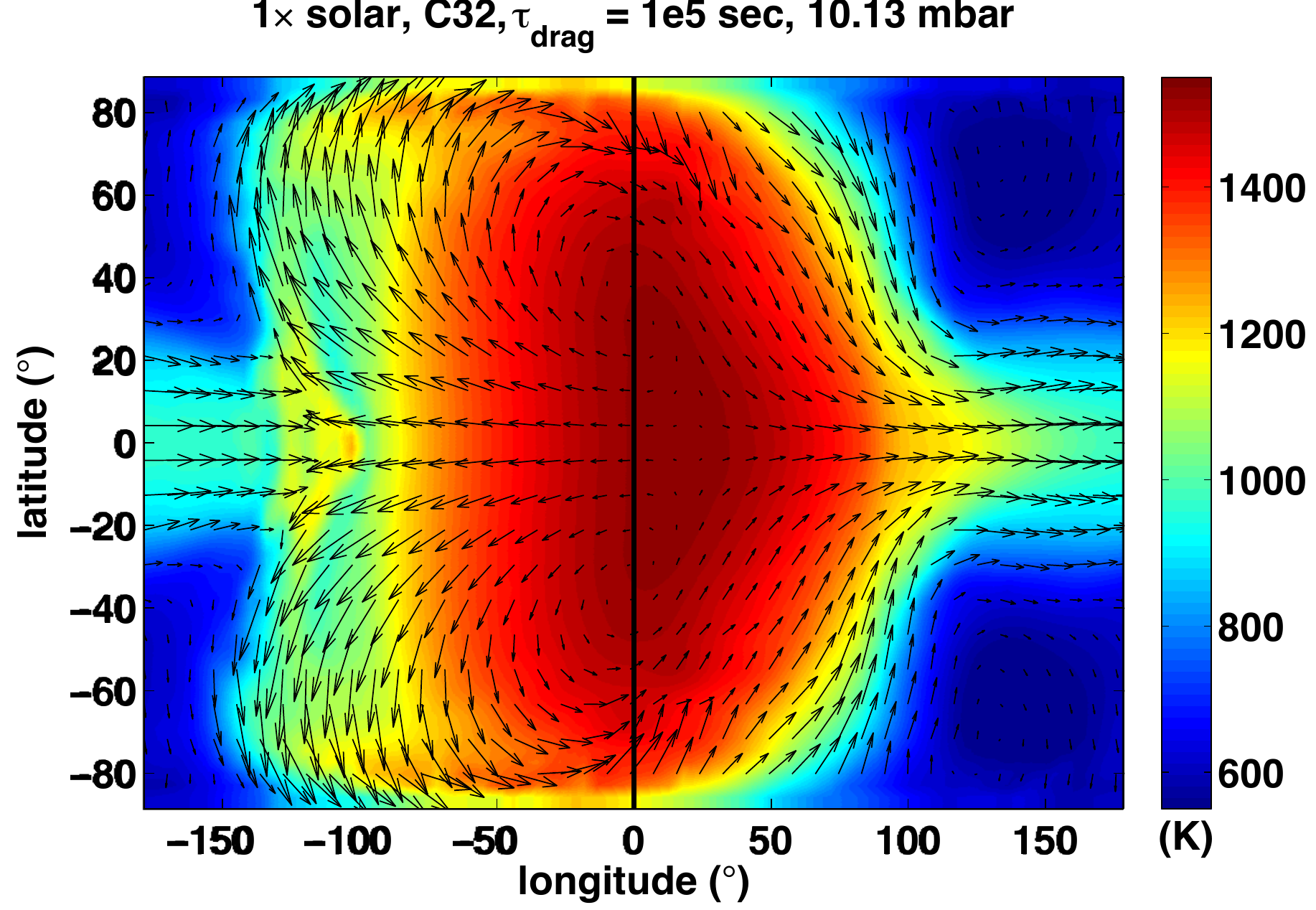}\\
\includegraphics[trim = 0.0in 0.0in 0.0in 0.0in, clip, width=0.25\textwidth]{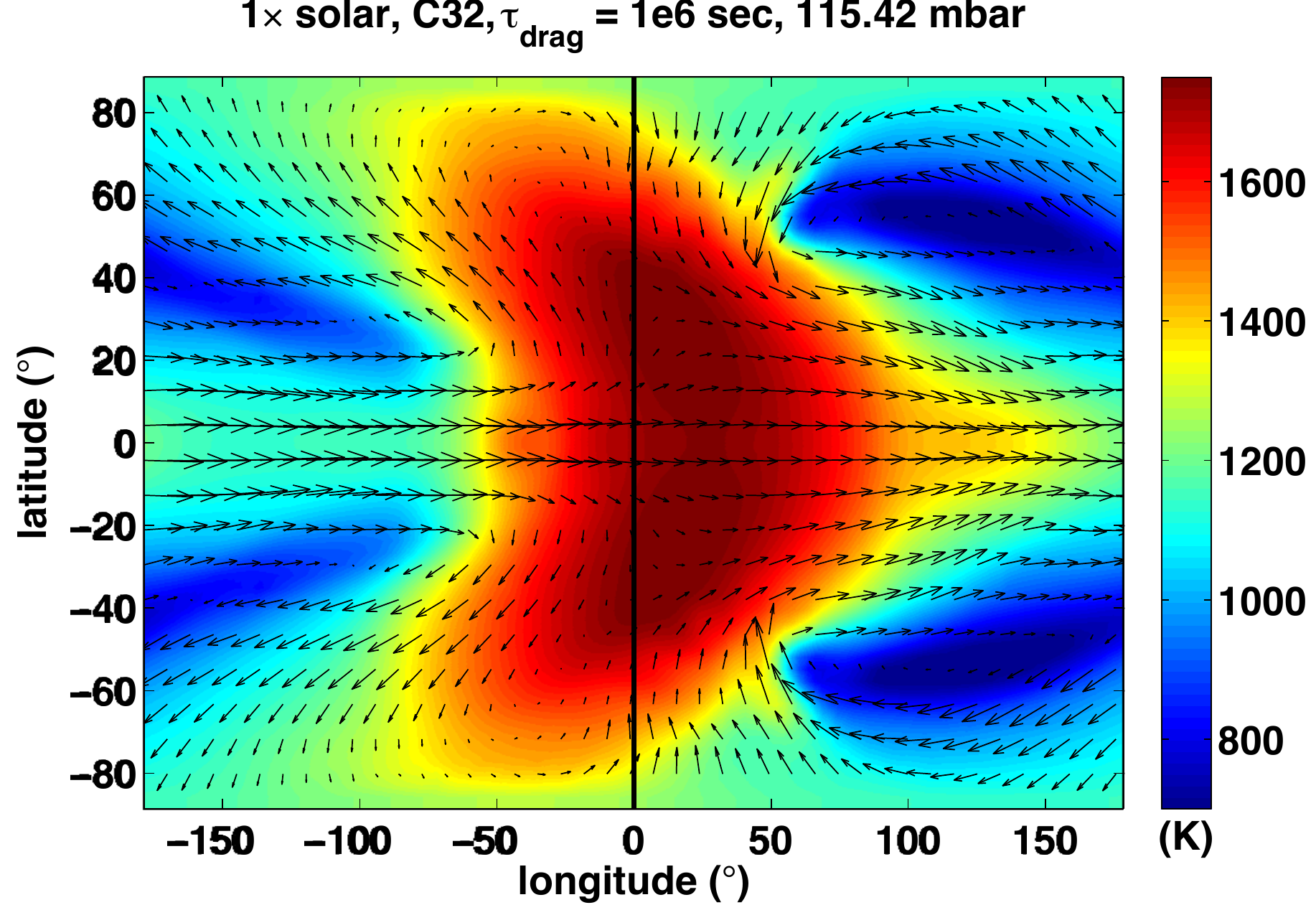}
\includegraphics[trim = 0.0in 0.0in 0.0in 0.0in, clip, width=0.25\textwidth]{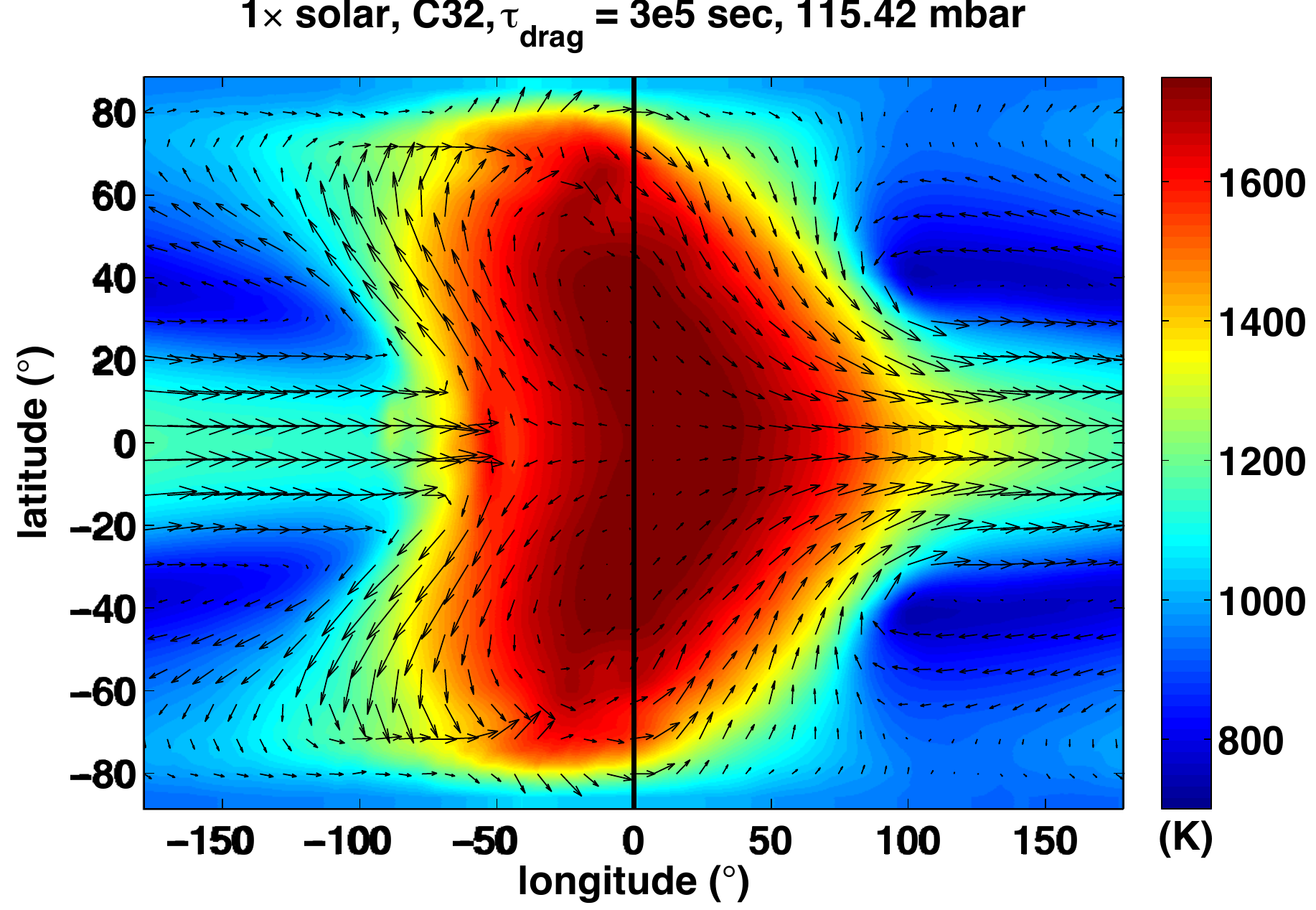}
\includegraphics[trim = 0.0in 0.0in 0.0in 0.0in, clip, width=0.25\textwidth]{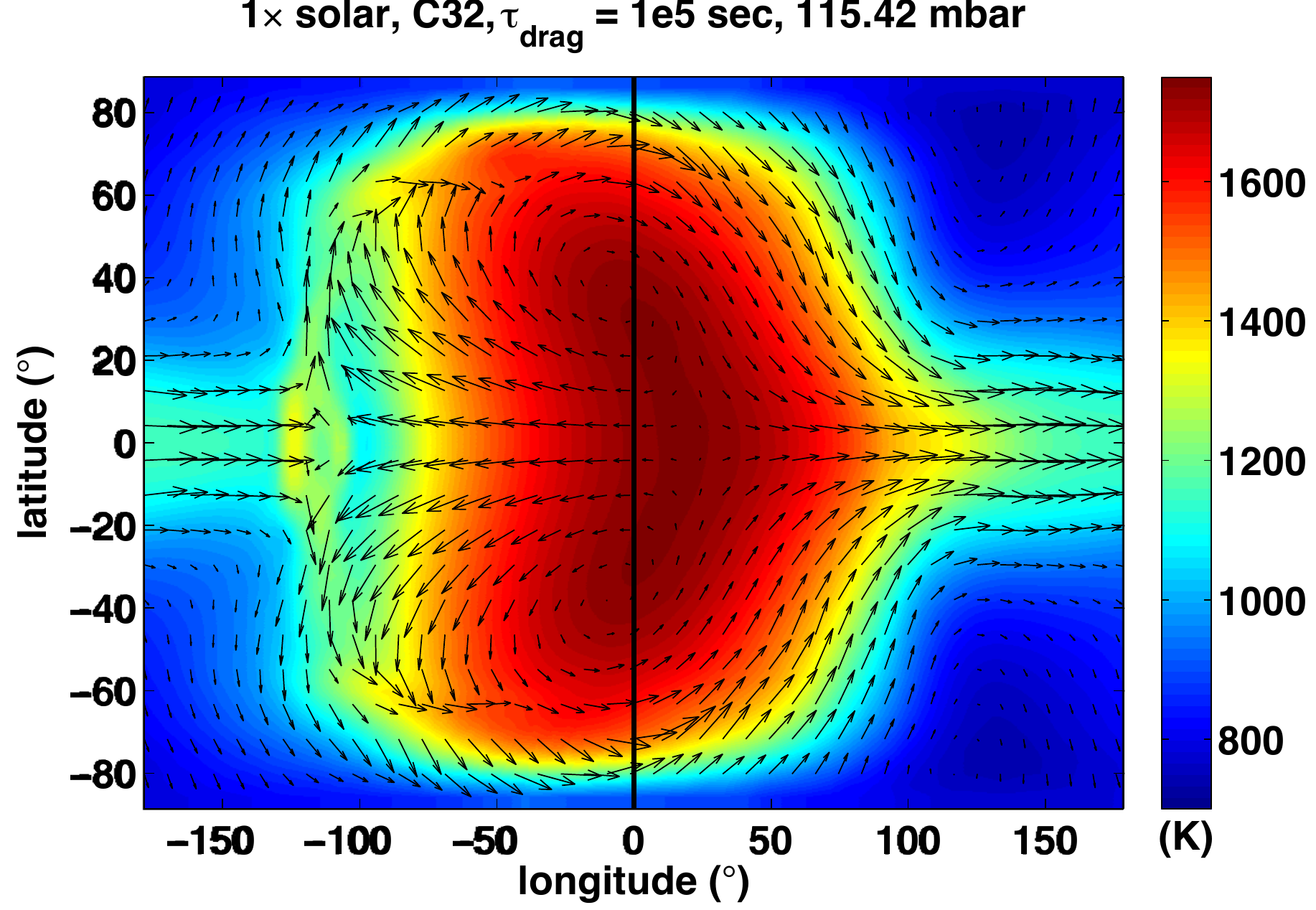}\\
\includegraphics[trim = 0.0in 0.0in 0.0in 0.0in, clip, width=0.25\textwidth]{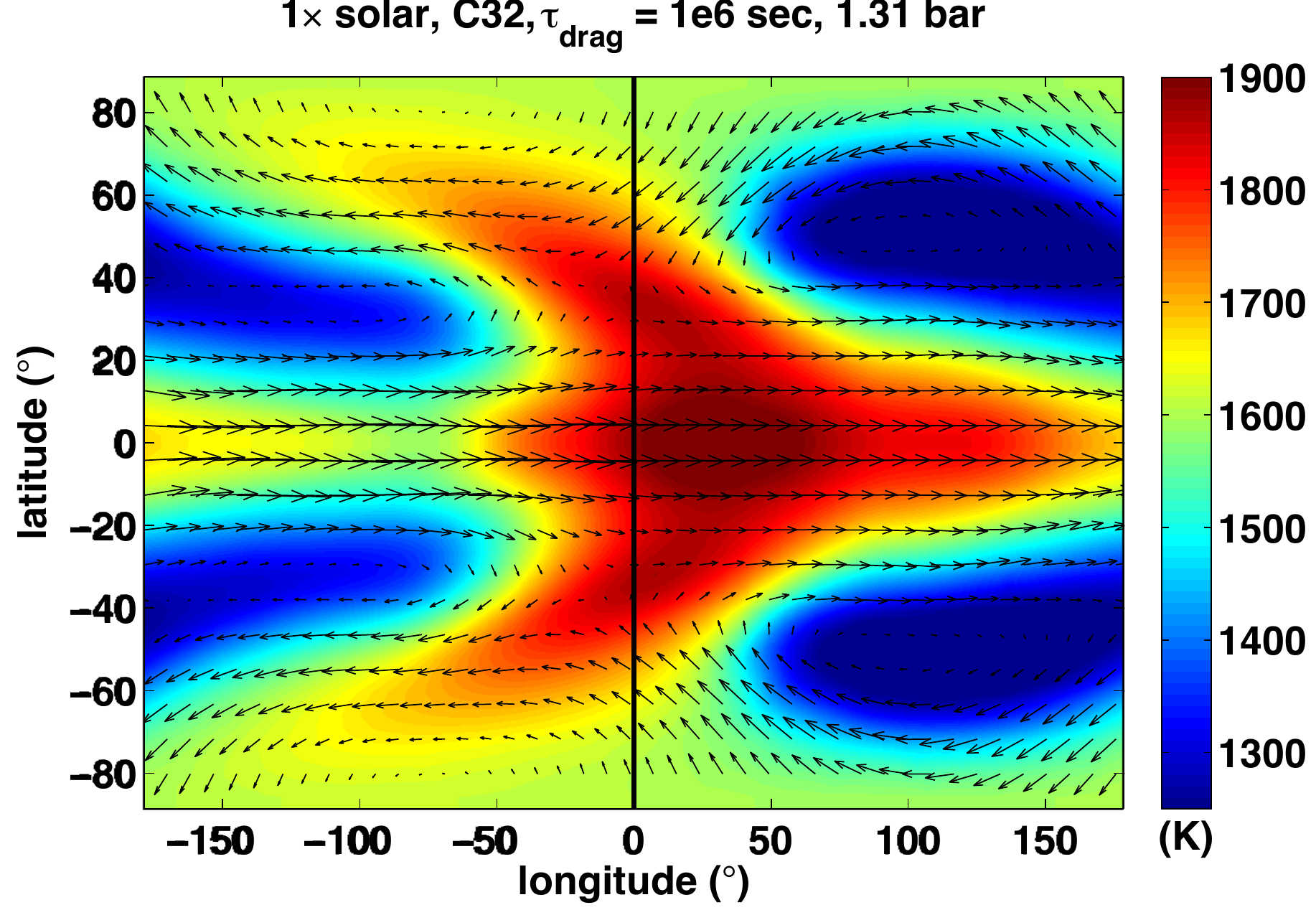}
\includegraphics[trim = 0.0in 0.0in 0.0in 0.0in, clip, width=0.25\textwidth]{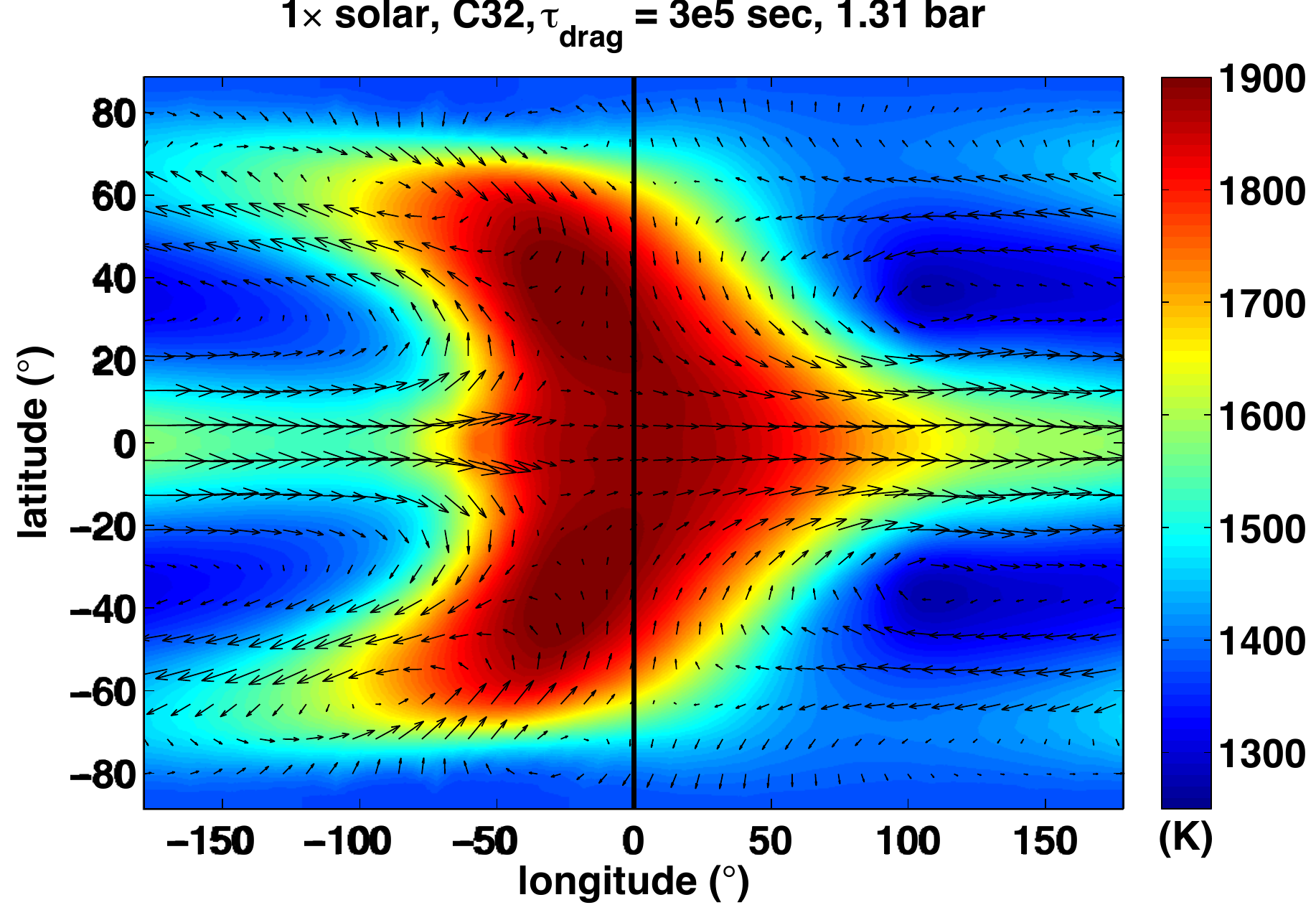}
\includegraphics[trim = 0.0in 0.0in 0.0in 0.0in, clip, width=0.25\textwidth]{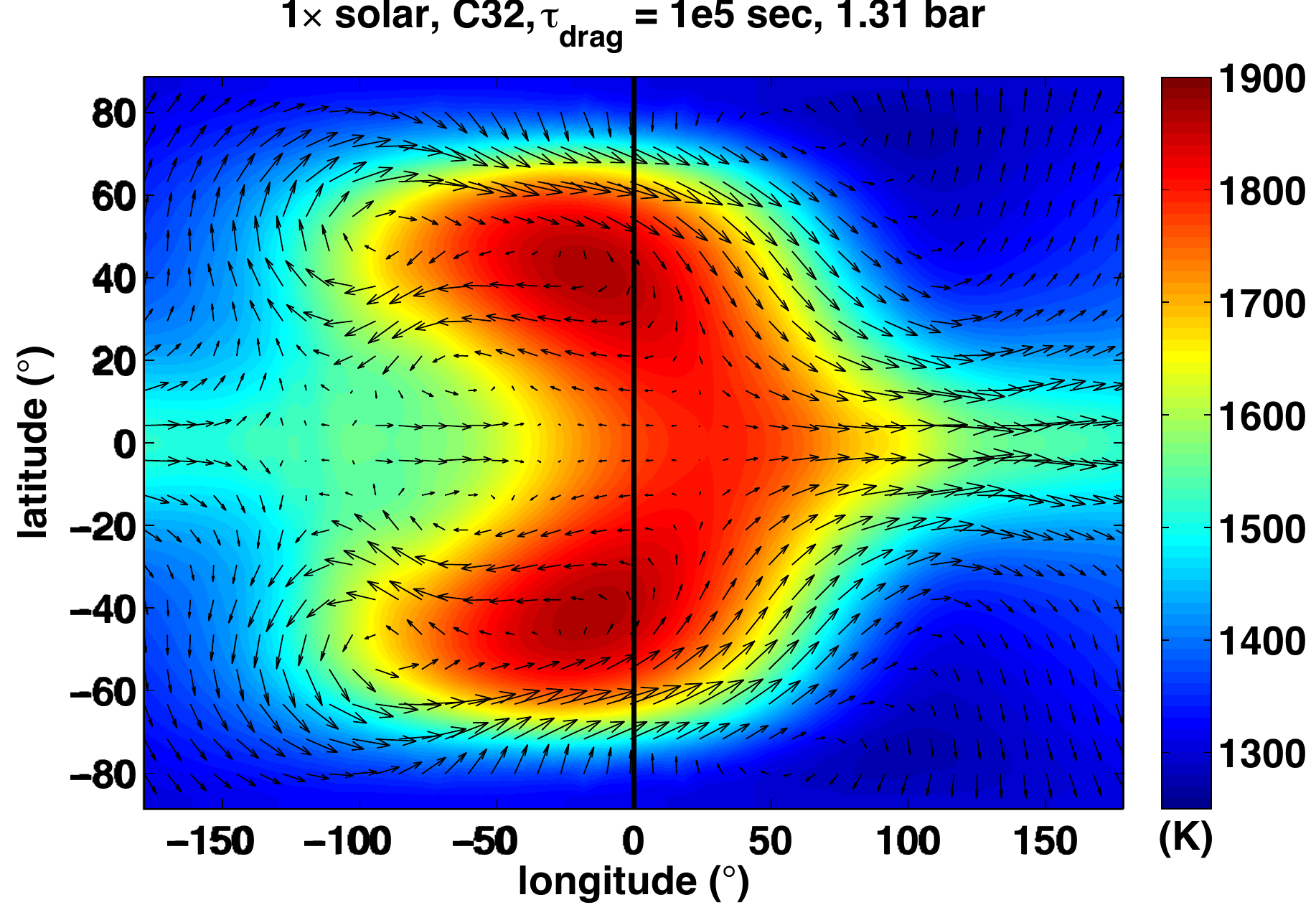}
\caption{Wind and temperature profiles of WASP-43b with an atmospheric composition of 1$\times$ solar and a resolution of C32 at varying degrees of frictional drag, increasing from left to right: $\tau_{drag}=1\times10^6$ s, left column; $\tau_{drag}=3\times10^5$ s, middle column; $\tau_{drag}=1\times10^5$ s, right column. The profiles are compared at four different pressure levels, from top to bottom: 1 mbar, 10 mbar, 100 mbar and 1 bar.  The black line in each profile denotes the longitude of the substellar point.  Each color scale is the same at each pressure level.}
\label{windtemp_drag}
\end{centering}
\end{figure*}

In comparing the wind and temperature profiles from weak to strong drag (Figure \ref{windtemp_drag}), we begin to see a transition from flow that is dominated by equatorial superrotation (left column) to one that is almost completely dominated by day-night flow (right column).  As a result, the eastward shift of the hotspot is reduced with increased drag.  This result has been predicted theoretically; if the drag timescale is shorter than the time it takes for Rossby and Kelvin waves to propagate over a planetary radius, zonal flow is damped and longitudinal phase shifts are reduced, and the circulation shifts from a regime dominated by the equatorial jet to a regime dominated by day-night flow \citep[e.g.,][]{showman+2013}. Indeed, the drag timescales we model here fall in the transitional timescale range necessary for such a shift (for typical hot Jupiter parameters, Showman et al. 2013).  Therefore, at a given pressure, one would expect that a strong-drag model would exhibit a smaller phase shift and larger phase variations than a weak-drag model.  

\subsection{Sensitivity to metallicity: 5$\times$ solar composition}

\begin{figure}
\begin{centering}
\epsscale{.80}
\includegraphics[trim = 0.0in 0.0in 0.0in 0.0in, clip, width=0.45\textwidth]{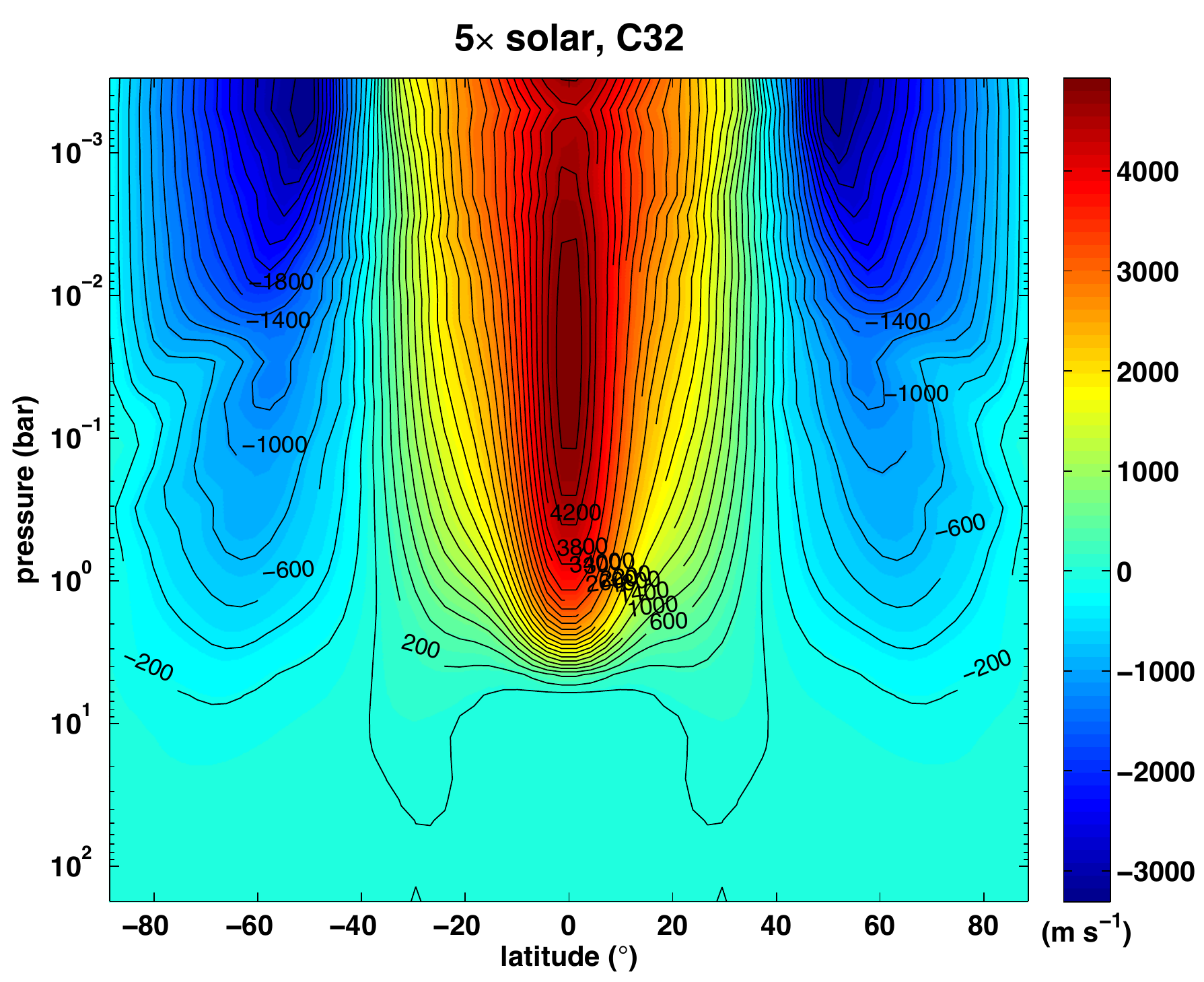}\\
\caption{Zonal-mean zonal wind profile of WASP-43b at a resolution of C32 and an atmospheric composition of 5$\times$ solar, averaged over an orbit.  }
\label{zonalwind_fivesolar}
\end{centering}
\end{figure}

\begin{figure}
\begin{centering}
\epsscale{.80}
\includegraphics[trim = 0.0in 0.0in 0.0in 0.0in, clip, width=0.4\textwidth]{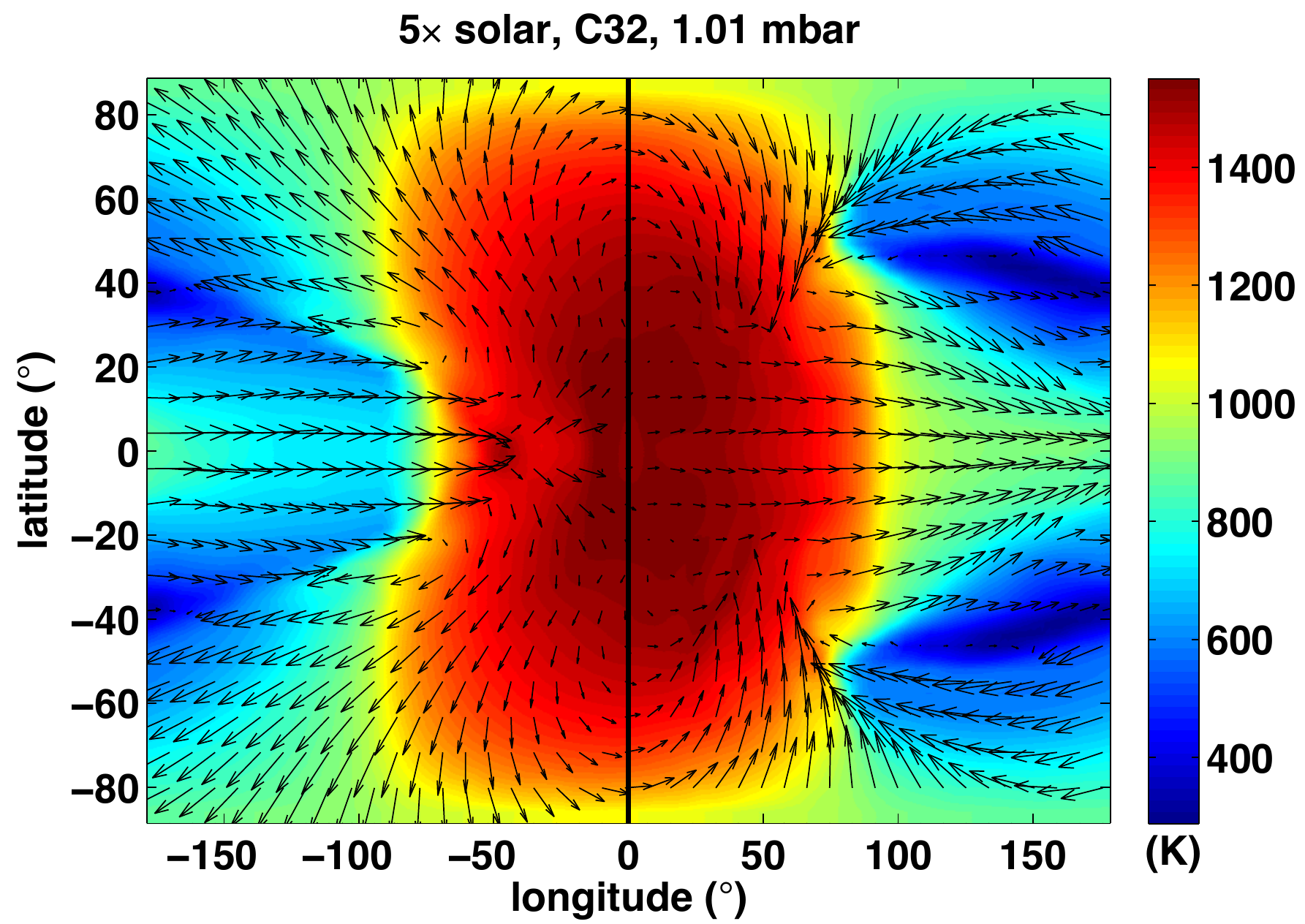}\\
\includegraphics[trim = 0.0in 0.0in 0.0in 0.0in, clip, width=0.4\textwidth]{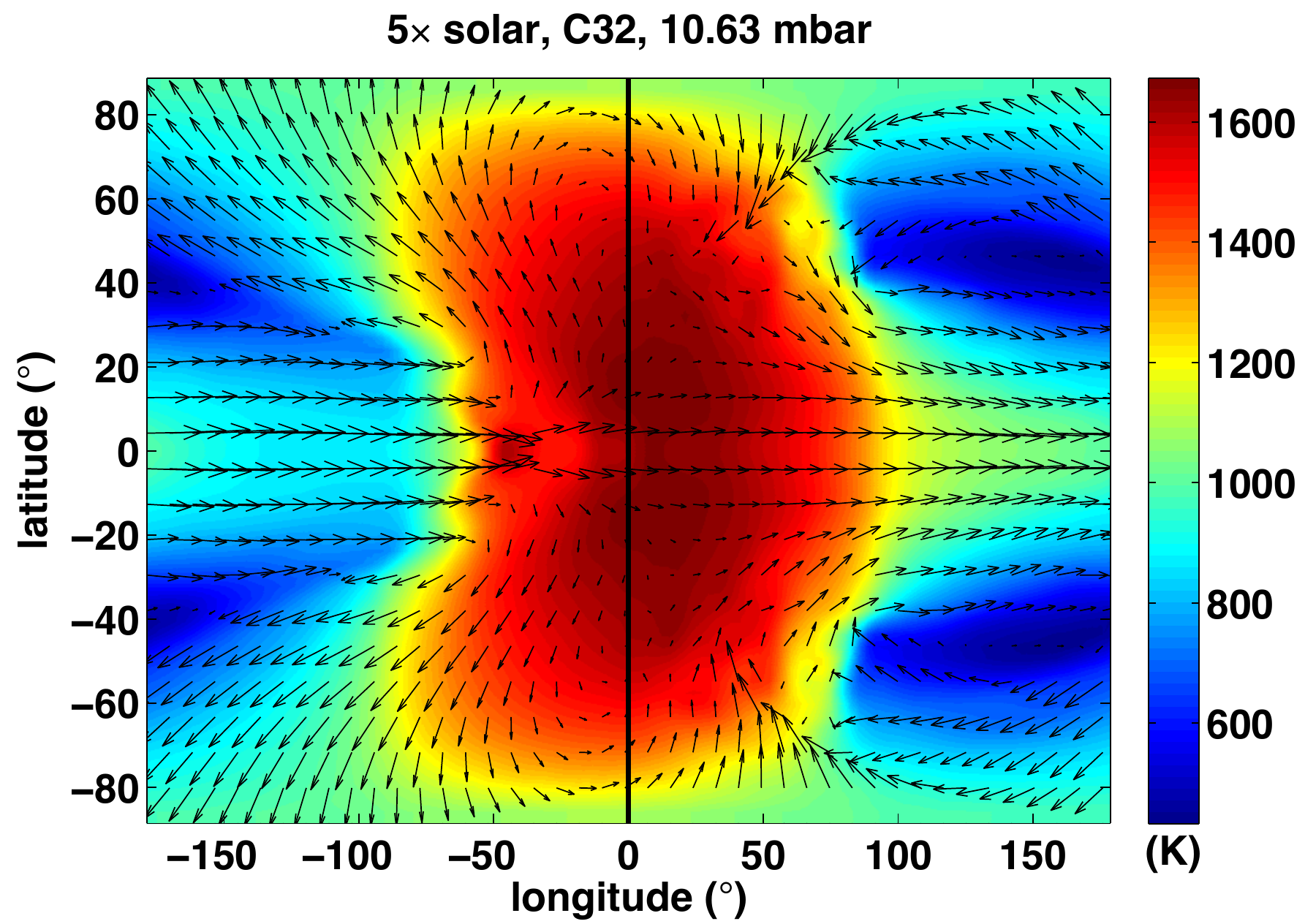}\\
\includegraphics[trim = 0.0in 0.0in 0.0in 0.0in, clip, width=0.4\textwidth]{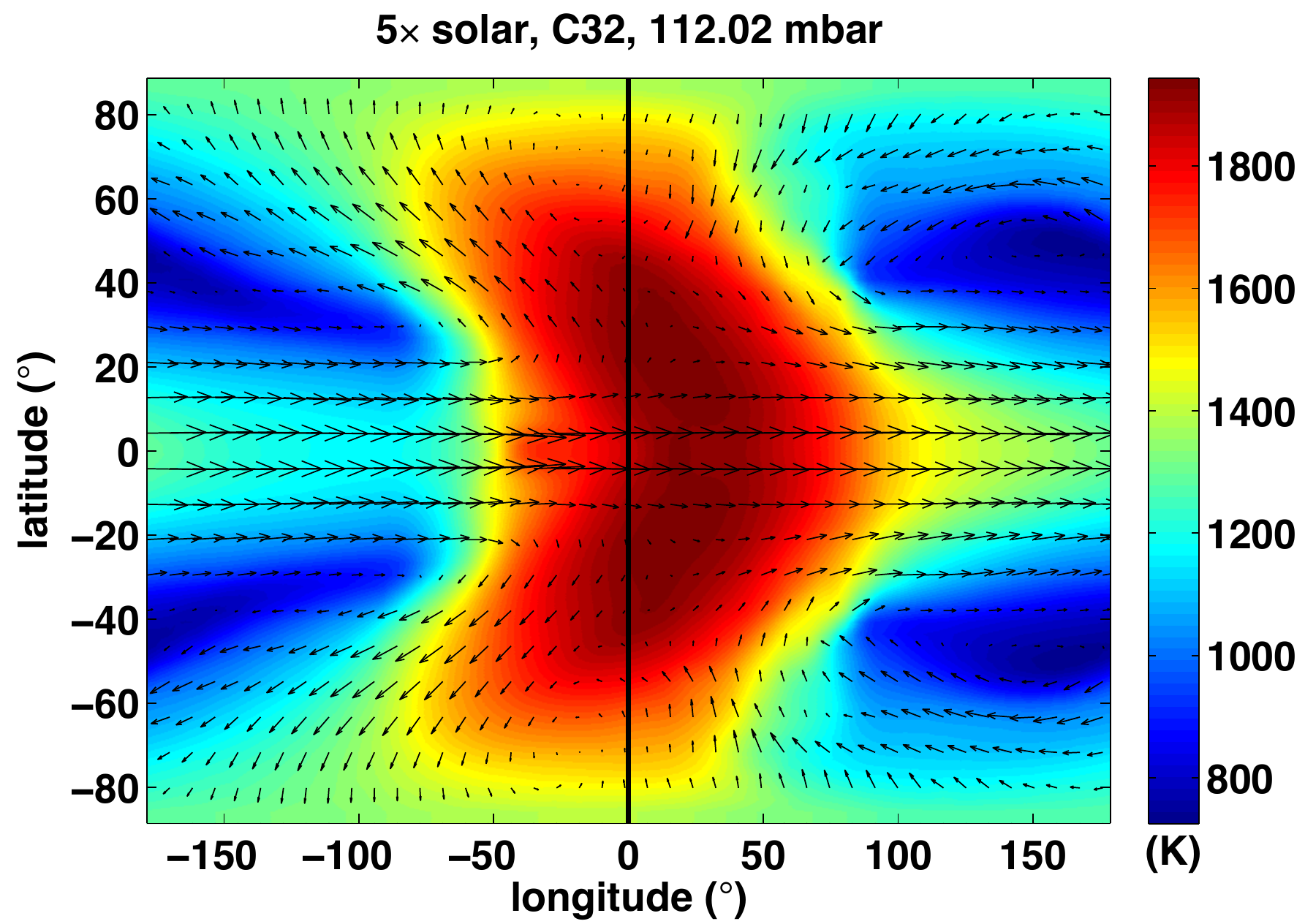}\\
\includegraphics[trim = 0.0in 0.0in 0.0in 0.0in, clip, width=0.4\textwidth]{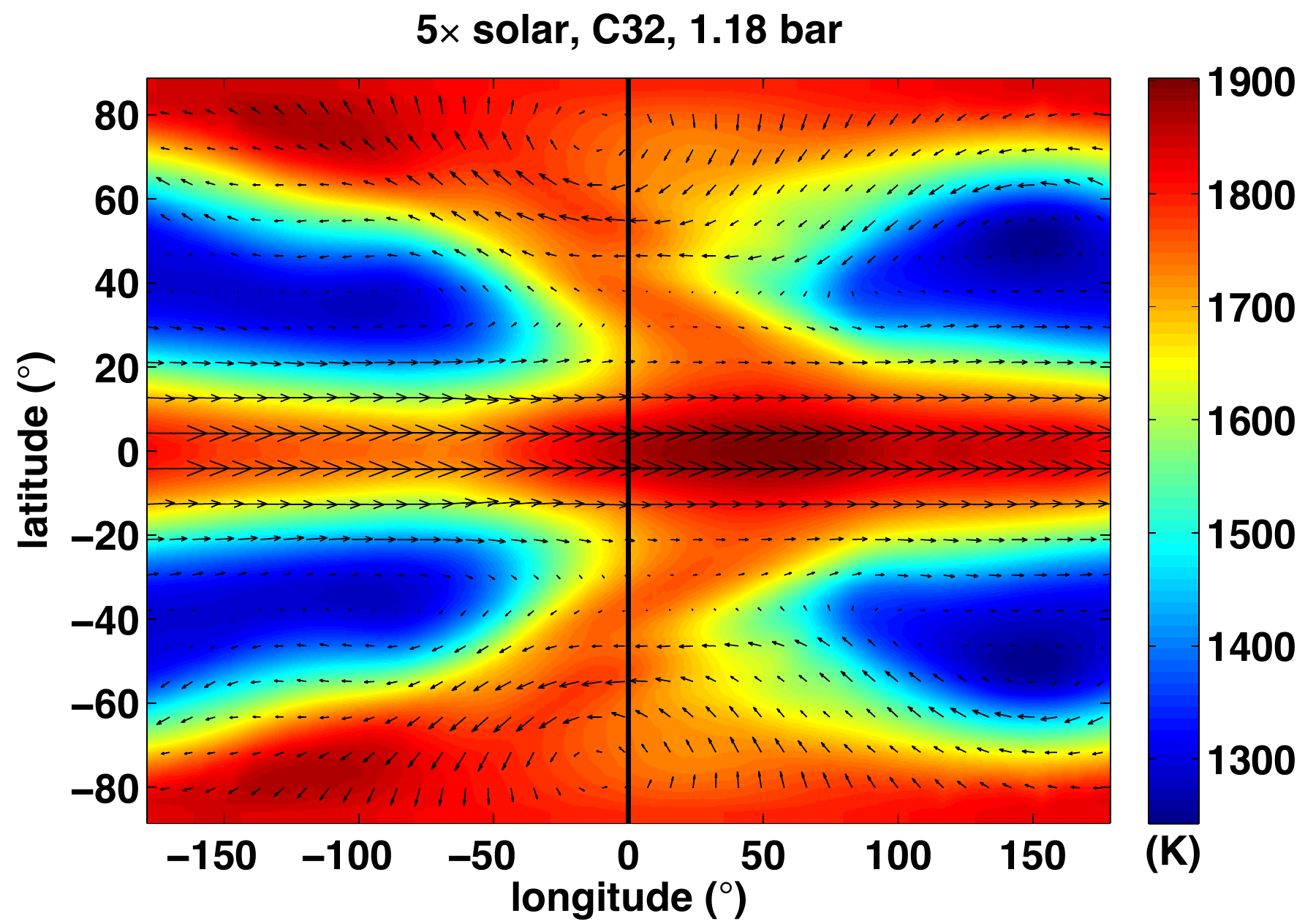}
\caption{Wind and temperature profiles of WASP-43b with an atmospheric composition of 5$\times$ solar, at a model resolution of C32. The profiles are shown at four different pressure levels, from top to bottom: 1 mbar, 10 mbar, 100 mbar and 1 bar.  The black line in each profile denotes the longitude of the substellar point. }
\label{windtemp_fivesolar}
\end{centering}
\end{figure}

Here we explore the sensitivity of the models to metallicity, as our group has previously conducted in studies of hot Jupiter HD 189733b \citep{showman+2009}, hot Neptune GJ 436b \citep{lewis+2010} and super Earth GJ 1214b \citep{kataria+2014}.  Figure \ref{zonalwind_fivesolar} plots the orbit-averaged zonal-mean zonal wind for a 5$\times$ solar model.  Consistent with previous metallicity sensitivity studies, as the metallicity is enhanced, the opacities are increased, stellar radiative energy is deposited higher in the atmosphere, and therefore the depth of the jet in the 5$\times$ solar model is shallower ($\sim$2 bars as compared to $\sim$ 10 bars in the 1$\times$ solar model).  Furthermore, the 5$\times$ solar model has a greater maximum jet speed; this is because the day-night forcing is larger.  This trend is reflected in the 5$\times$ solar wind and temperature profiles; at a given pressure level, the dayside and nightside temperature variations are larger than the 1$\times$ solar models (Figure \ref{windtemp_fivesolar}).  Therefore, one would expect observationally that the 5$\times$ solar model should exhibit larger phase variations than the 1$\times$ solar model.  Furthermore, the IR photosphere is shallower in the 5$\times$ solar model, where day-night phase shifts are less; therefore, one would also expect that the timing of peak IR flux should occur closer to secondary eclipse than the 1$\times$ solar model.  

\begin{figure}
\begin{centering}
\epsscale{.80}
\includegraphics[trim = 0.0in 0.0in 0.0in 0.0in, clip, width=0.45\textwidth]{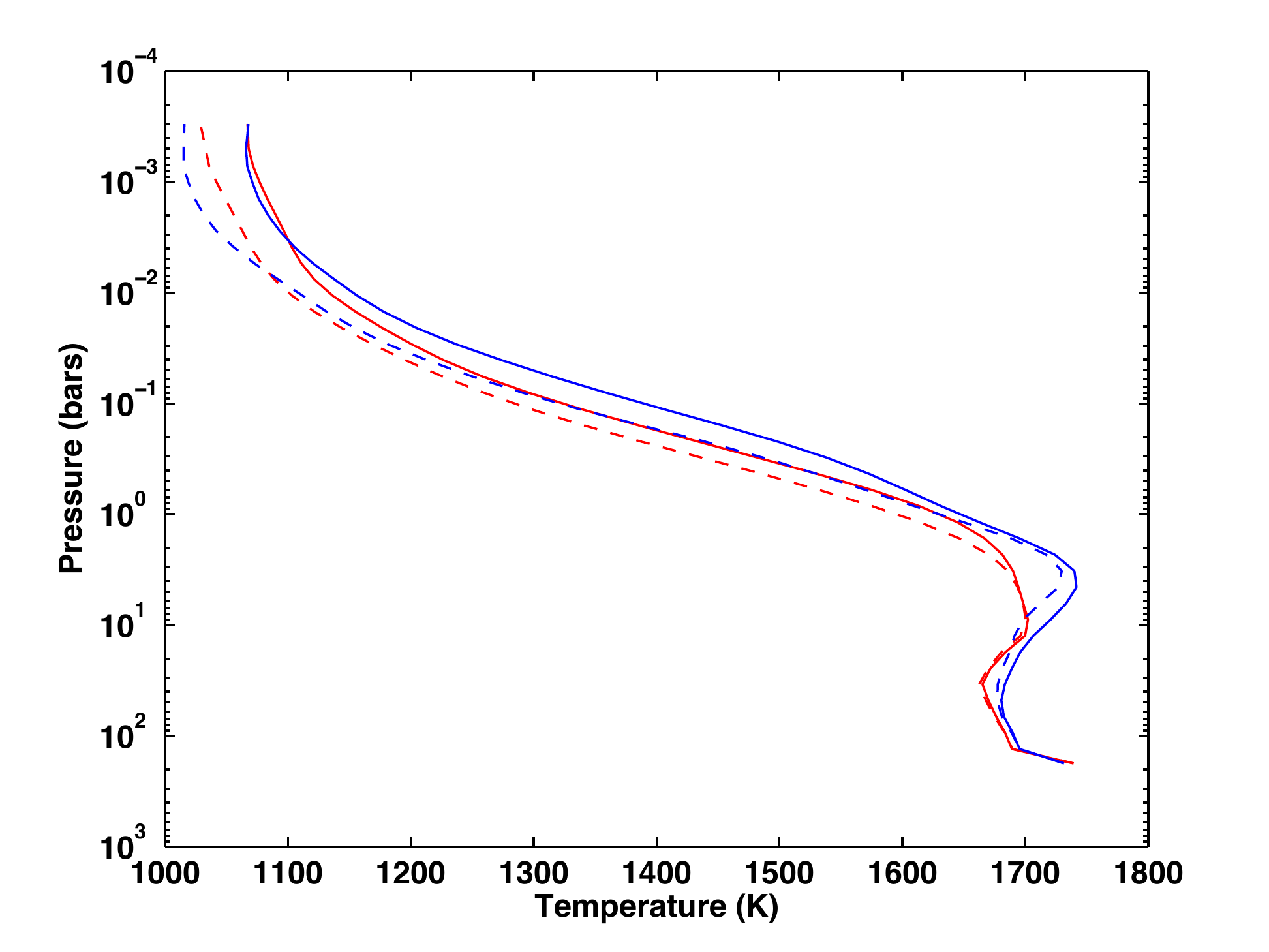}
\caption{Pressure-temperature profiles averaged in latitude (from pole to pole) at longitudes of the eastern (solid lines) and western (dashed lines) terminators for 1$\times$ (red) and 5$\times$ (blue) solar atmospheric compositions. Here, the eastern terminator corresponds to the 'dusk' terminator eastward of the substellar point, while the western terminator is the 'dawn' terminator westward of the substellar point.}
\label{pt_terminators}
\end{centering}
\end{figure}

Our model results can also lend insights to observations in transit, which probes the terminators of the planet.  Figure \ref{pt_terminators} plots pressure-temperature ($p$-$T$) profiles averaged in latitude, from pole to pole, for 1$\times$ (red) and 5$\times$ solar (blue) compositions of WASP-43b at the longitudes of the eastern terminator (or dusk terminator, east of the substellar point; solid profiles) and western terminator (dawn terminator, west of the substellar point; dashed profiles).  Because the circulation advects the hottest regions eastward of the substellar point, the eastern terminator profiles are moderately hotter than the western terminator profiles.  Above $\sim$5-6 mbar, the 5$\times$ solar profiles are cooler than the 1$\times$ solar model, while below those pressures the reverse is true.  These terminator profiles are overall much warmer than the scale height temperature retrieved from the transmission spectrum in \cite{kreidberg+2014b}.

\subsection{Models with TiO/VO}

Although observations of WASP-43b show that the planet does not have a dayside temperature inversion \citep{blecic+2014,line+2014,stevenson+2014}, we explore the dynamical response of such atmospheres by modeling 1$\times$ and 5$\times$ solar models with TiO and VO in chemical equilibrium.  Being strong visible absorbers, TiO and VO naturally lead to thermal inversions at typical pressures of $\sim$1 mbar \citep[e.g.,][]{hubeny+2003, fortney+2008, showman+2009}.  While doubts have been raised about the existence of TiO and VO in the atmospheres of hot Jupiters \citep[e.g.,][]{spiegel+2009, parmentier+2013}, any chemical species with very high visible opacity \citep[e.g., polysulfurs;][]{zahnle+2009} should lead to qualitatively similar effects, namely absorption of starlight at low pressure and generation of a thermal inversion.  We therefore include TiO and VO here simply as a straightforward way of including visible opacity and view them as a proxy for any strong visible-wavelength absorber.

\begin{figure*}
\begin{centering}
\epsscale{.80}
\includegraphics[trim = 0.0in 0.0in 0.0in 0.0in, clip, width=0.4\textwidth]{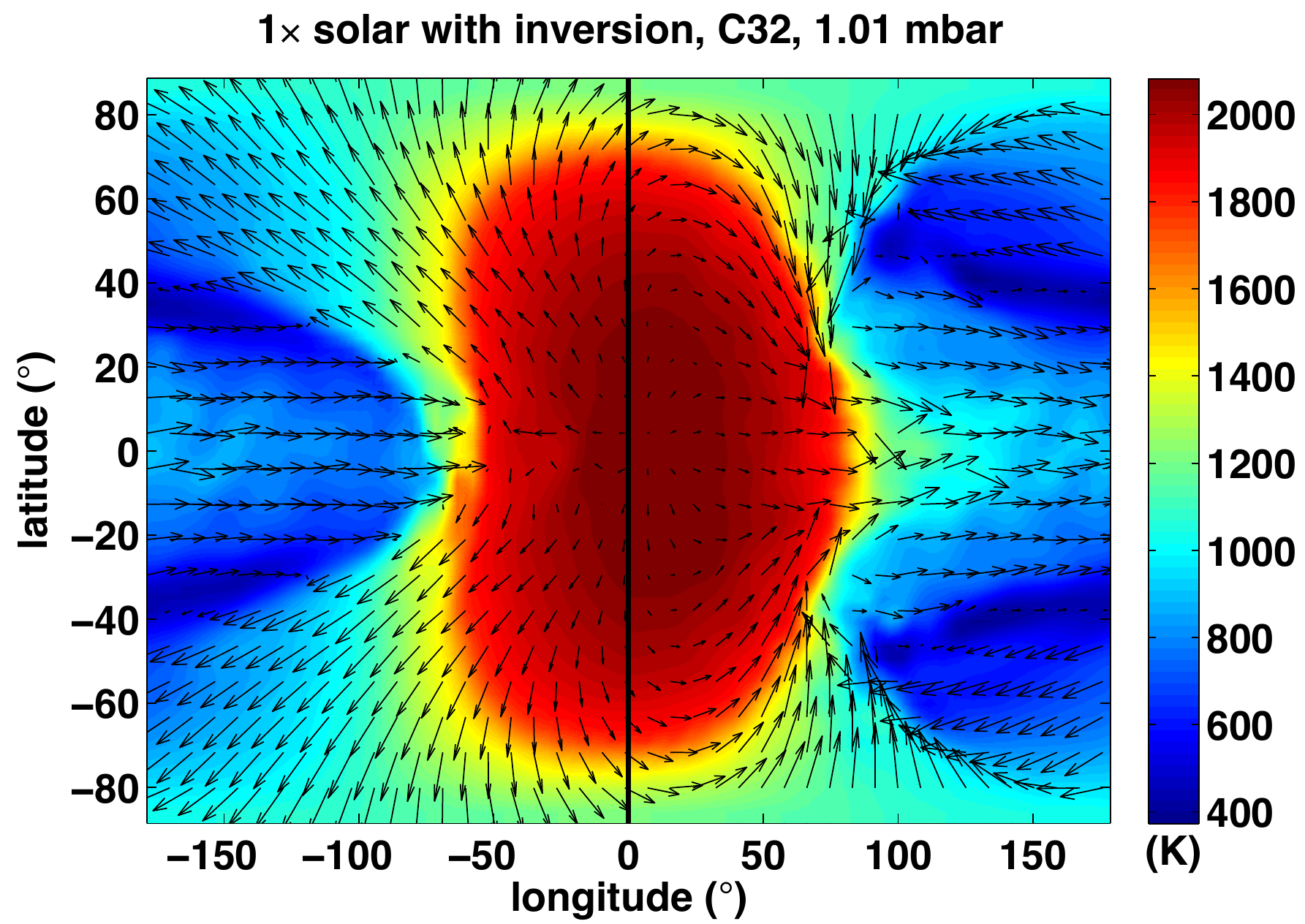}
\includegraphics[trim = 0.0in 0.0in 0.0in 0.0in, clip, width=0.4\textwidth]{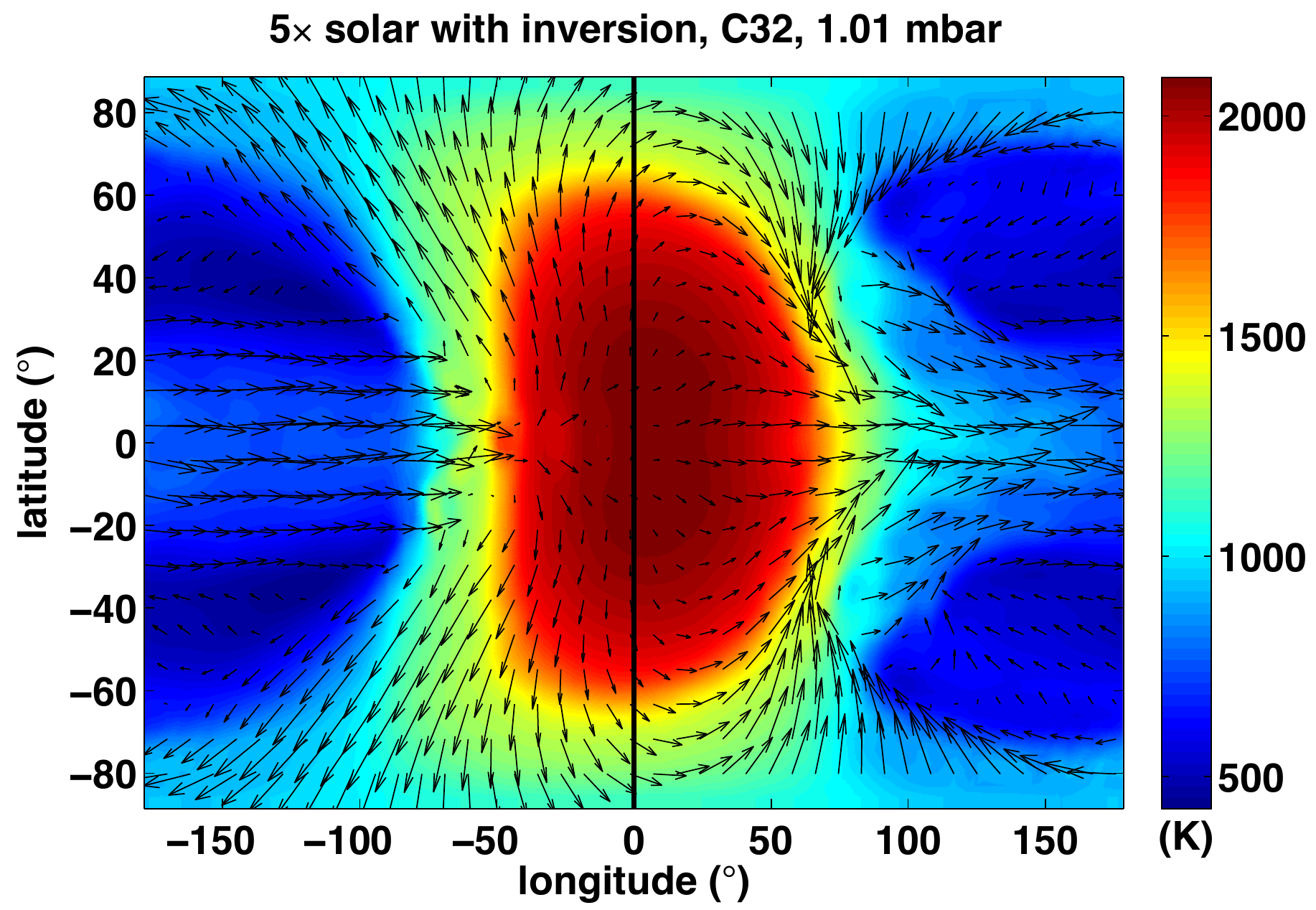}\\
\includegraphics[trim = 0.0in 0.0in 0.0in 0.0in, clip, width=0.4\textwidth]{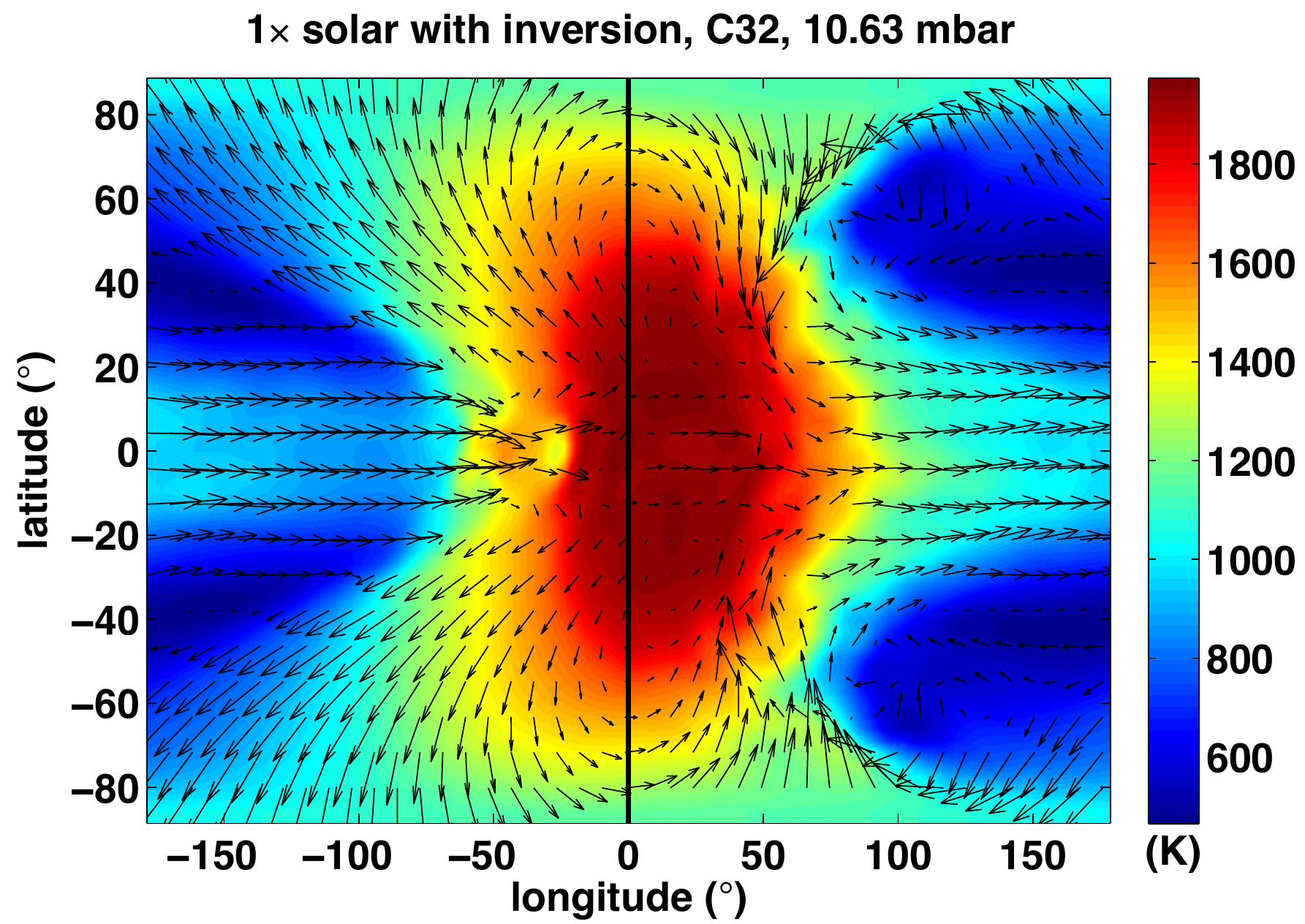}
\includegraphics[trim = 0.0in 0.0in 0.0in 0.0in, clip, width=0.4\textwidth]{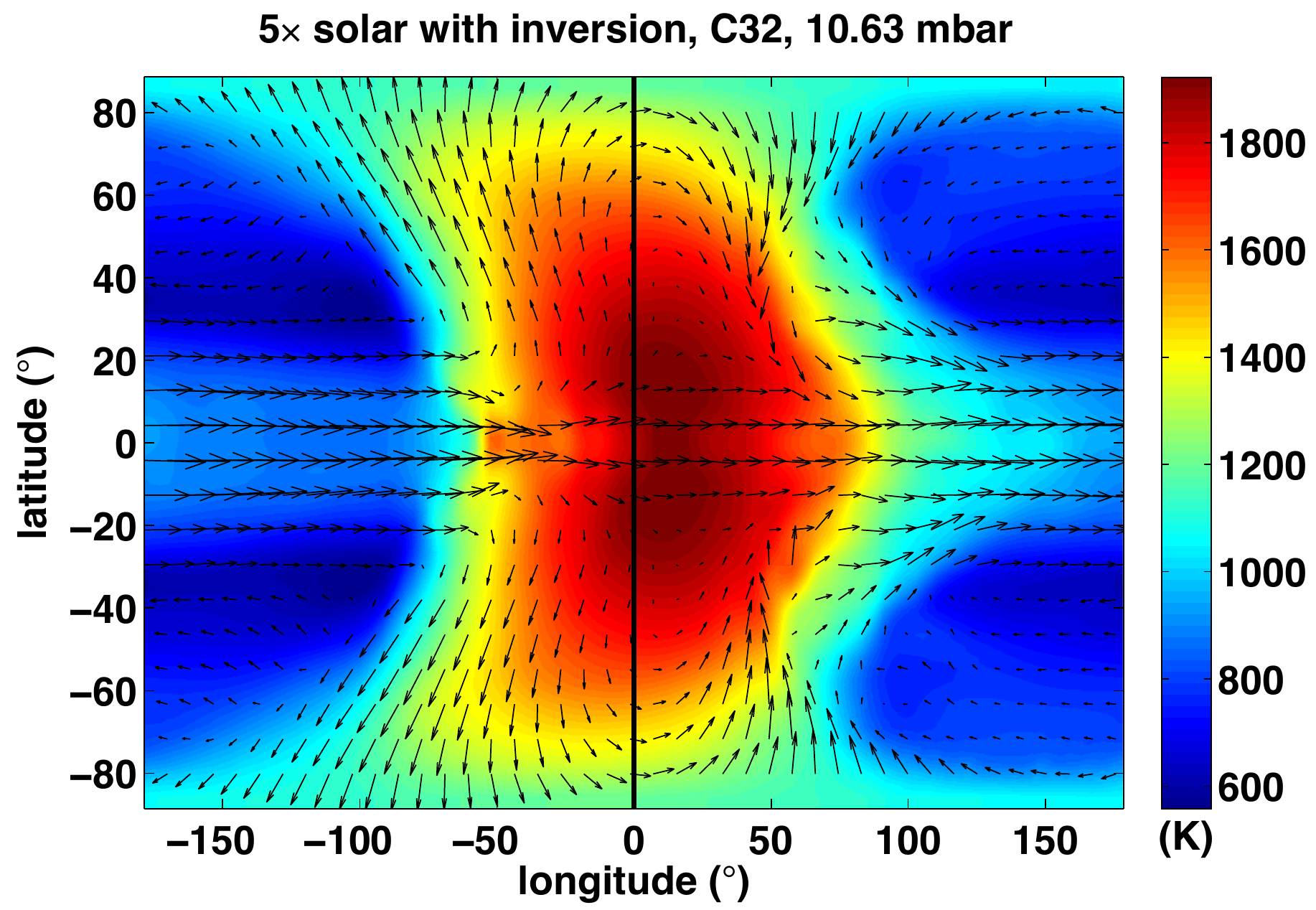}\\
\includegraphics[trim = 0.0in 0.0in 0.0in 0.0in, clip, width=0.4\textwidth]{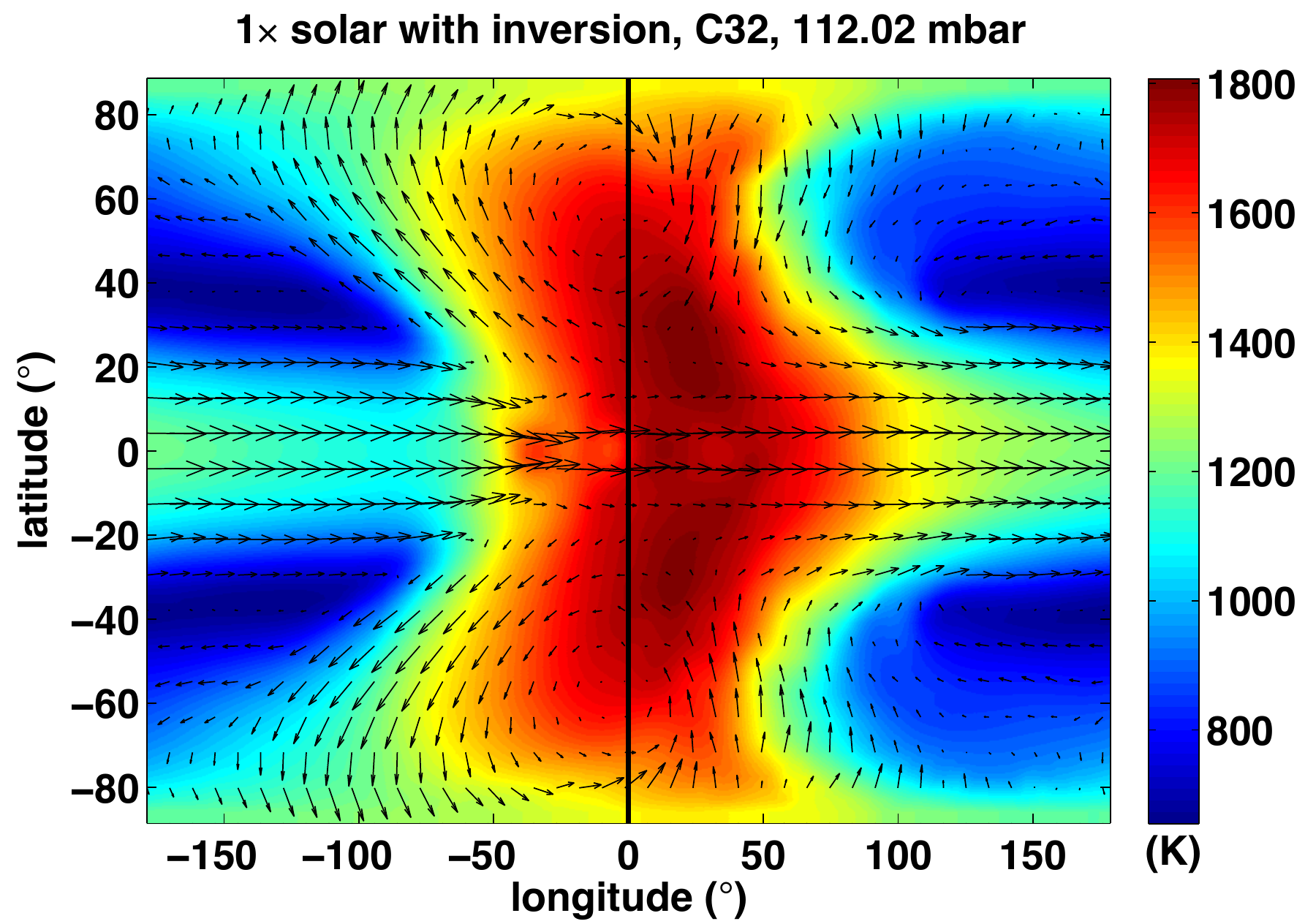}
\includegraphics[trim = 0.0in 0.0in 0.0in 0.0in, clip, width=0.4\textwidth]{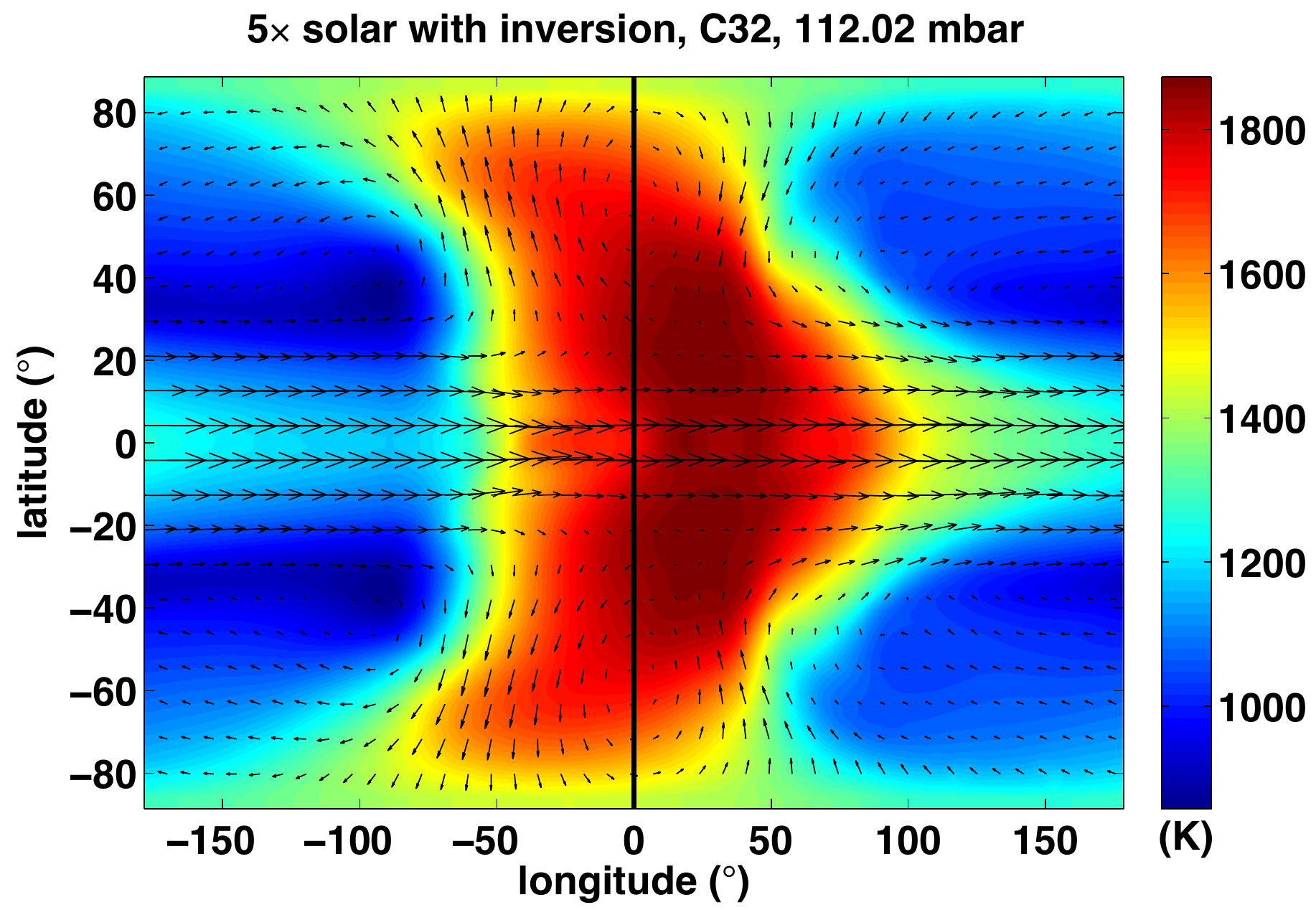}\\
\includegraphics[trim = 0.0in 0.0in 0.0in 0.0in, clip, width=0.4\textwidth]{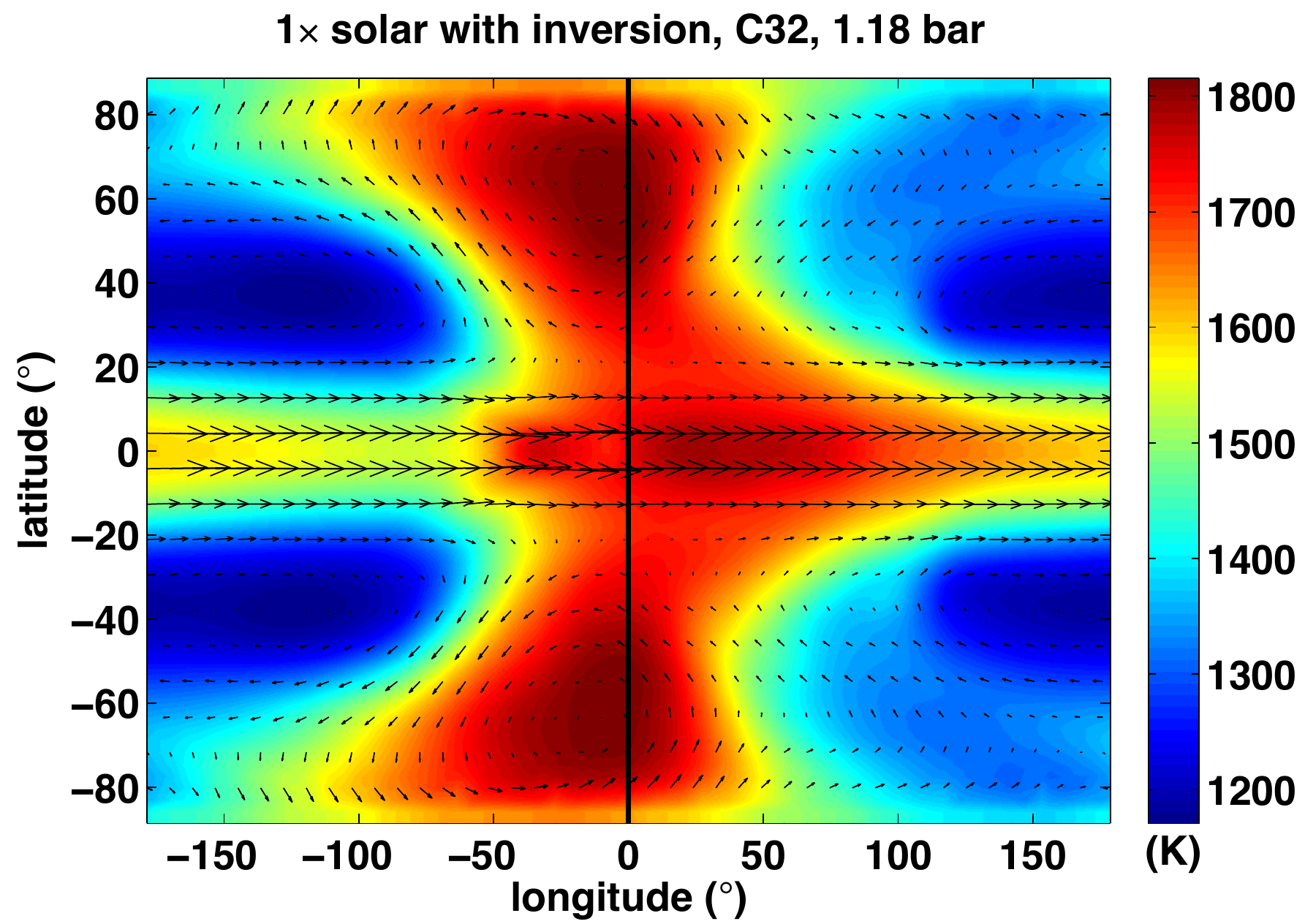}
\includegraphics[trim = 0.0in 0.0in 0.0in 0.0in, clip, width=0.4\textwidth]{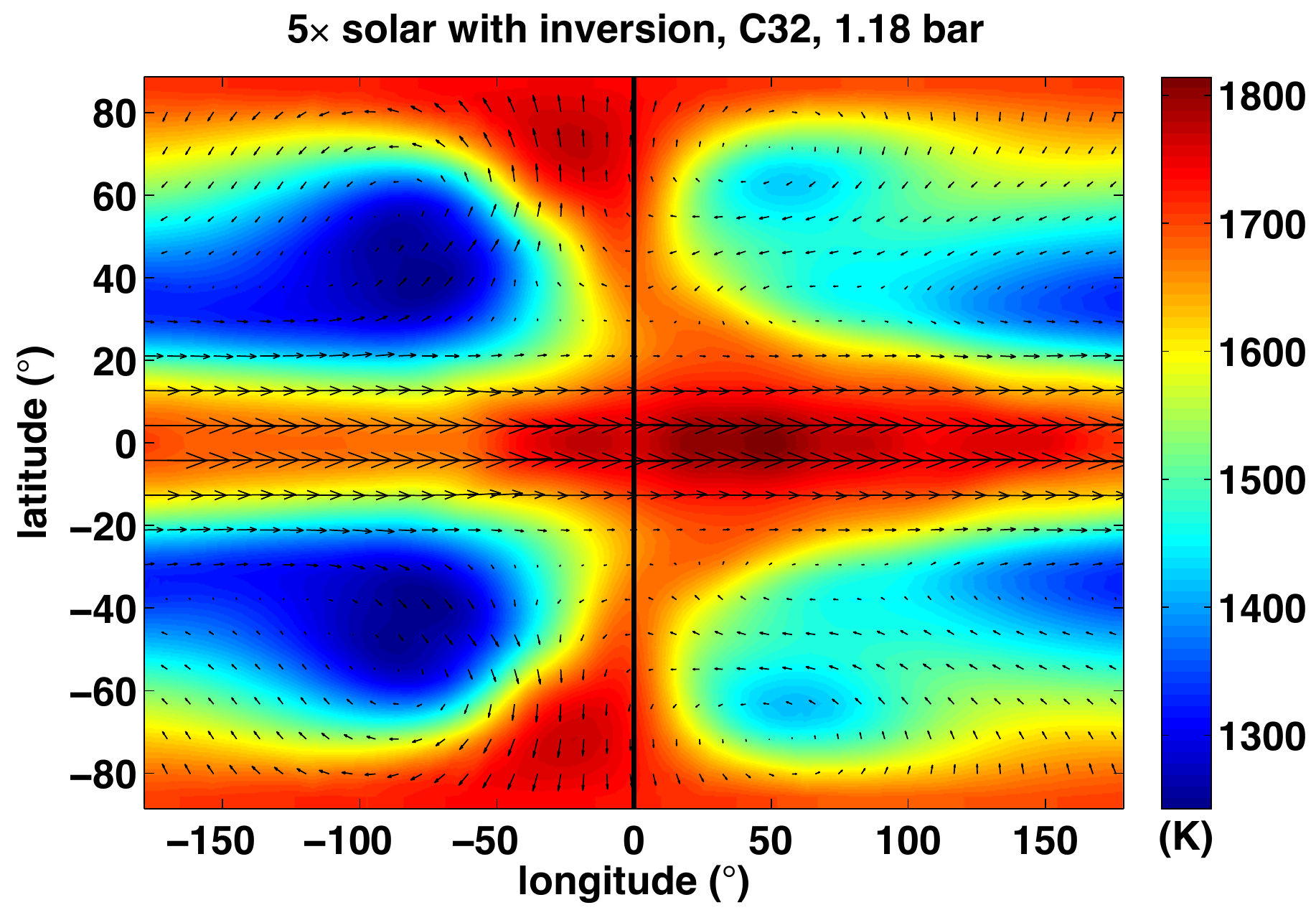}
\caption{Wind and temperature profiles of WASP-43b comparing 1$\times$ and 5$\times$ solar models with TiO/VO at a model resolution of C32. The profiles are shown at four different pressure levels, from top to bottom: 1 mbar, 10 mbar, 100 mbar and 1 bar.  The black line in each profile denotes the longitude of the substellar point. }
\label{windtemp_inversion}
\end{centering}
\end{figure*}

Similar to models of HD 209458b with TiO/VO \citep{showman+2009}, models of WASP-43b with TiO/VO generate localized temperature inversions within $\sim$60 degrees of longitude and latitude from the substellar point (Figure \ref{windtemp_inversion}).  This results in a vastly different temperature structure than non-TiO/VO models at pressures less than $\sim$100 mbar.  At these low pressures, the planet's maximum dayside temperature is higher than those in non-TiO/VO models (compare to Figures \ref{windtemp_onesolar} and \ref{windtemp_fivesolar}).  These high-temperature regions correspond to those areas where TiO and VO are in gaseous form, and therefore contribute another opacity source to the atmospheric heating.  While we do not explore the observational implications of these models in detail, we would expect that phase curves would exhibit higher phase variations and smaller phase shifts than non-TiO/VO models.

\section{Comparing models to HST WF3 observations}

Because the SPARC/MITgcm utilizes full radiative transfer, we can use our model outputs to generate synthetic spectra and light curves and directly compare them to observational data.  We compare our 1$\times$ and 5$\times$ solar models without TiO/VO as well as our 1$\times$ models with drag to the WFC3 observations, particularly in emission.  

\begin{figure}
\begin{centering}
\epsscale{.80}
\includegraphics[trim = 0.0in 0.0in 0.0in 0.0in, clip, width=0.45\textwidth]{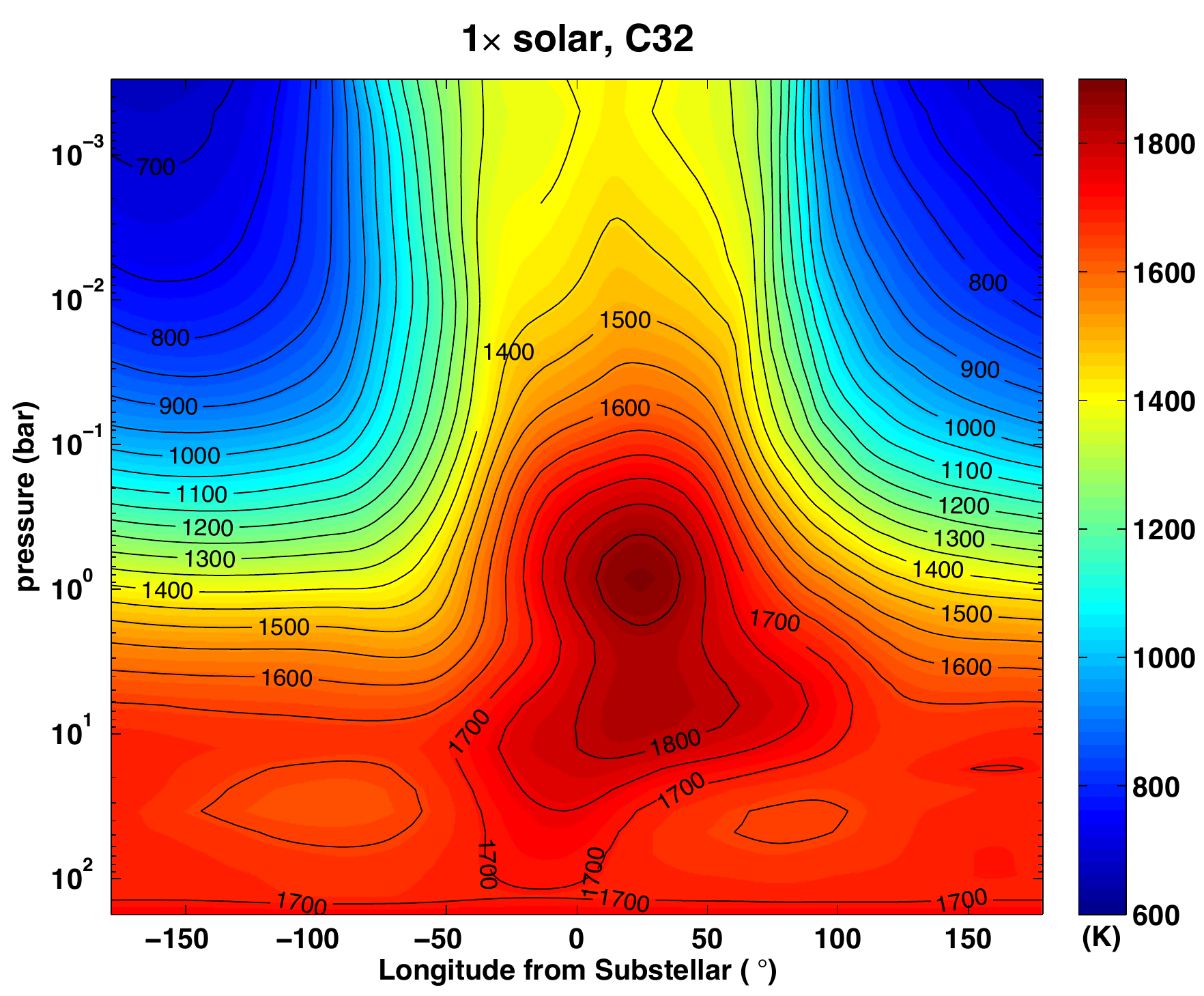}\\
\includegraphics[trim = 0.0in 0.0in 0.0in 0.0in, clip, width=0.45\textwidth]{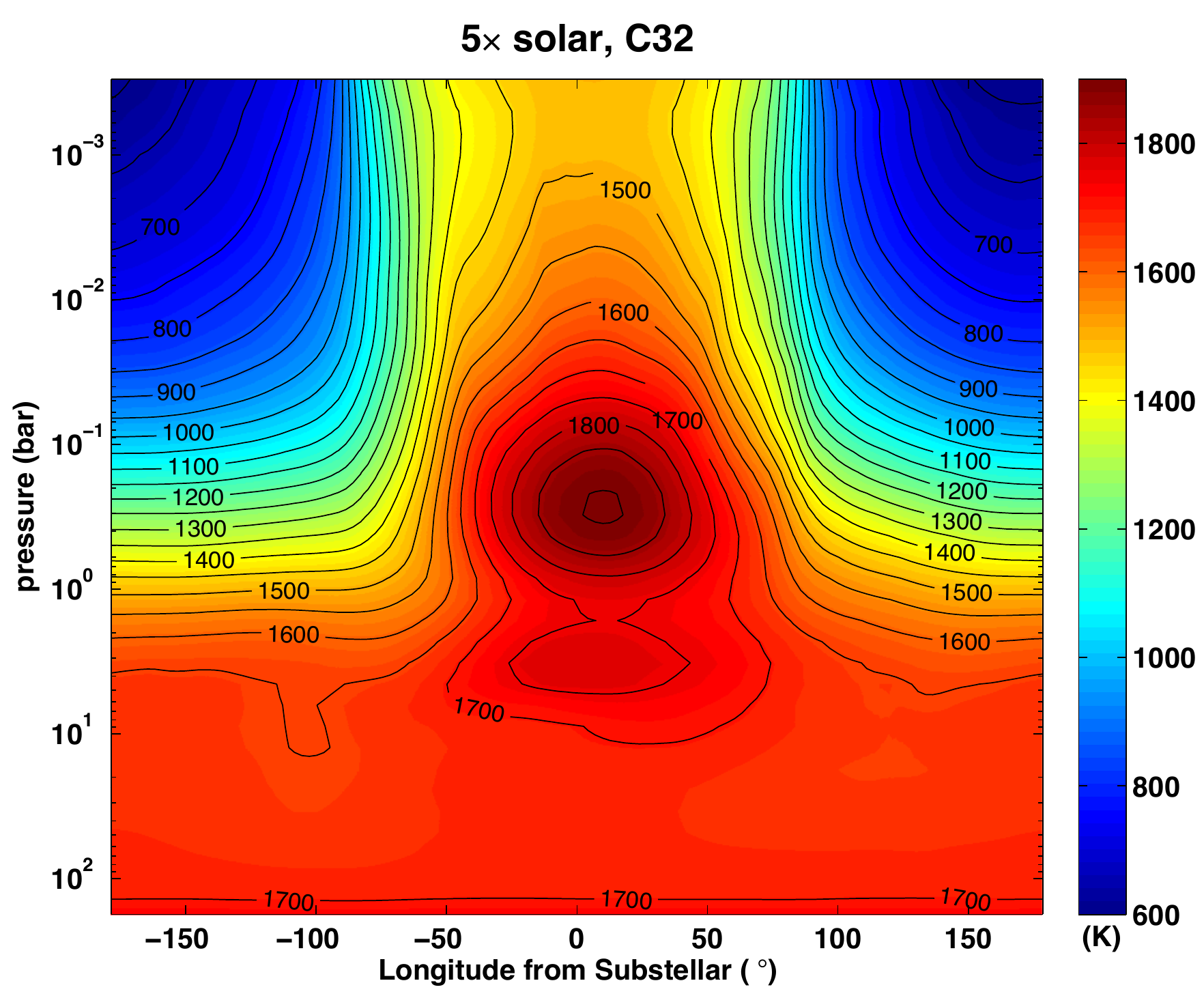}\\
\caption{Average temperature as a function of pressure and degrees from the substellar point for two atmospheric compositions: 1$\times$ solar (top) and 5$\times$ solar (bottom) without TiO/VO. These average temperatures are weighted by $\cos{\phi}$, where $\phi$ is latitude; this is equivalent to weighting each grid point by its projection angle toward an observer at the equator.   }
\label{templon_plots}
\end{centering}
\end{figure}

First, we can predict the magnitude of shifts in IR phase curves by plotting the average temperature as a function of pressure and longitude from the substellar point, as shown at C32 for the 1$\times$ and 5$\times$ solar models in Figure \ref{templon_plots}.  Here the temperature is averaged and weighted by $\cos \phi$, where $\phi$ is the latitude. At photospheric pressures, the hotspot in the 1$\times$ solar model is approximately 20-30 degrees from the substellar point, while the 5$\times$ solar model has a hotspot shift of approximately 10-20 degrees.  The latter is consistent with the phase shift calculated from the band-integrated ``white" light phase curve by \cite{stevenson+2014}; they observe a peak offset of 40 minutes before eclipse that corresponds to an eastward hotspot shift of 12.3 $\pm$ 1.0 degrees.

\begin{figure}
\begin{centering}
\epsscale{.80}
\includegraphics[trim = 0.0in 0.0in 0.0in 0.35in, clip, width=0.45\textwidth]{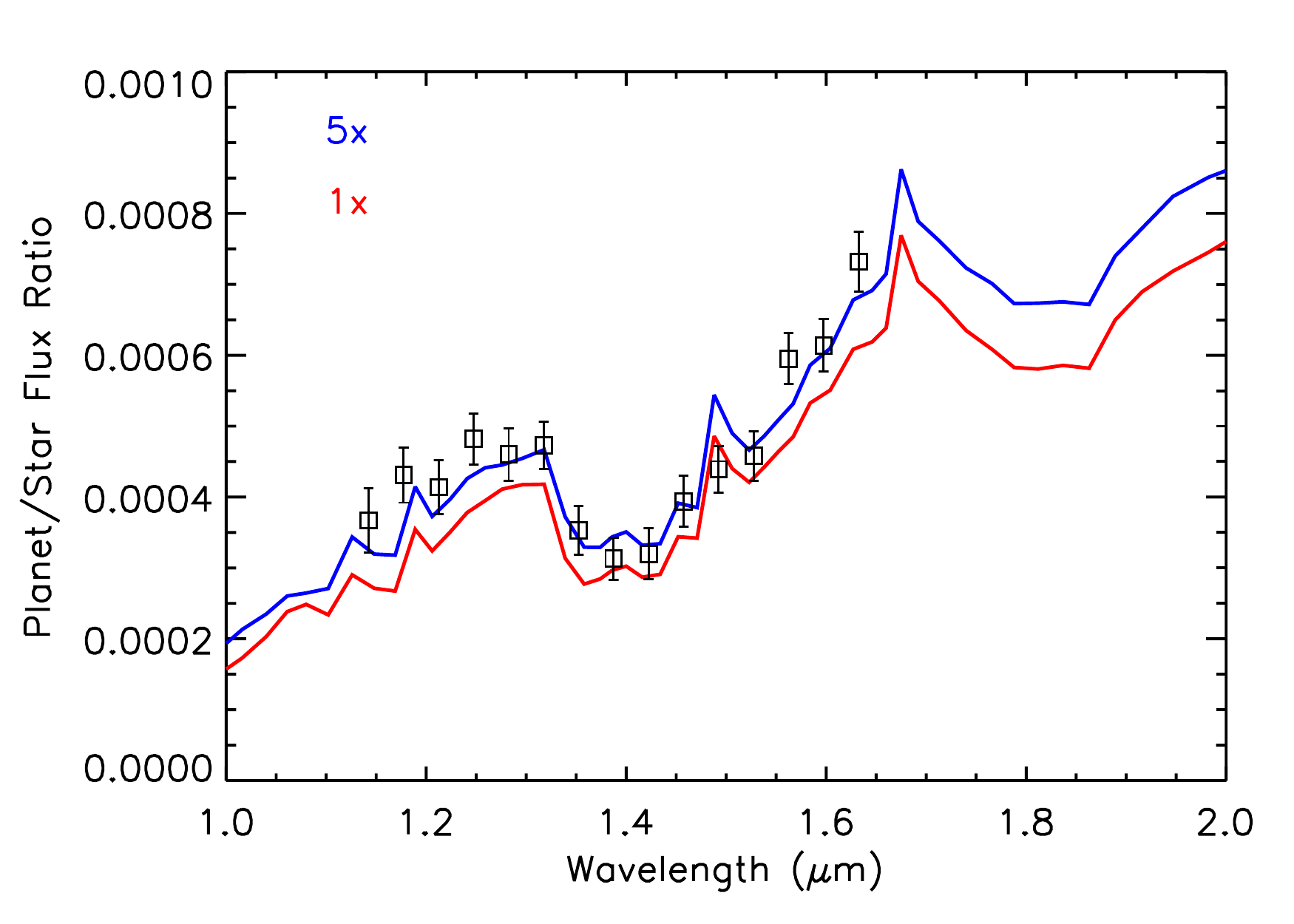}\\
\includegraphics[trim = 0.0in 0.0in 0.0in 0.35in, clip, width=0.45\textwidth]{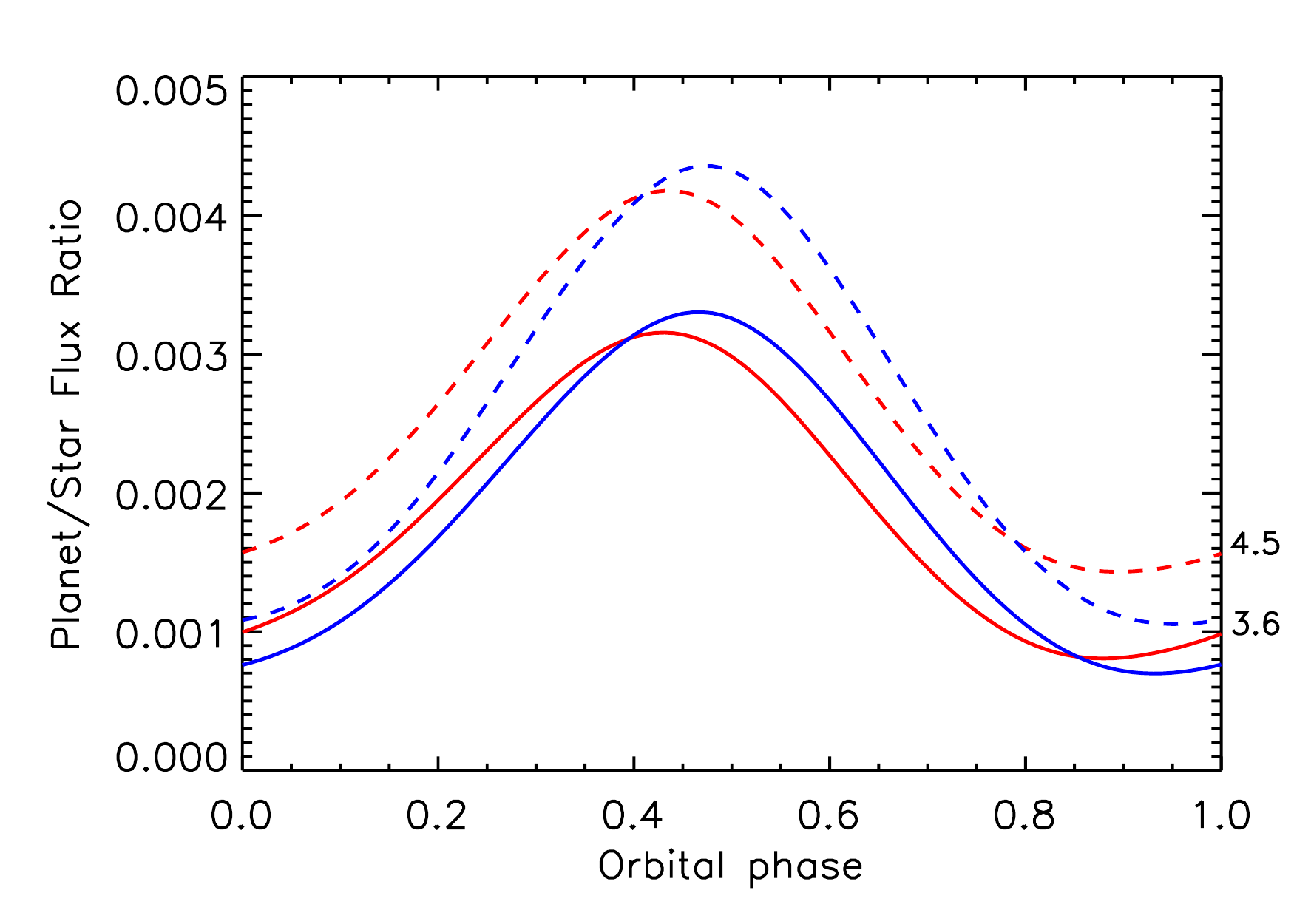}\\
\includegraphics[trim = 0.0in 0.0in 0.0in 0.35in, clip, width=0.45\textwidth]{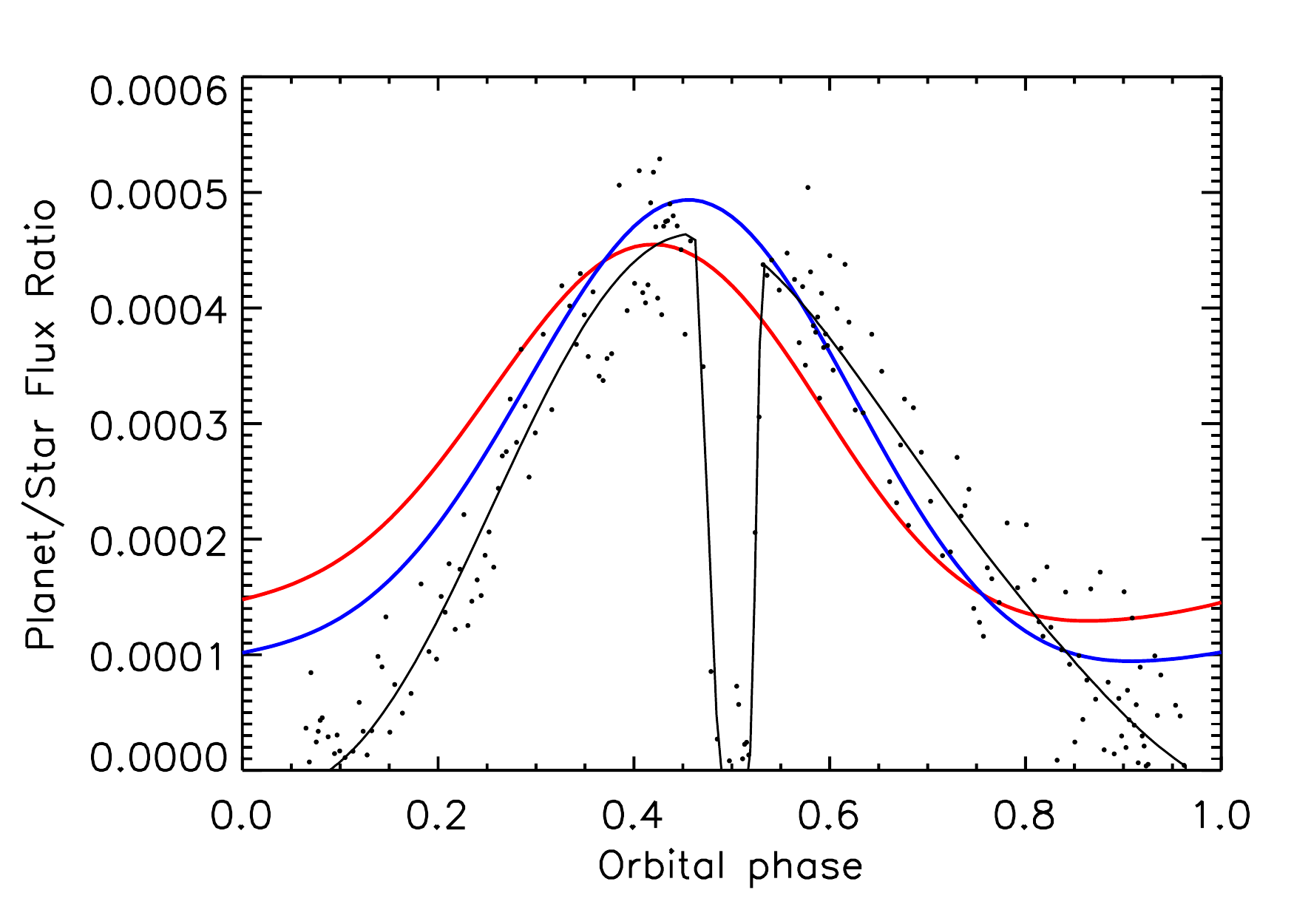}
\caption{Top panel: Predicted flux ratios versus wavelength for 1$\times$ solar (red profiles) and 5$\times$ solar (blue profiles) atmospheric compositions without TiO/VO.  Overplotted in black boxes (with error bars) is the WFC3 dayside spectrum obtained by \cite{stevenson+2014}. Middle panel: light curves for 1$\times$ and 5$\times$ solar models plotted as a function of orbital phase at the Spitzer IRAC bands: 3.6 microns (solid curves) and 4.5 microns (dashed curves). Bottom panel: 1$\times$ and 5$\times$ solar band-integrated ``white" light curves, plotted as a function of orbital phase.  Overplotted in black dots are the WFC3 binned light curve data, with the black line plotting the best fit. For the light curves in the middle and bottom panels, transit occurs at an orbital phase of 0.0, while secondary eclipse occurs at a orbital phase of 0.5.}
\label{datacompare_metallicity}
\end{centering}
\end{figure}

Next, we generate synthetic spectra and phase curves from our models to compare directly to the WFC3 data, using the methods described in \cite{fortney+2006} and \cite{showman+2009}.  Figure \ref{datacompare_metallicity} is comprised of three panels that compare the WFC3 data to the 1$\times$ solar (red profiles) and 5$\times$ solar (blue profiles) models.  The top panel plots the planet to star flux ratio on the dayside as a function of wavelength, with the WFC3 dayside emission spectrum overplotted in black squares (with error bars).  It is evident that the predicted dayside flux ratios of the 5$\times$ solar model are an excellent match to each of the dayside WFC3 points.  

In the middle and bottom panels of Figure \ref{datacompare_metallicity} we plot the flux ratio as a function of orbital phase at three wavelengths: the Spitzer IRAC 3.6 and 4.5 micron bands (middle panel) and in WFC3 ``white" light (band-integrated across WFC3 wavelengths, bottom panel).  In the bottom panel, the WFC3 binned white light curve is overplotted in black dots, with the solid black line corresponding to the best fit.  As predicted, the peak IR flux in the 5$\times$ solar model occurs later (i.e., closer to secondary eclipse) than in the 1$\times$ solar model, as the observations probe shallower pressures where the day-night phase shift is less.  The amplitude of phase variations for the 5$\times$ solar model is also larger.  Here again, our 5$\times$ solar model better matches the WFC3 observations; the peak IR flux of the 1$\times$ solar model occurs too early as compared to the data and the amplitude of the phase variations is too small.  

\begin{figure}
\begin{centering}
\epsscale{.80}
\includegraphics[trim = 0.0in 0.0in 0.0in 0.35in, clip, width=0.45\textwidth]{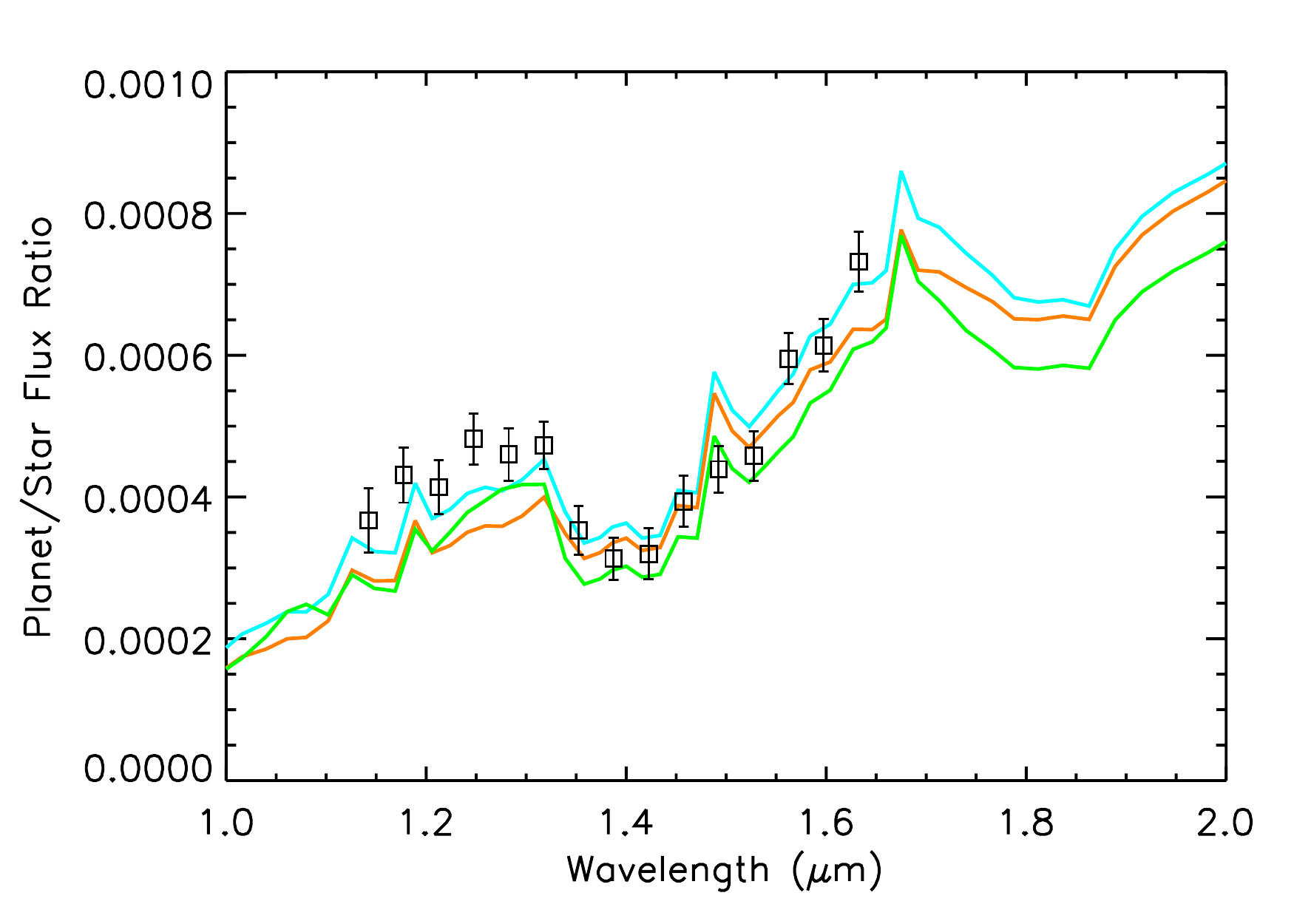}\\
\includegraphics[trim = 0.0in 0.0in 0.0in 0.35in, clip, width=0.45\textwidth]{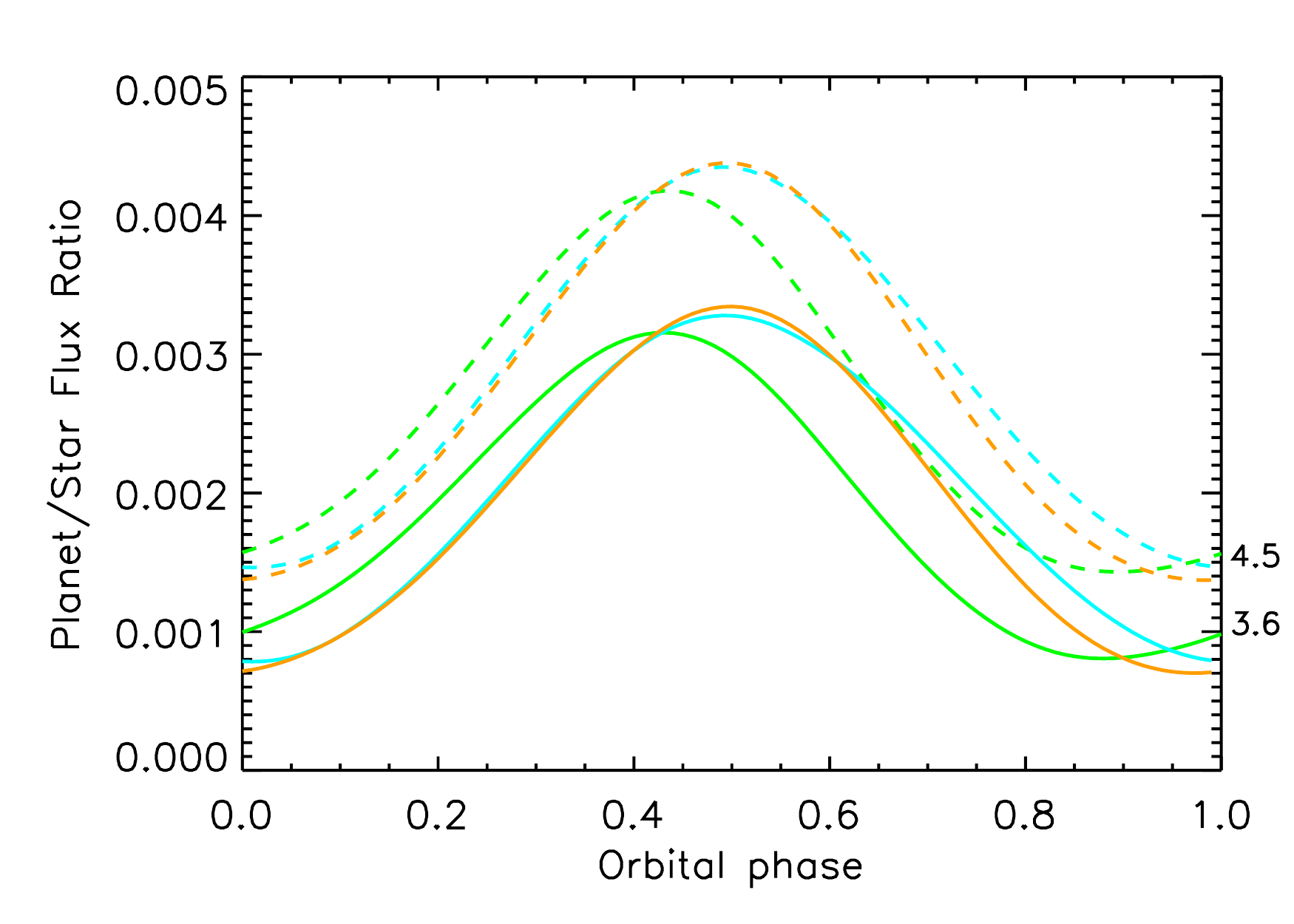}\\
\includegraphics[trim = 0.0in 0.0in 0.0in 0.35in, clip, width=0.45\textwidth]{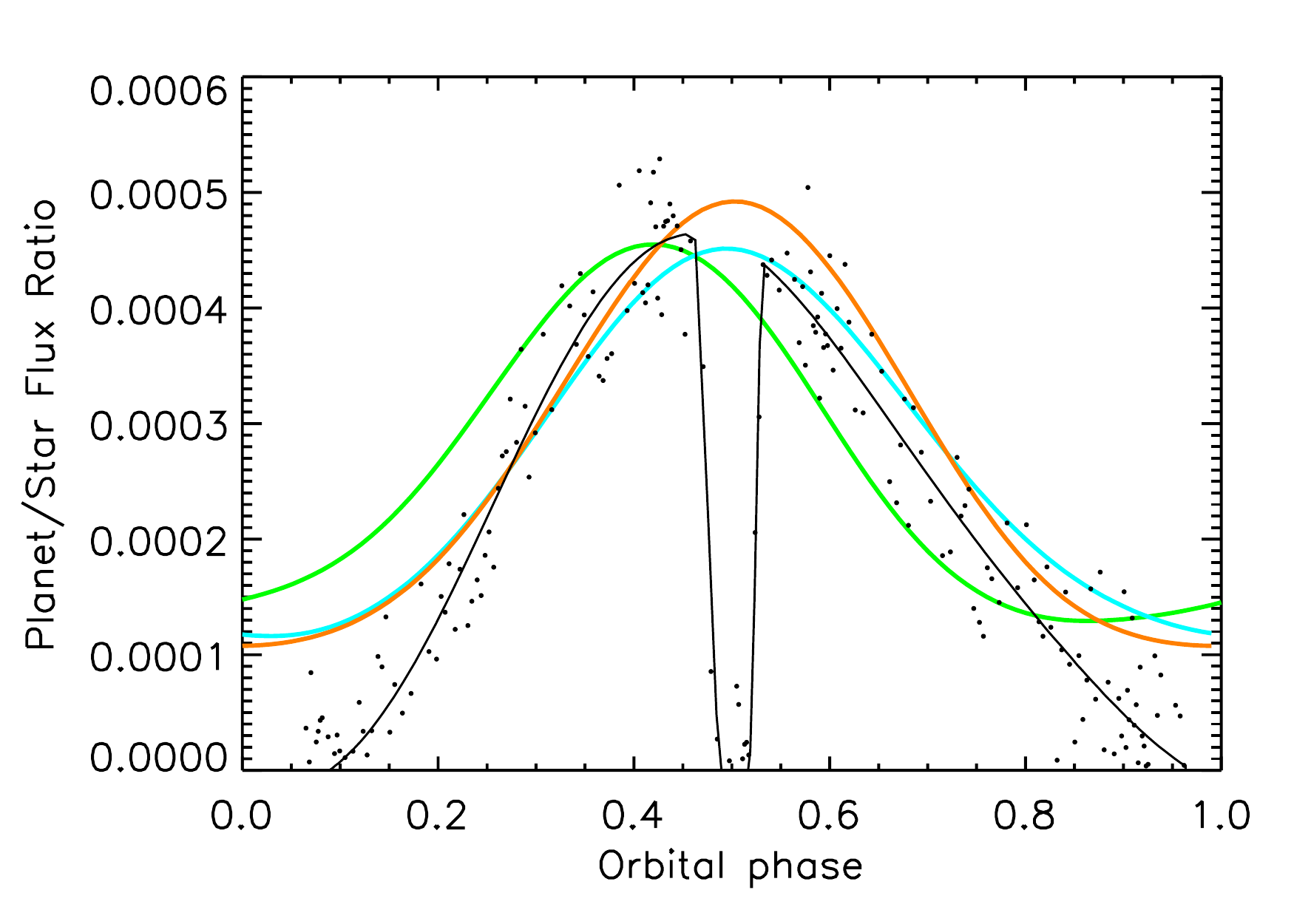}
\caption{Top panel: Predicted flux ratios versus wavelength for 1$\times$ solar models without TiO/VO with varying degrees of frictional drag: $\mathrm{\tau_{drag}}=0$ s (green line), $\mathrm{\tau_{drag}}=3\times10^5$ s (orange line) and $\mathrm{\tau_{drag}}=1\times10^5$ s (blue line).  Overplotted in black boxes is the WFC3 dayside spectrum obtained by \cite{stevenson+2014}.  Middle panel: Light curves for the same frictional drag models plotted as a function of orbital phase at the Spitzer IRAC bands: 3.6 microns (solid curves) and 4.5 microns (dashed curves).  Bottom panel: Band-integrated ``white" light curves for each frictional drag model, plotted as a function of orbital phase.  Overplotted in black dots are the WFC3 binned light curve data, with the black line plotting the best fit. For the light curves in the middle and bottom panels, transit occurs at an orbital phase of 0.0, while secondary eclipse occurs at a orbital phase of 0.5.}
\label{datacompare_drag}
\end{centering}
\end{figure}

Next we compare our frictional drag models to the WFC3 data (Figure~\ref{datacompare_drag}). Like our nominal 1$\times$ and 5$\times$ solar models without upper-level large-scale drag, our models with frictional drag produce a dayside spectrum that matches the observations reasonably well, especially when the drag time constant is shortest ($10^5$ s).  Nevertheless, despite overall similar shapes, the synthetic dayside spectra trend slightly lower in flux than the observations.  As expected theoretically, stronger drag reduces the day-night circulation efficiency and therefore makes the dayside hotter, leading to brighter dayside spectra for shorter drag time constants (Figure 12, top panel).   Examining the phase curves (Figure 12, middle and bottom), the nightside in models with strong frictional drag is likewise cooler than models without such drag, again as predicted theoretically. Nevertheless, the effect is modest, and even in our models with strong drag ($10^5$ s), the nightside remains significantly warmer than the observations indicate.  It would thus appear that, at least over the range of drag strength explored here, the addition of drag does not provide a ready solution to the model-data discrepancy on the nightside. Strong drag also has the effect of reducing the offset of the flux peak from secondary eclipse, again as predicted by basic theoretical considerations.  

\begin{figure*}
\begin{centering}
\epsscale{.80}
\includegraphics[trim = 0.6in 2.3in 0.6in 1.5in, clip, width=0.95\textwidth]{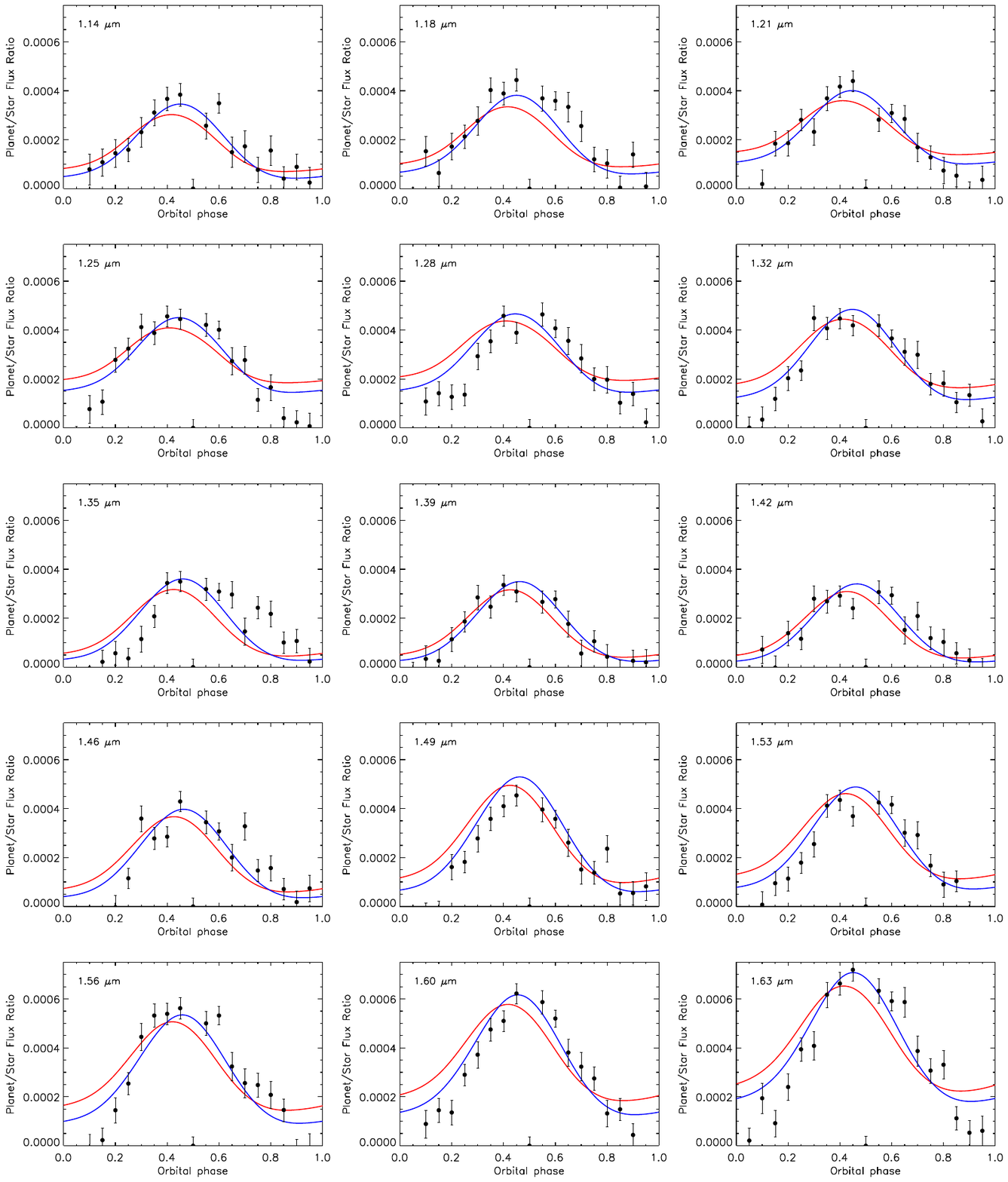}%Fig 13
\caption{Comparison of our 1$\times$ and 5$\times$ solar models of WASP-43b (red and blue lines, respectively) with the spectroscopic phase curves obtained by WFC3 at each binned wavelength (black points, with error bars).  For each curve, transit occurs at an orbital phase of 0.0 and secondary eclipse occurs at a phase of 0.5. }
\label{lc_all}
\end{centering}
\end{figure*}

\begin{figure}
\begin{centering}
\epsscale{.80}
\includegraphics[trim = 0.3in 0in 0.0in 0.0in, clip, width=0.48\textwidth]{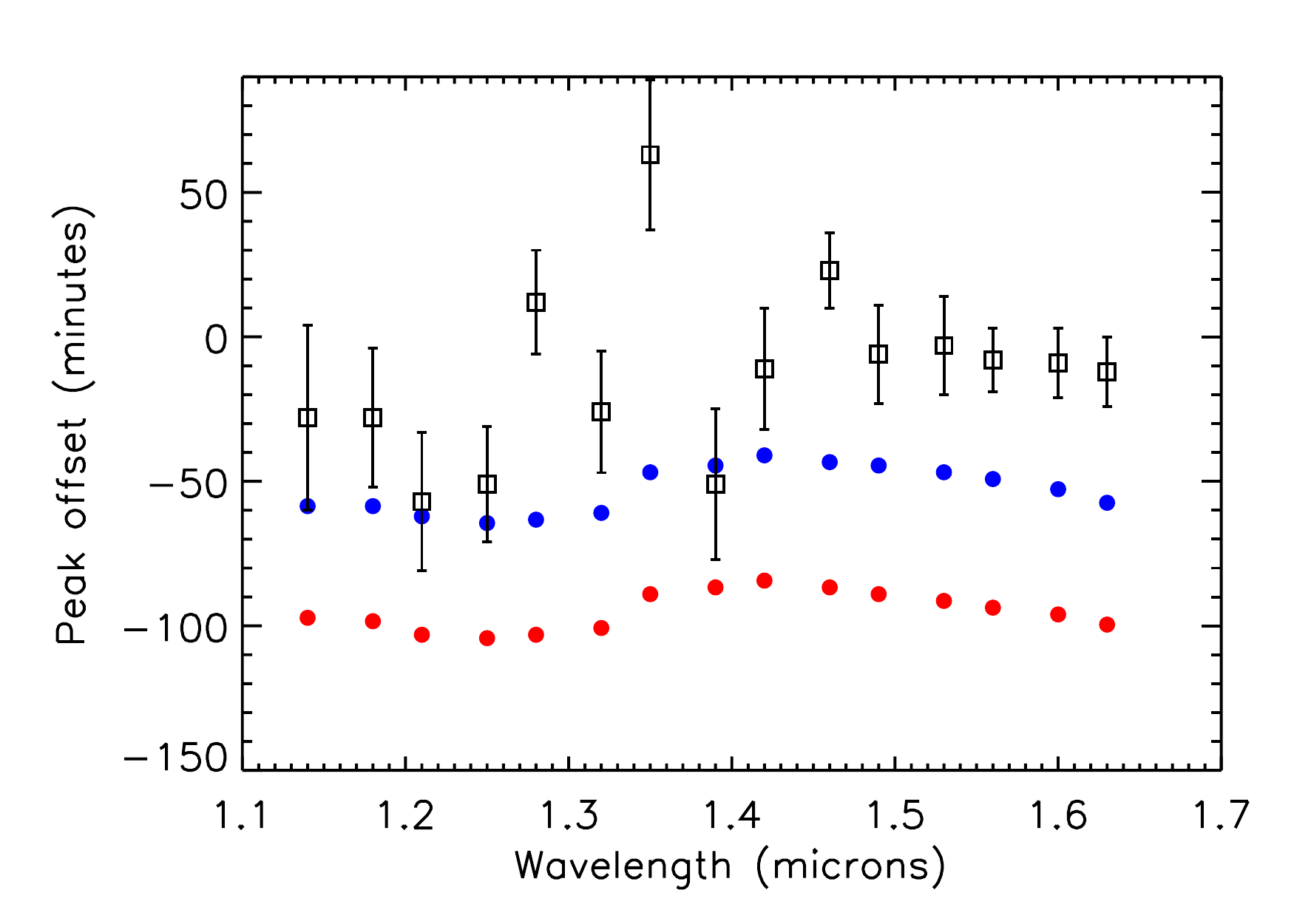} %Fig 14
\caption{Model phase curve peak offsets at each binned WFC3 wavelength for 1$\times$ (red) and 5$\times$ solar (blue) atmospheric compositions.  Peak offsets measured from the WFC3 phase curves are overplotted in black squares with error bars. }
\label{peakoffset_plot}
\end{centering}
\end{figure}

\begin{figure*}
\begin{centering}
\epsscale{.80}
\includegraphics[trim = 0.6in 2.3in 0.6in 1.5in, clip, width=0.95\textwidth]{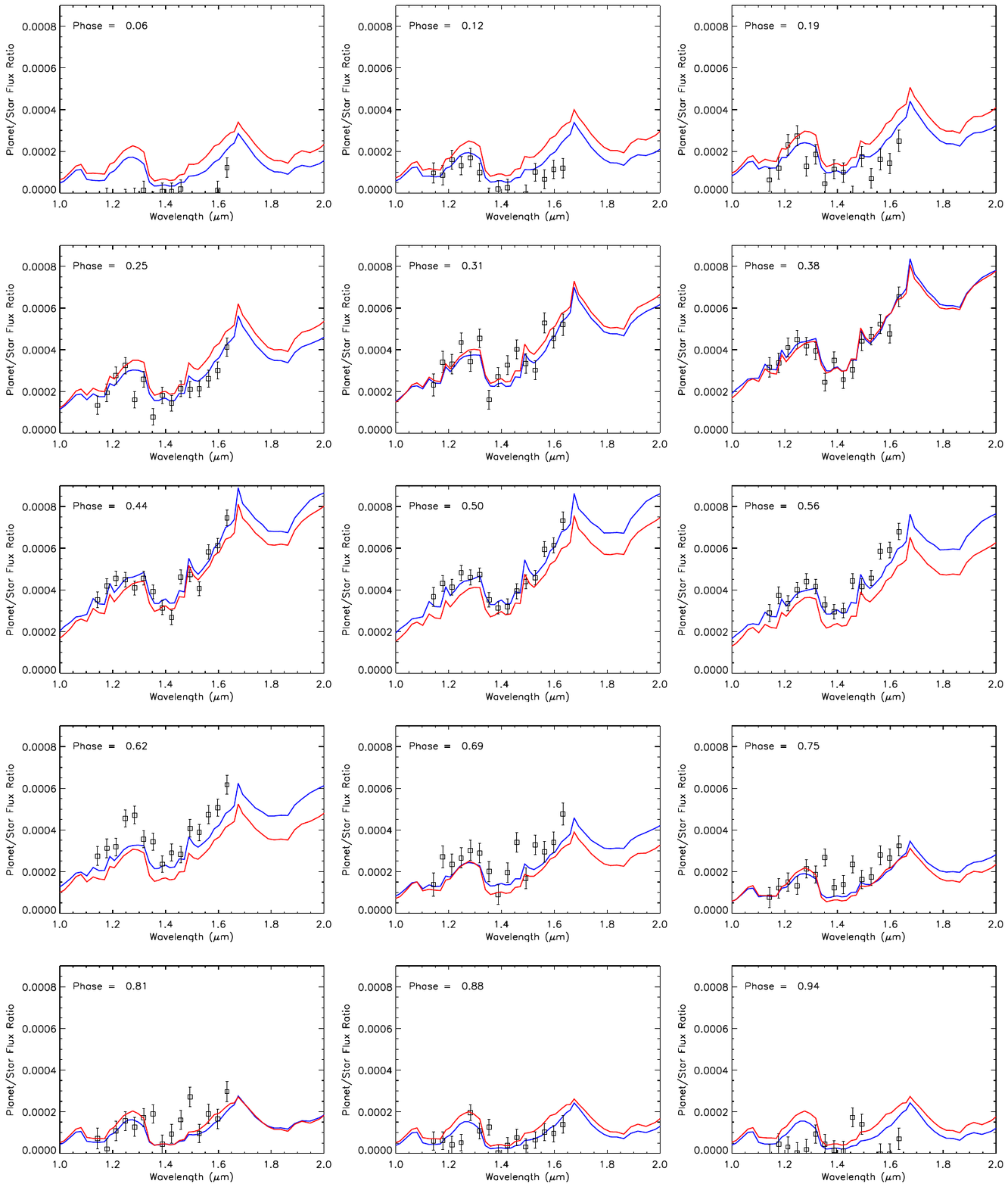} %Fig 15
\caption{Comparison of our predicted flux ratios of WASP-43b at 1$\times$ and 5$\times$ solar (red and blue lines, respectively), with the WFC3 emission spectrum at each binned phase (black squares, with error bars).}
\label{emer_all}
\end{centering}
\end{figure*}

\subsection{Model comparisons at each binned wavelength and phase}

Because the WFC3 dataset contains phase information at multiple wavelengths, we can compare our models to the data at each binned wavelength and phase; this is shown in Figures \ref{lc_all}, \ref{peakoffset_plot} and \ref{emer_all}.  In Figure \ref{lc_all} we compare our 1$\times$ (red) and 5$\times$ (blue) model phase curves at each binned wavelength to the WFC3 phase curves (black dots, with error bars). For each wavelength, we still see that the 5$\times$ solar model better matches both the timing of peak IR flux and also the flux amplitude.  We re-express this comparison in wavelength and phase shift in Figure 14. The phase curve peak offsets for the 1$\times$ and 5$\times$ solar models are plotted at the binned wavelengths shown in Figure \ref{lc_all}, with the WFC3 measurements plotted in black squares.  Aside from the $\sim$50 min outlier at 1.35 $\mu$m, the relative peak offsets obtained from the models are consistent with the trend of the measured values, particularly those for the 5$\times$ solar model.

Figure \ref{emer_all} shows how well we are matching the flux on the dayside, plotting the flux ratios as a function of wavelength at each binned orbital phase from the nightside (phases 0.06, 0.12, and 0.19) to the dayside (phases $0.25-0.75$) and again to the nightside (phases 0.81, 0.88, and 0.94).  Comparing the model spectra alone, as the phase moves from nightside to dayside, the amplitude of the 1$\times$ and 5$\times$ solar models flip in magnitude; the 1$\times$ solar model emits more flux on the nightside and less flux on the dayside than the 5$\times$ solar model.  This stems from the fact that the 1$\times$ solar model is recirculating heat more efficiently to the nightside, as evident in the smaller variations in flux with orbital phase.  Comparing the models to the observed emission spectrum at each binned phase, we see again that there is better agreement between the models and spectra on the dayside than on the nightside.  

The brighter nightside fluxes derived from our models yield recirculation efficiencies that are larger than the value calculated from the retrieved WFC3 dayside and nightside fluxes \citep{stevenson+2014}.  Using the expression from \cite{stevenson+2014}, $\mathcal{F}=\frac{1}{2}(1+\frac{F_{night}}{F_{day}})$, we calculate a recirculation efficiency of 0.67 and 0.61 for the 1$\times$ and 5$\times$ solar models, respectively.  Compared to the recirculation efficiency of 0.503 derived from the WFC3 observations, our models suggest a more efficient recirculation of heat from the dayside to the nightside. This is re-emphasized in a comparison of the retrieved dayside pressure-temperature profile from the WFC3 data compared to model profiles on the dayside (Figure \ref{dayside_pt}).  The retrieved profile (black line) falls within the hottest dayside profiles (red and blue solid lines) and not the dayside average or globally-averaged pressure-temperature profiles (dashed and dash-dot lines).  

\subsection{Constraining the nightside flux}

Overall, while our 5$\times$ solar model does the best job of matching the WFC3 data on the dayside, we find much poorer agreement on the nightside at all wavelengths.  One might expect that the nightside at the WFC3 wavelengths should appear much cooler.  With dayside temperatures in the 1600-1800 K range and nightside temperatures in the 800-1000 K range, the WFC3 band falls near the peak of the Planck function on the dayside and shortward of the peak on the nightside.  Therefore, at WFC3 wavelengths we are sampling a smaller fraction of the flux on the nightside, and one might expect much larger flux variations as compared to Spitzer wavelengths, which sample more of the nightside flux.

As \cite{burrows+1997} and others have shown, the reason for this supposed discrepancy is the fact that WASP-43b is emitting more flux at low $\mathrm{T_{eff}}$ (on the nightside) in the near-IR than what would be expected for a simple blackbody.  Figure \ref{emer_flux} illustrates this point by plotting the emergent flux of WASP-43b as a function of wavelength for our 1$\times$ solar atmospheric model without TiO and VO.  In the first panel, the values are plotted as a function of wavelength, and in the second, the fluxes are divided by the emergent flux at transit.  For both panels we present the emergent flux at six orbital phases: transit, when nightside is visible (black line); 60$^{\circ}$ after transit (red line); 120$^{\circ}$ after transit (green); secondary eclipse, when the dayside is visible (dark blue); 60$^{\circ}$ after secondary eclipse (light blue); and 120$^{\circ}$ after secondary eclipse (magenta). Overplotted in the top panel in dashed lines are blackbody curves for temperatures ranging from 700-1700 K, and both panels plot the normalized WFC3 and Spitzer bandpasses in dash-dot and dotted curves, respectively.  Indeed, on the nightside (black, red, and magenta profiles) the planet is emitting much more flux than a simple blackbody profile.  Since mid-IR water opacity is pushing flux in the blue end of the IR, the fluxes in this range rise and fall with the near-IR fluxes (bottom panel).  Still, because we fall shortward of the peak of the Planck function on the nightside at WFC3 wavelengths, the nightside effective temperature is difficult to constrain.  
\begin{figure}
\begin{centering}
\epsscale{.80}
\includegraphics[trim = 0.0in 0.0in 0.0in 0.0in, clip, width=0.48\textwidth]{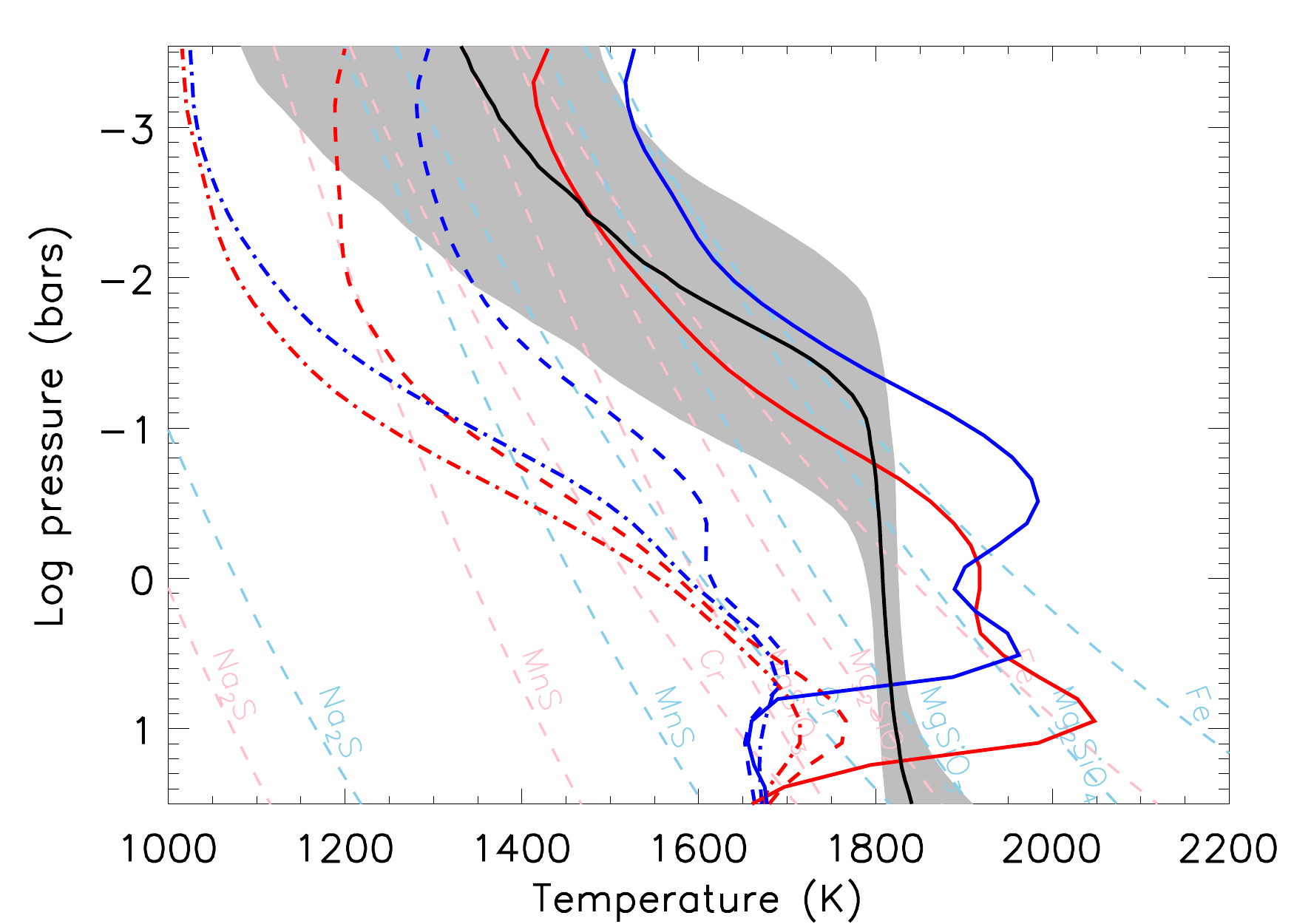} %Fig 16
\caption{Comparison of model dayside pressure-temperature profiles to the retrieved temperature profile from the WASP-43b secondary-eclipse data.  Solid red and blue lines show the hottest pressure-temperature profiles on the dayside for the 1$\times$ and 5$\times$ solar models, respectively.  Dashed lines and dot-dash lines represent the dayside average and globally-averaged pressure-temperature profiles for those same models.  The retrieved WFC3 pressure-temperature profile is overplotted in the solid black line, with 1$\sigma$ uncertainties in shaded grey.  Overplotted in pink and light blue dashed lines are condensation curves for $ \rm Na_2S$, MnS, Cr, $\rm MgSiO_3$, $\rm Mg_2SiO_4$, and Fe; these materials are expected to condense in planetary and brown dwarf atmospheres.}
\label{dayside_pt}
\end{centering}
\end{figure}

\begin{figure}
\begin{centering}
\epsscale{.80}
\includegraphics[trim = 0.0in 0.0in 0.0in 0.0in, clip, width=0.45\textwidth]{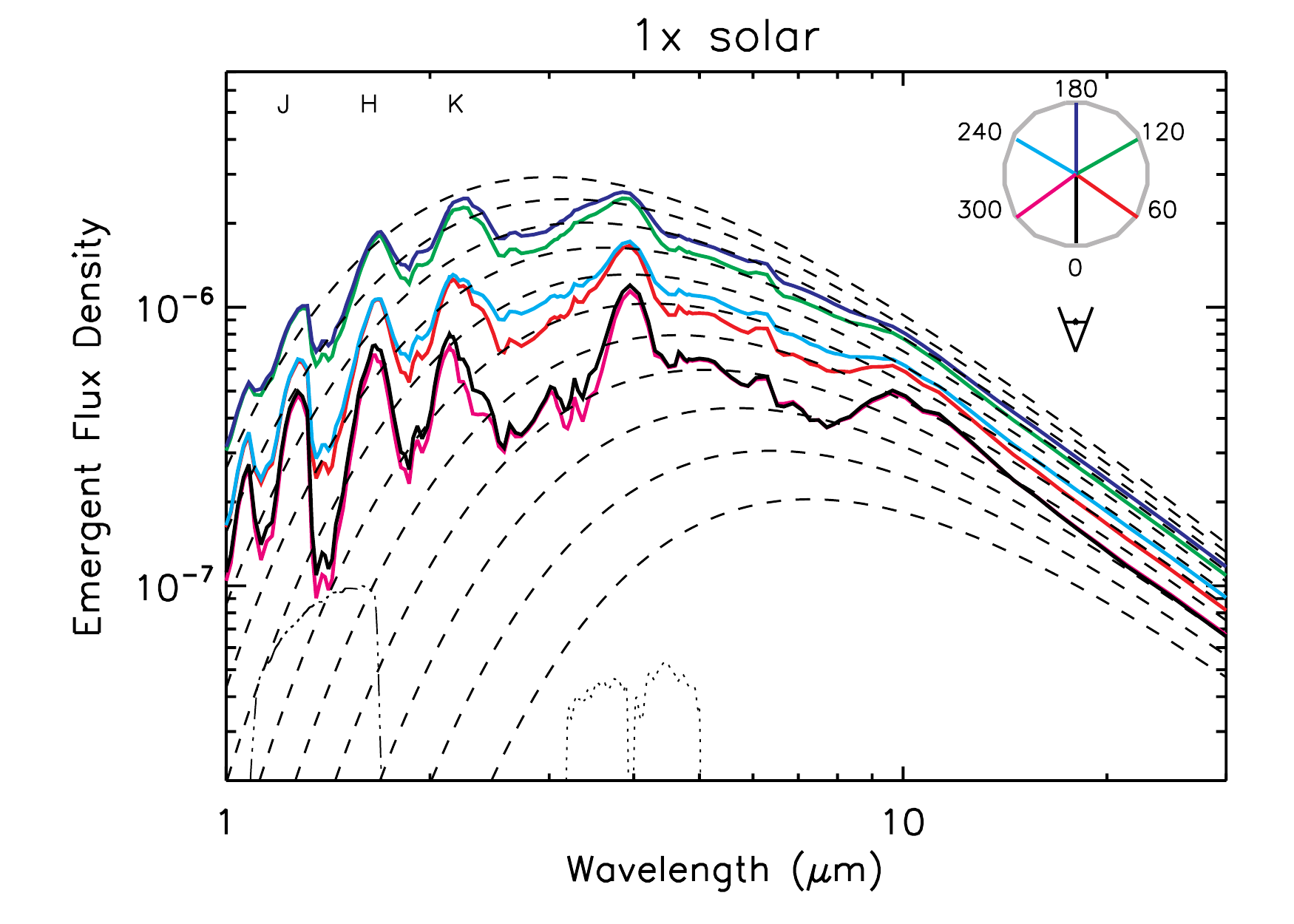} %Fig 17A
\includegraphics[trim = 0.0in 0.0in 0.0in 0.0in, clip, width=0.45\textwidth]{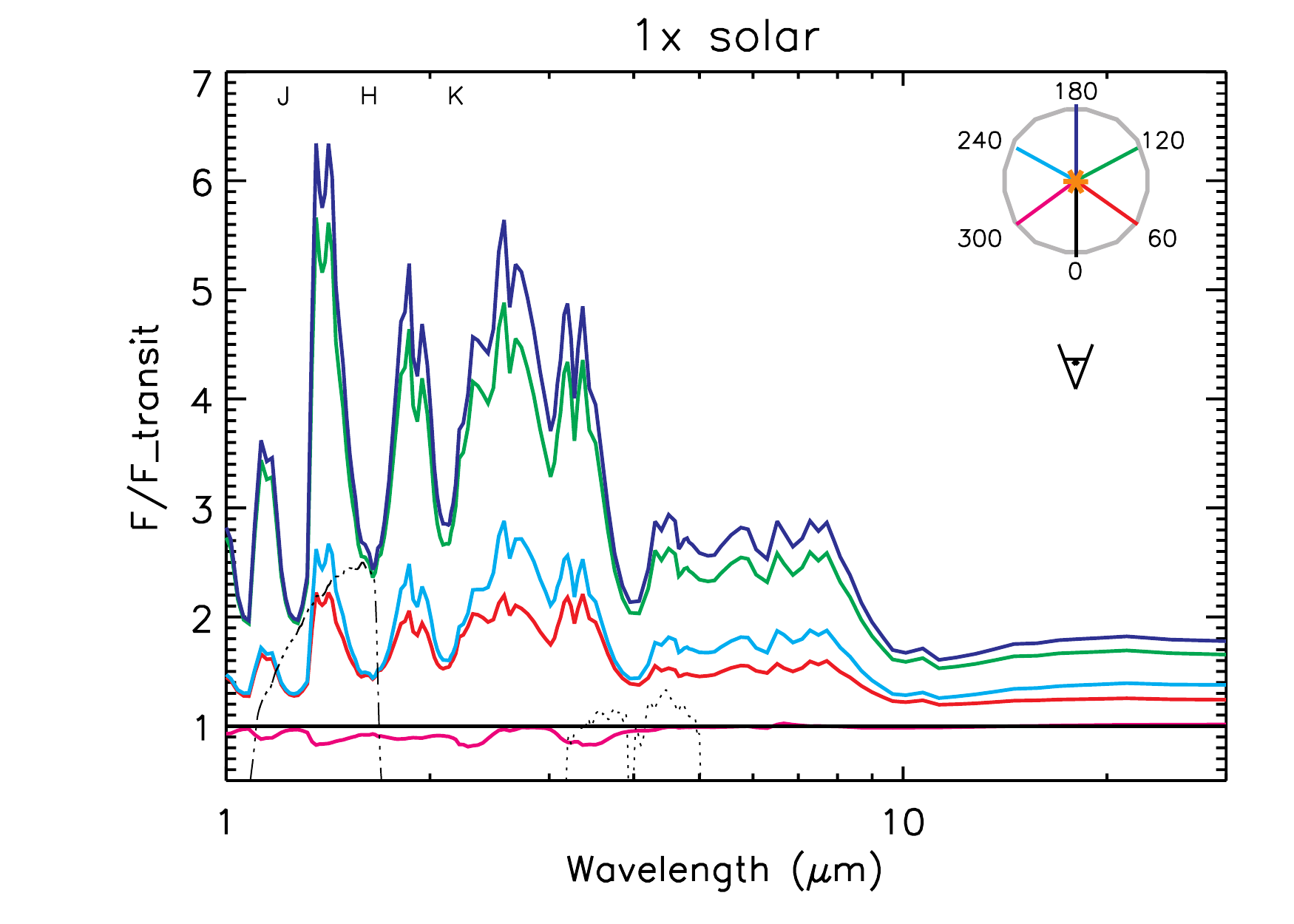} %Fig 17B
\caption{Top: Emergent flux density (in units of ergs$^{-1}$s$^{-1}$cm$^{-2}$Hz$^{-1}$) for 1$\times$ solar composition without TiO/VO as a function of wavelength at six orbital phases: transit, when nightside is visible (black line); 60$^{\circ}$ after transit (red line); 120$^{\circ}$ after transit (green); secondary eclipse, when the dayside is visible (dark blue); 60$^{\circ}$ after secondary eclipse (light blue); and 120$^{\circ}$ after secondary eclipse (magenta).  Overplotted in dashed lines are blackbody curves for 700-1700 K, in increments of 100 K.  Bottom: The same emergent flux density divided by the emergent flux at transit, at the same phases plotted above.  The phases are illustrated in the inset figure, shown in the top right of each panel.  Overplotted in both panels are the normalized WFC3 ``white-light" bandpass (dash-dotted curve) and the two Spitzer IRAC bandpasses (dashed curves).  The letters at the top of both panels indicate the locations of the J, H and K wavelength bands.}
\label{emer_flux}
\end{centering}
\end{figure}

The disagreement on the nightside flux could also be attributed to a thick nightside cloud at low pressures, which has been suggested in studies such as \cite{showman+guillot_2002}.  In this scenario, our observations would probe lower pressures where the day-night temperature difference is larger, causing the nightside emergent fluxes to be low.  We highlight possible cloud species by plotting condensation curves in Figure~\ref{dayside_pt}, shown as light pink and light blue dashed lines.  These curves correspond to the pressures and temperatures where the vapor pressure of each element or molecule is equal to its saturation vapor pressure.  The cloud base would be located where the condensation curve intersects with a given $p$-$T$ profile.  Comparing the globally-averaged $p$-$T$ profiles (dash-dot curves) to the condensation curves, we see that for both 1$\times$ and 5$\times$ solar compositions, MnS and Cr are possible condensibles at pressures probed by the WFC3 observations.
  Given the nightside temperatures quoted above, $\rm Na_2S$ clouds are possible as well.  However, it is likely that these cloud types would not be optically thick enough to modify the emission spectrum at WFC3 wavelengths \citep{morley+2012}.  Condensation of the more-abundant silicate minerals at greater pressure, and vertical lofting to high altitude, may be another possible source of cloud material.  Nevertheless, the apparent detection of molecular features in transit spectra \citep{kreidberg+2014b} might suggest that any nightside clouds do not extend all the way to the planetary terminators.

Lastly, we might find better agreement in the nightside flux by imposing a much shorter drag time constant (much less than $10^5$ s) throughout the domain, or instead allowing the drag time constant to vary both horizontally and vertically.  In the latter case, studies such as \cite{perna+2010} have shown that the frictional drag time constant could be many orders of magnitude larger on the dayside than the nightside, and could increase significantly with altitude. Both of these scenarios could yield larger day-night contrasts and smaller phase offsets, which would better match the WFC3 data.  Future work by our group will further explore this parameter space and its effect on hot Jupiter circulation, with much shorter drag time timescales (e.g., $\rm 10^3$ s; Komacek \& Showman in prep).

\begin{figure}
\begin{centering}
\epsscale{.80}
\includegraphics[trim = 0.0in 0.0in 0.0in 0.0in, clip, width=0.45\textwidth]{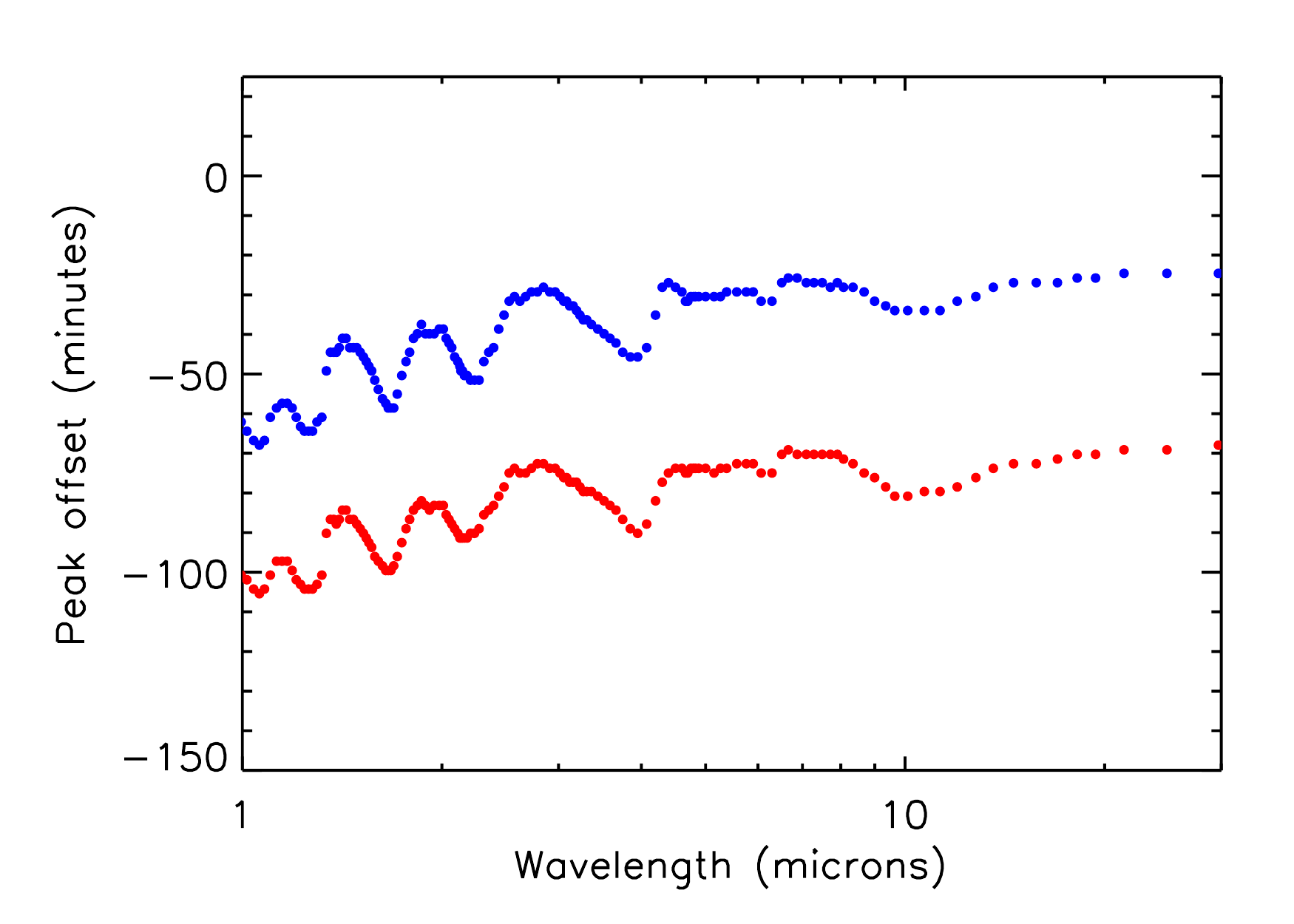} %Fig 18
\caption{Phase curve peak offset as a function of wavelength (196 points in total) for our 1$\times$ (red) and 5$\times$ solar (blue) models.  }
\label{peak_all}
\end{centering}
\end{figure}

To better constrain the planet's nightside flux, it would be useful to observe the nightside at other wavelengths.  It could be that the total (wavelength-averaged) nightside flux is the same for both the models and data, and not as low as the WFC3 data might suggest.  Future observations by JWST would help constrain this value, as the telescope will have the capability to obtain high-precision, spectrophotometric phase curves of exoplanets over a wide range of wavelengths (0.6-28 $\mu$m).  While we do not present the full range of phase curve observations possible with JWST in this paper, Figure \ref{emer_flux} shows the expected range in day-night flux variations at wavelengths longward of 1 $\mu$m.  We also plot the expected phase curve offsets at those same wavelengths (Figure \ref{peak_all}).  The trends in peak offsets seen in Figure \ref{emer_flux} correlate with the flux variations shown in Figure~\ref{peak_all}; at wavelengths where the planet exhibits a large day-night contrast, smaller phase offsets are expected.  These figures can be used to predict the phase curve peak offsets and phase curve amplitudes that could be observed with JWST and other future telescopes.

\section{Discussion}

The overall good agreement between our models and spectrophotometric phase curve observations of WASP-43b--in particular, the fact that the flux peak precedes secondary eclipse and that our phase curve amplitudes are similar to those observed--indicates that our model is in the correct dynamical regime and allows us to make inferences about WASP-43b itself.  In particular, in our model, the shifted flux peak results from the eastward displacement of the dayside hot spot due to an eastward equatorial jet, and this provides evidence indicating that WASP-43b thus exhibits a superrotating equatorial jet.  While such equatorial superrotation and its consequences for lightcurves have long been predicted \citep{showman+guillot_2002} and are now largely understood theoretically \citep{showman+polvani_2011,showman+2013,tsai+2014}, robust data-model comparisons have thus far been limited to only a handful of hot Jupiters.  As such, WASP-43b joins a small but growing list of hot Jupiters--in particular, HD 189733b and HD 209458b--where such flux-peak offsets have been both observed and explicitly predicted in planet-specific GCMs as being due to equatorial superrotation.  Thus, for this trio of planets, we can attribute the observed IR phase curve behavior to superrotation with confidence.  Interestingly, a recent observational-model comparison for the eccentric hot Jupiter HAT-P-2b even suggests that this planet exhibits equatorial superrotation \citep{lewis+2013,lewis+2014}.  Future WFC3 observations, analysis of archival Spitzer data, and continued dynamical modeling may soon significantly expand this list.

By comparing our models of WASP-43b and previous models of hot Jupiter HD 209458b (Showman et al. 2009; N.~K. Lewis, private communication), we can make general statements about the atmospheric circulation of hot Jupiters, particularly as a function of rotation rate, at constant stellar flux.  As the rotation rate on a hot Jupiter is increased, we can expect decreases in equatorial jet width, as the width of the equatorial jet on WASP-43b is 75$\%$ narrower than on HD 209458b, a result of the planet's smaller Rossby deformation radius.  Additional work by our group explores in further detail the effects of rotation rate and also orbital distance on the dynamics \citep{showman+2014}.  

The good agreement between our dayside 5$\times$ solar model and the WFC3 dayside emission spectra \citep{stevenson+2014} is particularly remarkable given the absence of tuning in our model integrations.  Generally, one-dimensional (1D) models of hot Jupiters are 
extensively tuned to match available dayside photometry; in particular, 1D models require an {\it assumption} of the day-night circulation efficiency (either by choosing to adopt the global-mean or dayside-only stellar flux, and/or choose a redistribution efficiency and the pressure range over which it is applied, a treatment that involves a minimum of three free parameters).  1D models are often also tuned with assumptions about an arbitrary visible absorber (leading to at least another three free parameters--the strength of the visible opacity and the pressure range over which it exists).  Recent 1D models of WASP-43b, for example, not only make assumptions about redistribution, but include haze and further vary the carbon-to-oxygen (C/O) ratio in an effort to match $K_s$ band secondary eclipse measurements \citep{zhou+2014}.  In contrast, the circulation efficiency in our self-consistent 3D models is a natural property of the solution, requiring no free parameters.

\section{Conclusions}
We present three-dimensional atmospheric circulation models coupled to full radiative transfer for the hot Jupiter WASP-43b, a 2 Jupiter-mass, 1 Jupiter-radius planet in a 0.81 day orbit around a K7 type star.  We investigate the circulation as a function of three parameters: model resolution, composition (specifically opacity changes that arise from increased metallicity and the inclusion of TiO and VO) and frictional drag.   Overall, we find that the dynamical regime of WASP-43b is not unlike that of other canonical hot Jupiters including HD 189733b and HD 209458b, with large day-night temperature variations at photospheric pressures and equatorial superrotation that leads to eastward offsets of the hottest regions from the substellar point.   We capture these bulk features of the circulation at both high and low resolution.  In increasing the atmospheric metallicity, we find that the equatorial jet becomes both shallower and faster, as energy is deposited higher in the atmosphere where the day-night forcing is greater.  By including TiO and VO, we find localized temperature inversions on the dayside at pressures less than $\sim$100 mbar, which represent the regions where TiO and VO can become gaseous and become an additional opacity source.  As we increase frictional drag in our 1$\times$ solar models, we find that the flow transitions from one that is dominated by equatorial superrotation to one that is almost completely dominated by day-night flow.  
 
We utilize our range of model results to interpret the HST WFC3 observations of WASP-43b presented in \cite{stevenson+2014}.  In generating theoretical lightcurves and spectra from our models, we find that the 5$\times$ solar model (without TiO/VO) is the best match to the observations, particularly on the dayside.  This result is purely a model output, and includes no additional tuning that is often done when using one-dimensional models to interpret observational data.  While our model nightside appears to be too bright, this is not an unexpected result, as the planet's emergent flux deviates significantly from a simple blackbody at shorter wavelengths.  Future observations with both Spitzer and JWST should further clarify the observed differences in nightside fluxes between our models and the WFC3 data.  Flux ratios for the 1$\times$ and 5$\times$ solar models at Spitzer wavelengths are expected to differ by only a few percent, so it may prove difficult to constrain the metallicity from broadband photometry alone.  However, like the WFC3 lightcurves, we would expect that the 5$\times$ solar model at 3.6 and 4.5 $\mu$m should exhibit larger flux variations than the 1$\times$ solar model, and that the timing of peak IR flux should occur closer to secondary eclipse (Figure \ref{datacompare_metallicity}, middle panel).    

Taken together, the circulation studies presented by our group to date using the SPARC/MITgcm have explored exoplanet circulation over a wide range of planetary properties, including eccentricity, orbital distance, rotation rate, mass, gravity, composition, metallicity, and stellar flux.  Even in this large phase space, we continue to find that superrotation is a robust dynamical feature for close-in exoplanets, with only a few exceptions.  As exoplanet surveys continue to discover new planets, circulation modeling will help characterize their atmospheres, and continue to probe further dynamical regimes.

\acknowledgments

The authors were supported by NASA through a grant from the Space Telescope Science Institute (program GO-13467).  T.K. acknowledges support from the Harriet P. Jenkins Pre-Doctoral Fellowship Program (JPFP).  A.P.S. and T.K. were supported by Origins grant NNX12AI79G..  K.B.S. acknowledges support from the Sagan Fellowship Program, supported by NASA and administered by the NASA Exoplanet Science Institute (NExScI).  L.K. acknowledges support from the NSF Graduate Research Fellowship Program.  J. B. acknowledges support from the Alfred P. Sloan Foundation.  Resources supporting this work were provided by the NASA High-End Computing (HEC) Program through the NASA Advanced Supercomputing (NAS) Division at Ames Research Center.  We thank the anonymous referee for their helpful comments and suggestions.

%\bibliography{bibliography}

\begin{thebibliography}{10}
\bibitem[Adcroft et al.(2004)]{adcroft+2004} Adcroft, A., Campin, J.-M., Hill, C., \& Marshall, J.\ 2004, Monthly Weather Review, 132, 2845
\bibitem[Apai et al.(2013)]{apai+2013} Apai, D., Radigan, J., Buenzli, E., et al.\ 2013, \apj, 768, 121 
\bibitem[Barstow et al.(2014)]{barstow+2014} Barstow, J.~K., Aigrain, S., Irwin, P.~G.~J., et al.\ 2014, \apj, 786, 154 
\bibitem[Blecic et al.(2014)]{blecic+2014} Blecic, J., Harrington, J., Madhusudhan, N., et al.\ 2014, \apj, 781, 116
\bibitem[Buenzli et al.(2012)]{buenzli+2012} Buenzli, E., Apai, D., Morley, C.~V., et al.\ 2012, \apjl, 760, L31
\bibitem[Burrows et al.(1997)]{burrows+1997} Burrows, A., Marley, M., Hubbard, W.~B., et al.\ 1997, \apj, 491, 856 
\bibitem[Charbonneau et al.(2002)]{charbonneau+2002} Charbonneau, D., Brown, T.~M., Noyes, R.~W., \& Gilliland, R.~L.\ 2002, \apj, 568, 377
\bibitem[Charbonneau et al.(2008)]{charbonneau+2008} Charbonneau, D., Knutson, H.~A., Barman, T., et al.\ 2008, \apj, 686, 1341 %hd189 
\bibitem[Crossfield et al.(2012)]{crossfield+2012} Crossfield, I.~J.~M., Knutson, H., Fortney, J., et al.\ 2012, \apj, 752, 81 %hd209
\bibitem[Deming et al.(2013)]{deming+2013} Deming, D., Wilkins, A., McCullough, P., et al.\ 2013, \apj, 774, 95
\bibitem[D{\'e}sert et al.(2008)]{desert+2008} D{\'e}sert, J.-M., Vidal-Madjar, A., Lecavelier Des Etangs, A., et al.\ 2008, \aap, 492, 585 %hd209
\bibitem[Dobbs-Dixon \& Agol(2013)]{dobbsdixon+agol_2013} Dobbs-Dixon, I., \& Agol, E.\ 2013, \mnras, 435, 3159 
\bibitem[Ehrenreich et al.(2014)]{ehrenreich+2014} Ehrenreich, D., Bonfils, X., Lovis, C., et al.\ 2014, arXiv:1405.1056
\bibitem[Fortney et al.(2005)]{fortney+2005} Fortney, J.~J., Marley, M.~S., Lodders, K., Saumon, D., \& Freedman, R.\ 2005, \apj, 627, L69
\bibitem[Fortney et al.(2006)]{fortney+2006} Fortney, J.~J., Cooper, C.~S., Showman, A.~P., Marley, M.~S., \& Freedman, R.~S.\ 2006, \apj, 652, 746
\bibitem[Fortney et al.(2008)]{fortney+2008} Fortney, J.~J., Lodders, K., Marley, M.~S., \& Freedman, R.~S.\ 2008, \apj, 678, 1419
\bibitem[Gillon et al.(2012)]{gillon+2012} Gillon, M., Triaud, A.~H.~M.~J., Fortney, J.~J., et al.\ 2012, \aap, 542, A4
\bibitem[Goody et al.(1989)]{goody+1989} Goody, R., West, R., Chen, L., \& Crisp, D.\ 1989, \jqsrt, 42, 539
%\bibitem[Chen et al.(2014)]{2014A&A...563A..40C} Chen, G., van Boekel, R., Wang, H., et al.\ 2014, \aap, 563, A40
%\bibitem[Czesla et al.(2013)]{2013A&A...560A..17C} Czesla, S., Salz, M., Schneider, P.~C., \& Schmitt, J.~H.~M.~M.\ 2013, \aap, 560, A17
\bibitem[Hellier et al.(2011)]{hellier+2011} Hellier, C., Anderson, D.~R., Collier Cameron, A., et al.\ 2011, \aap, 535, L7
\bibitem[Heng et al.(2011)]{heng+2011} Heng, K., Frierson, D.~M.~W., \& Phillipps, P.~J.\ 2011, \mnras, 418, 2669 
\bibitem[Hubeny et al.(2003)]{hubeny+2003} Hubeny, I., Burrows, A., \& Sudarsky, D.\ 2003, \apj, 594, 1011 
\bibitem[Kataria et al.(2013)]{kataria+2013} Kataria, T., Showman, A.~P., Lewis, N.~K., et al.\ 2013, \apj, 767, 76
\bibitem[Kataria et al.(2014)]{kataria+2014} Kataria, T., Showman, A.~P., Fortney, J.~J., Marley, M.~S., \& Freedman, R.~S.\ 2014, \apj, 785, 92
\bibitem[Knutson et al.(2007)]{knutson+2007} Knutson, H.~A., et al.\ 2007, \nat, 447, 183
\bibitem[Knutson et al.(2008)]{knutson+2008} Knutson, H.~A., Charbonneau, D., Allen, L.~E., Burrows, A., \& Megeath, S.~T.\ 2008, \apj, 673, 526
\bibitem[Knutson et al.(2014a)]{knutson+2014a} Knutson, H.~A., Benneke, B., Deming, D., \& Homeier, D.\ 2014, \nat, 505, 66
\bibitem[Knutson et al.(2014b)]{knutson+2014b} Knutson, H.~A., Dragomir, D., Kreidberg, L., et al.\ 2014, arXiv:1403.4602
\bibitem[Konopacky et al.(2013)]{konopacky+2013} Konopacky, Q.~M., Barman, T.~S., Macintosh, B.~A., \& Marois, C.\ 2013, Science, 339, 1398
\bibitem[Kreidberg et al.(2014a)]{kreidberg+2014a} Kreidberg, L., Bean, J.~L., D{\'e}sert, J.-M., et al.\ 2014, \nat, 505, 69
\bibitem[Kreidberg et al.(2014b)]{kreidberg+2014b} Kreidberg, L., Bean, J.~L., D{\'e}sert, J.-M., et al.\ 2014, \apjl, 793, L27 
\bibitem[Lewis et al.(2010)]{lewis+2010} Lewis, N.~K., Showman, A.~P., Fortney, J.~J., Marley, M.~S., Freedman, R.~S., \& Lodders, K.\ 2010, \apj, 720, 344
\bibitem[Lewis et al.(2013)]{lewis+2013} Lewis, N.~K., Knutson, H.~A., Showman, A.~P., et al.\ 2013, \apj, 766, 95 
\bibitem[Lewis et al.(2014)]{lewis+2014} Lewis, N.~K., Showman, A.~P., Fortney, J.~J., Knutson, H.~A., \& Marley, M.~S.\ 2014, arXiv:1409.5108 
%\bibitem[Maciejewski et al.(2013)]{2013IBVS.6082....1M} Maciejewski, G., Puchalski, D., Saral, G., et al.\ 2013, Information Bulletin on Variable Stars, 6082, 1
\bibitem[Line et al.(2013)]{line+2013} Line, M.~R., Knutson, H., Deming, D., Wilkins, A., \& Desert, J.-M.\ 2013, \apj, 778, 183
\bibitem[Line et al.(2014)]{line+2014} Line, M.~R., Knutson, H., Wolf, A.~S., \& Yung, Y.~L.\ 2014, \apj, 783, 70 
\bibitem[Liu \& Showman(2013)]{liu+showman_2013} Liu, B., \& Showman, A.~P.\ 2013, \apj, 770, 42
\bibitem[Lodders(2003)]{lodders2003} Lodders, K.\ 2003, \apj, 591, 1220
\bibitem[Majeau et al.(2012)]{majeau+2012} Majeau, C., Agol, E., \& Cowan, N.~B.\ 2012, \apjl, 747, L20 
\bibitem[Mandell et al.(2013)]{mandell+2013} Mandell, A.~M., Haynes, K., Sinukoff, E., et al.\ 2013, \apj, 779, 128
\bibitem[Marley \& McKay(1999)]{marley+mckay_1999} Marley, M.~S., \& McKay, C.~P.\ 1999, \icarus, 138, 268
\bibitem[Morley et al.(2012)]{morley+2012} Morley, C.~V., Fortney, J.~J., Marley, M.~S., Visscher, C., Saumon, D., \& Leggett, S.~K., et al.\ 2012, \apj, 756, 172
%\bibitem[Murgas et al.(2014)]{2014A&A...563A..41M} Murgas, F., Pall{\'e}, E., Zapatero Osorio, M.~R., et al.\ 2014, \aap, 563, A41
\bibitem[Parmentier et al.(2013)]{parmentier+2013} Parmentier, V., Showman, A.~P., \& Lian, Y.\ 2013, \aap, 558, A91
\bibitem[Perez-Becker \& Showman(2013)]{perezbecker+showman_2013} Perez-Becker, D., \& Showman, A.~P.\ 2013, \apj, 776, 134
\bibitem[Perna et al.(2010)]{perna+2010} Perna, R., Menou, K., \& Rauscher, E.\ 2010, \apj, 719, 1421 
\bibitem[Pont et al.(2008)]{pont+2008} Pont, F., Knutson, H., Gilliland, R.~L., Moutou, C., \& Charbonneau, D.\ 2008,
\mnras, 385, 109
\bibitem[Quintana et al.(2014)]{quintana+2014} Quintana, E.~V., Barclay, T., Raymond, S.~N., et al.\ 2014, Science, 344, 277
\bibitem[Ranjan et al.(2014)]{ranjan+2014} Ranjan, S., Charbonneau, D., D{\'e}sert, J.-M., et al.\ 2014, \apj, 785, 148
\bibitem[Rauscher \& Menou(2012)]{rauscher+menou_2012} Rauscher, E., \& Menou, K.\ 2012, \apj, 750, 96 
\bibitem[Rogers \& Showman(2014)]{rogers+showman_2014} Rogers, T.~M., \& Showman, A.~P.\ 2014, \apjl, 782, L4 
\bibitem[Showman \& Guillot(2002)]{showman+guillot_2002} Showman, A.~P., \& Guillot, T.\ 2002, \aap, 385, 166 
\bibitem[Showman et al.(2009)]{showman+2009} Showman, A.~P., Fortney, J.~J., Lian, Y., Marley, M.~S., Freedman, R.~S., Knutson, H.~A., \& Charbonneau, D.\ 2009, \apj, 699, 564
\bibitem[Showman \& Polvani(2011)] {showman+polvani_2011}Showman, A.~P., \& Polvani, L.~M.\ 2011, \apj, 738, 71 
\bibitem[Showman et al.(2013)]{showman+2013} Showman, A.~P., Fortney, J.~J., Lewis, N.~K., \& Shabram, M.\ 2013, \apj, 762, 24 
\bibitem[Showman et al.(2014)]{showman+2014} Showman, A.~P., Lewis, N.~K., \& Fortney, J.~J.\ 2014, arXiv:1411.4731 
\bibitem[Sing et al.(2008)]{sing+2008} Sing, D.~K., Vidal-Madjar, A., D{\'e}sert, J.-M., Lecavelier des Etangs, A., \& Ballester, G.\ 2008, \apj, 686, 658 %hd209 
\bibitem[Snellen et al.(2014)]{snellen+2014} Snellen, I.~A.~G., Brandl, B.~R., de Kok, R.~J., et al.\ 2014, \nat, 509, 63
\bibitem[Spiegel et al.(2009)]{spiegel+2009} Spiegel, D.~S., Silverio, K., \& Burrows, A.\ 2009, \apj, 699, 1487 
\bibitem[Stevenson et al.(2014)]{stevenson+2014} Stevenson, K.~B., D\'esert, J-M, Line, M.~R., et al.\ 2014, Science, in press
\bibitem[Tsai et al.(2014)]{tsai+2014} Tsai, S.-M., Dobbs-Dixon, I., \& Gu, P.-G.\ 2014, \apj, 793, 141 
\bibitem[Wakeford et al.(2013)]{wakeford+2013} Wakeford, H.~R., Sing, D.~K., Deming, D., et al.\ 2013, \mnras, 435, 3481
\bibitem[Wang et al.(2013)]{wang+2013} Wang, W., van Boekel, R., Madhusudhan, N., et al.\ 2013, \apj, 770, 70 
\bibitem[Wilkins et al.(2014)]{wilkins+2014} Wilkins, A.~N., Deming, D., Madhusudhan, N., et al.\ 2014, \apj, 783, 113
\bibitem[Zahnle et al.(2009)]{zahnle+2009} Zahnle, K., Marley, M.~S., Freedman, R.~S., Lodders, K., 
\& Fortney, J.~J.\ 2009, \apjl, 701, L20 
%\bibitem[Wang et al.(2013)]{2013ApJ...770...70W} Wang, W., van Boekel, R., Madhusudhan, N., et al.\ 2013, \apj, 770, 70
\bibitem[Zhou et al.(2014)]{zhou+2014} Zhou, G., Bayliss, D.~D.~R., Kedziora-Chudczer, L., et al.\ 2014, arXiv:1409.2775 
\end{thebibliography}

\end{document}